\documentclass{SCIS2021}
\DeclareGraphicsExtensions{.eps,.pdf,.jpeg,.png}
\usepackage{epstopdf}
\usepackage{makecell} 

\newcommand{\ba}{\begin{array}}
\newcommand{\ea}{\end{array}}
\newcommand{\be}{\begin{displaymath}}
\newcommand{\ee}{\end{displaymath}}
\newcommand{\ben}{\begin{equation}}
\newcommand{\een}{\end{equation}}
\newcommand{\bena}{\begin{eqnarray}}
\newcommand{\eena}{\end{eqnarray}}
\newcommand{\beqa}{\begin{eqnarray*}}
\newcommand{\enqa}{\end{eqnarray*}}

\newcommand{\bc}{\begin{center}}
\newcommand{\ec}{\end{center}}
\newcommand{\bi}{\begin{itemize}}
\newcommand{\ei}{\end{itemize}}
\newcommand{\benu}{\begin{enumerate}}
\newcommand{\eenu}{\end{enumerate}}
\newcommand{\bdes}{\begin{description}}
\newcommand{\edes}{\end{description}}
\newcommand{\bt}{\begin{tabular}}
\newcommand{\et}{\end{tabular}}

\newcommand \thetabf{{\mbox{\boldmath$\theta$\unboldmath}}}
\newcommand{\Phibf}{\mbox{${\bf \Phi}$}}

\newcommand \alphabf{\mbox{\boldmath$\alpha$\unboldmath}}

\newcommand \phibf{\mbox{\boldmath$\phi$\unboldmath}}

\newcommand \Thetabf{\hbox{$\bf \Theta$}}

\newcommand \abf{{\bf a}}

\newcommand \jbf{{\bf j}}

\newcommand \nbf{{\bf n}}

\newcommand \qbf{{\bf q}}

\newcommand \sbf{{\bf s}}

\newcommand \vbf{{\bf v}}
\newcommand \wbf{{\bf w}}
\newcommand \xbf{{\bf x}}

\newcommand \Cbf{{\bf C}}

\newcommand \Hbf{{\bf H}}
\newcommand \Ibf{{\bf I}}
\newcommand \Jbf{{\bf J}}

\newcommand \Nbf{{\bf N}}

\newcommand \Pbf{{\bf P}}

\newcommand \Rbf{{\bf R}}
\newcommand \Sbf{{\bf S}}

\newcommand \Xbf{{\bf X}}

\newcommand{\circlambda}{\mbox{$\Lambda$
             \kern-.85em\raise1.5ex
             \hbox{$\scriptstyle{\circ}$}}\,}

\renewcommand \thetabf{\boldsymbol{\theta}}
\renewcommand \Phibf{\boldsymbol{\Phi}}

\renewcommand \alphabf{\boldsymbol{\alpha}}

\renewcommand \phibf{\boldsymbol{\phi}}

\renewcommand \Thetabf{\boldsymbol{\Theta}}

\begin{document}
\ArticleType{RESEARCH PAPER}
\Year{2020}
\Month{}
\Vol{}
\No{}
\DOI{}
\ArtNo{}
\ReceiveDate{}
\ReviseDate{}
\AcceptDate{}
\OnlineDate{}

\title{Multichannel adaptive signal detection: Basic theory and literature review}{Multichannel adaptive signal detection: Basic theory and literature review}

\author[1]{Weijian LIU}{}
\author[2]{Jun LIU}{}
\author[3]{Chengpeng HAO}{}
\author[4]{Yongchan GAO}{}
\author[1]{Yong-Liang WANG}{ylwangkjld@163.com}
\AuthorMark{Liu W}

\AuthorCitation{Liu W, Liu J, Hao C, et al}


\address[1]{Wuhan Electronic Information Institute, Wuhan {\rm 430019}, China}
\address[2]{Department of Electronic Engineering and Information Science, University of Science and Technology of China, Hefei {\rm 230027}, China}
\address[3]{State Key Laboratory of Acoustics, Institute of Acoustics, Chinese Academy of Sciences, Beijing {\rm 100190}, China}
\address[4]{Xidian University, Xi'an {\rm 710071}, China}
\abstract{Multichannel adaptive signal detection jointly uses the test and training data to form an adaptive detector, and then make a decision on whether a target exists or not. Remarkably, the resulting adaptive detectors usually possess the constant false alarm rate (CFAR) properties, and hence no additional CFAR processing is needed. Filtering is not needed as a processing procedure either, since the function of filtering is embedded in the adaptive detector. Moreover, adaptive detection usually exhibits better detection performance than the filtering-then-CFAR detection technique. Multichannel adaptive signal detection has been more than 30 years since the first multichannel adaptive detector was proposed by Kelly in 1986. However, there are fewer overview articles on this topic. In this paper we give a tutorial overview of multichannel adaptive signal detection, with emphasis on Gaussian background. We present the main deign criteria for adaptive detectors, investigate the relationship between adaptive detection and filtering-then-CFAR detection, relationship between adaptive detectors and adaptive filters, summarize typical adaptive detectors, show numerical examples, give comprehensive literature review, and discuss some possible further research tracks. }


\maketitle

\section{Introduction}


Signal detection in noise is a fundamental problem in various areas, such as radar, sonar, communications, optical image, hyperspectral imagery, remote sensing, medical imaging, subsurface prospecting, and so on.
Taking the radar system for example, the received data for early radar systems are of single channel, and hence, the data are scalar-valued. In contrast, with the applications of pulsed Doppler techniques and/or multiple transmit/receive (T/R) modules, along with the increase in computation power and advances in hardware design, the received data for modern radar systems are usually multichannel, namely, vector-valued or even matrix-valued.
Moreover, the frequency diversity, polarization diversity, or waveform diversity can also lead to the multichannel form of the received data.
The multichannel data contain more information, compared with the single-channel data. On one hand, using the multichannel data, 
we have more degrees of freedom (DOFs) to design adaptive processors. 
On the other hand, using the multichannel data model, it is more convenient to characterize the correlated properties between data in different channels. Using these correlated properties, one can design a filter, whose output signal-to-noise (SNR) is often higher than that for a single-channel data. Similarly, utilizing the data correlation, one can devise a detector, which has superior detection performance to a detector for single-channel data. 

Remarkably, noise is ubiquitous, which, in a general sense, usually includes thermal noise and clutter.
For multichannel data in the cell under test (also called primary data), the noise covariance matrix is unknown and needs to be estimated. A common strategy is using the training data (also called secondary data) to form appropriate estimator.
It is pointed out in \cite{GiniGreco99b} that modern strategy for radar detection should include the following three features: 1) being adaptive 
to the noise spectral density or its probability density function (PDF), 2) maintaining constant false alarm rate (CFAR) property, and 3) having a relatively simple processing scheme.
Multichannel adaptive signal detection is a kind of this strategy. It jointly utilizes the test data and training data to design adaptive detectors, which usually possess the CFAR property. 
The resulting adaptive detector is then compared with a certain detection threshold, set to ensure a fixed probability of false alarm (PFA). Finally, a target is declared to be present (absent) if the threshold is exceeded (not exceeded). 

Two points are worth to be emphasized. One is that the word ``adaptive'' in the first feature above indicates that the spectrum character of the noise 
is unknown in advance or is changing in the operational environment, and hence adaptive techniques are needed.
The other is that the CFAR property or the CFARness\footnote{CFARness is an important property required by an effective detector in practice, because the PFA may be dramatically raised to an unaffordable value if a detector does not maintain CFARness and the noise changes severely.}, which, for single-channel signal, means that the detection threshold of a detector is independent of the noise power. Equivalently, the statistical property of the detector is functionally independent of the noise power under the signal-absence hypothesis. In contrast, for multichannel signal detection, the CFARness means that the statistical property of the detector is also functionally independent of the structure of the noise covariance matrix under the signal-absence hypothesis. This kind of CFARness is referred to as the matrix CFAR in \cite{ChongPascal10JSTSP} and covariance matrix-CFAR in \cite{GinolhacForster14}. 

Multichannel adaptive signal detection was first investigated by Kelly in 1986. In the seminal paper \cite{Kelly86}, Kelly proposed the famous detector, i.e., Kelly's GLRT (KGLRT) for detecting a rank-one signal in homogeneous environment (HE). The rank-one signal has a known steering vector but unknown amplitude. For the HE model, the noise in the training and test data is both subject to mean-zero circularly complex Gaussian distribution, with the same covariance matrix. 

There is more than three decades since Kelly proposed the famous KGLRT in 1986. Multichannel adaptive signal detection has been adopted in various areas. Based on different design criteria, numerous detectors have been proposed for different problems. Recently, an important book is edited by De Maio and Greco \cite{DeMaioGreco16Book}.
However, there are seldom survey papers on multichannel signal detection. In particular, references \cite{GiniFarina02b} and \cite{SangstonFarina16} gave overview of signal detection in compound-Gaussian clutter for subspace signals and rank-one signals, respectively. These two references are mainly on known clutter or known noise covariance matrix. Moreover, the target is point-like and no signal mismatch is considered.
Different from the above two references, in this paper we give a review of multichannel adaptive signal detection in unknown noise, with emphasis on Gaussian background.

In this paper, we give a tutorial on multichannel adaptive signal detection, and present a brief survey of the state of the art.
For brevity, ``adaptive detection'' always means ``multichannel adaptive signal detection'' in the following.
In Section 2, we present the basic theory for adaptive detection, including data model, main detector design criteria, relationship between adaptive detection and filter-then-CFAR detection, and relationship between adaptive detection and adaptive filtering.
In Section 3, we give comprehensive literature review.
In Section 4, we analyse and compare the detection performance of some typical adaptive detectors.
Finally, Section 5 summarizes this paper and gives some further research tracks in adaptive detection. 

\section{Basic theory}
\subsection{Main detector design criteria}

The GLRT, Rao test, and Wald test are three main detector design criteria\footnote{There are also some other often used criteria, such as the gradient test \cite{Lemonte16}, Durbin test \cite{Durbin70}, test based on maximal invariant statistic \cite{Scharf91}, multifamily likelihood ratio test \cite{Kay05}, and other modifications of the likelihood ratio test \cite{Kay05}, which are utilized for adaptive detector design, e.g., \cite{AbramovichSpencer07a,CarotenutoDeMaio15GRS,CarotenutoDeMaio15GRSL,CarotenutoDeMaio16GRS,CiuonzoCarotenuto17, DeMaioDeNicola09a,DeMaioOrlando17GRS,DeMaioHan18,GerlachSteiner99,DeMaio02b,DeMaioFarina07,AubryDeMaio14b, AubryDeMaio15TSP,RongAubry20TSP}.}.
These three criteria are referred to as ``the Holy Trinity'' in statistical inference \cite{Rao05ScoreTest}.
Before listing these criteria, we need to formulate a binary hypothesis mathematically.
A binary hypothesis has two possible cases, namely, the null (signal-absence) hypothesis and alternative (signal-presence) hypothesis.  Hence, a binary hypothesis test can be written as
\begin{equation}
\label{ch02_1}
\left\{ {\begin{array}{l}
	\text{H}_0 :{ {{ \xbf}}}={ { \nbf}} , ~\xbf_{\text{e},l}=\nbf_{\text{e},l},~ l=1,2,\cdots,L,\\
	\text{H}_1 :{ {{\xbf}}}= {\sbf}+{ { \nbf}} , ~\xbf_{\text{e},l}=\nbf_{\text{e},l},~ l=1,2,\cdots,L,\\ 
	\end{array}} \right.
\end{equation}
where $\text{H}_0$ denotes the null hypothesis, $\text{H}_1$ denotes the alternative hypothesis, $ {\xbf}$ is the test data, $ {\sbf}$ is the signal to be detected,  $ {\nbf}$ is the noise in the test data, whose covariance matrix, denoted as $\Rbf$, is generally unknown,  $\{\xbf_{\text{e},l}\}_{l=1}^{L}$ are $L$ training data, used to estimate the unknown $\Rbf$.

For the detection problem in \eqref{ch02_1}, the GLRT is \cite{Kay98}
\begin{equation}
\label{ch02_GLRT}
t_{\normalfont{\text{GLRT}}}=\frac{\mathop {\max }\limits_{\Thetabf_1}~f_1({  {\mathbf {x},\Xbf_L}})}{\mathop {\max }\limits_{\Thetabf_0}~f_0({  {\mathbf {x},\Xbf_L}})},
\end{equation}
where  ${\Thetabf_1}$ and ${\Thetabf_0}$ denote the unknown parameters under hypotheses ${\text{H}}_1$ and ${\text{H}}_0$, respectively, $f_1(\xbf,\Xbf_L)$ and $f_0(\xbf,\Xbf_L)$ are the joint PDFs of the test data $\xbf$ and training data  $\Xbf_L=[\xbf_\text{e,1},\xbf_\text{e,2},\cdots,\xbf_{\text{e},L}]$ under hypotheses $\text{H}_1$ and ${\text{H}}_0$, respectively.

To derive the Rao and Wald tests, we need the Fisher information matrix (FIM), which, for circularly symmetric random parameters, is defined as \cite{PagadaraiWyglinski11}
\begin{equation}
\label{ch02_4}
{\mathbf {I}}(\Thetabf )=\text{E}\left[ { {\dfrac{\partial \ln f(\mathbf{x},\Xbf_L)}{\partial \Thetabf ^\ast }}  {\dfrac{\partial \ln f(\mathbf{x},\Xbf_L)}{\partial \Thetabf ^T}} } \right].
\end{equation}
For convenience, the FIM is usually partitioned as
\begin{equation}
\label{ch02_5}
{\mathbf {I}}(\Thetabf )=\left[ {\begin{array}{*{20}c}{\mathbf {I}_{\Thetabf_{\text{r}} ,\Thetabf_{\text{r}} } (\Thetabf )} \hfill & {\mathbf {I}_{\Thetabf_{\text{r}} ,\Thetabf_{\text{s}}}(\Thetabf )} \hfill \\
	{\mathbf {I}_{\Thetabf_{\text{s}} ,\Thetabf_{\text{r}} } (\Thetabf )} \hfill & {\mathbf {I}_{\Thetabf_{\text{s}} ,\Thetabf_{\text{s}} }(\Thetabf )}
	\end{array} } \right],
\end{equation}
where
\begin{equation}
\label{ch02_Theta}
\Thetabf =[\Thetabf_{\text{r}}^T ,\Thetabf_{\text{s}}^T ]^T,
\end{equation}
\begin{subequations}
	\label{ch02_6}
\begin{equation}
\label{}
\mathbf {I}_{\Thetabf_{\text{r}} ,\Thetabf_{\text{r}} } (\Thetabf )=\text{E}\left[ {\dfrac{\partial \ln f(\mathbf{x},\Xbf_L)}{\partial\Thetabf_{\text{r}}^\ast}
	\dfrac{\partial \ln f(\mathbf{x},\Xbf_L)}{\partial \Thetabf_{\text{r}}^T}
} \right] ,
\end{equation}
\begin{equation}
\label{}
\mathbf {I}_{\Thetabf_{\text{r}} ,\Thetabf_{\text{s}} } (\Thetabf )=\text{E}\left[ {
	\dfrac{\partial \ln f(\mathbf{x},\Xbf_L)}{\partial \Thetabf_{\text{r}}^\ast}
	\dfrac{\partial \ln f(\mathbf{x},\Xbf_L)}{\partial \Thetabf_{\text{s}}^T}
} \right] ,
\end{equation}
\begin{equation}
\label{}
\mathbf {I}_{\Thetabf_{\text{s}} ,\Thetabf_{\text{r}} } (\Thetabf )=\text{E}\left[ {
	\dfrac{\partial \ln f(\mathbf{x},\Xbf_L)}{\partial \Thetabf_{\text{s}}^\ast}
	\dfrac{\partial \ln f(\mathbf{x},\Xbf_L)}{\partial \Thetabf_{\text{r}}^T}
} \right] ,
\end{equation}
\begin{equation}
\label{}
\mathbf {I}_{\Thetabf_{\text{s}} ,\Thetabf_{\text{s}} } (\Thetabf )=\text{E}\left[ {
	\dfrac{\partial \ln f(\mathbf{x},\Xbf_L)}{\partial \Thetabf_{\text{s}}^\ast}
	\dfrac{\partial \ln f(\mathbf{x},\Xbf_L)}{\partial \Thetabf_{\text{s}}^T}  } \right].
\end{equation}
\end{subequations}
In \eqref{ch02_Theta}, $\Thetabf_{\text{r}} $ is the relevant parameter, such as the signal amplitude, $\Thetabf_{\text{s}}$ is the nuisance parameter, e.g., the noise covariance matrix.
Note that if $\ln f({\mathbf {x}},\Xbf_L)$ is twice differential with respect to $\Thetabf$, then the FIM in \eqref{ch02_4}, under the regularity condition, can be calculated by \cite{LiuWang14}
\begin{equation}
\label{ch02_7}
{\mathbf {I}}(\Thetabf )=-\text{E}\left[ {\dfrac{\partial ^2\ln f(\mathbf{x},\Xbf_L)}{\partial \Thetabf^\ast \partial \Thetabf ^T}} \right],
\end{equation}
which is often more easier to be derived.

Then, the Rao and Wald tests for complex-valued signals are \cite{LiuWang14}\footnote{The complex-valued Rao test is also given in \cite{KayZhu16TSP} which is a generalization of the one in \eqref{ch02_thrm:RaoWald} and suitable of non-circularly symmetric random parameters.}
\begin{equation}
\label{ch02_thrm:RaoWald}
\begin{aligned}
t_{\normalfont{\text{Rao}}} =\left. {\dfrac
	{\partial \ln f_1 ({\mathbf {x}},\Xbf_L)}{\partial \Thetabf_{\text{r}} }}
\right|_{\Thetabf =\hat{\Thetabf }_0 }^T
 [\mathbf {I}^{-1}(\hat{\Thetabf }_0
)]_{\Thetabf_{\text{r}} ,\Thetabf_{\text{r}} } \left. {\dfrac{\partial \ln f_1 ({\mathbf {x},\Xbf_L} )}{\partial \Thetabf_{\text{r}}^\ast }} \right|_{\Thetabf =\hat{\Thetabf }_0 },
\end{aligned}
\end{equation}
and
\begin{equation}
\label{ch02_11}
t_{\normalfont{}\text{Wald}} =(\hat{\Thetabf }_{\text{r}_1} -\Thetabf _{\text{r}_0} )^H\left\{ {[\mathbf {I}^{-1}(\hat{\Thetabf }_1 )]_{\Thetabf_{\text{r}} ,\Thetabf_{\text{r}} } }
\right\}^{-1}(\hat{\Thetabf }_{\text{r}_1} -\Thetabf _{\text{r}_0} ),
\end{equation}
respectively, where $\hat{\Thetabf }_0 $ and $\hat{\Thetabf }_1 $ are the maximum likelihood estimates (MLEs) of $\Thetabf $ under hypotheses $\normalfont{\text{H}}_0$ and $\normalfont{\text{H}}_1$, respectively, $\hat{\Thetabf }_{\text{r}_1} $ is the MLE of $\Thetabf_{\text{r}} $ under hypothesis $\normalfont{\text{H}}_1$, $\Thetabf _{\text{r}_0} $ is the value of $\Thetabf_{\text{r}} $ under hypothesis $\normalfont{\text{H}}_0$, and $\left\{ {[\mathbf {I}^{-1}(\Thetabf )]_{\Thetabf_{\text{r}},\Thetabf_{\text{r}} } } \right\}^{-1}$ is the Schur complement of $\mathbf {I}_{\Thetabf_{\text{s}} ,\Thetabf_{\text{s}} } (\Thetabf )$, namely,
\begin{equation}
\label{ch02_12}
\begin{aligned}
\left\{ {[\mathbf {I}^{-1}(\Thetabf )]_{\Thetabf_{\text{r}} ,\Thetabf_{\text{r}} } } \right\}^{-1}= \mathbf {I}_{\Thetabf_{\text{r}} ,\Thetabf_{\text{r}} } (\Thetabf )
 -\Ibf_{\Thetabf_{\text{r}} ,\Thetabf_{\text{s}} } (\Thetabf )~\mathbf {I}_{\Thetabf_{\text{s}}
	,\Thetabf_{\text{s}} }^{-1} (\Thetabf )~\mathbf {I}_{\Thetabf_{\text{s}} ,\Thetabf_{\text{r}} } (\Thetabf ).
\end{aligned}
\end{equation}
In some cases the relevant parameter $\Thetabf_{\text{r}} $ and/or the nuisance parameter $\Thetabf_{\text{s}}$ may be known. Obviously, in these cases we use these true values, and do not need to derive their MLEs.

It is worthy pointing out that the two-step variations of the three design criteria are also adopted. Precisely, the GLRT, Rao test, or Wald test is first derived under the assumption that the noise covariance matrix is known or its structure is known. Then the noise covariance matrix in the corresponding detector is replaced by a proper estimate by using the training data. For example, the two-step GLRT (2S-GLRT) can be mathematically expressed as
\begin{equation}
\label{ch02_2SGLRT}
t_{\normalfont{\text{2S-GLRT}}}=\left.\left[\frac{\mathop {\max }\limits_{\Thetabf_1^\prime}~f_1({  {\mathbf {x},\Xbf_L}})} {\mathop {\max }\limits_{\Thetabf_0^\prime}~f_0({  {\mathbf {x},\Xbf_L}})}\right]\right|_{\Rbf=\hat{\Rbf}},
\end{equation}
where $\Thetabf_1^\prime$ and $\Thetabf_0^\prime$ denote the unknown parameters except for $\Rbf$ under hypotheses ${\text{H}}_1$ and ${\text{H}}_0$, respectively, and $\hat{\Rbf}$ is an appropriate estimation of $\Rbf$.

From the three detector design criteria in \eqref{ch02_GLRT}, \eqref{ch02_thrm:RaoWald} and \eqref{ch02_11}, we know that one of the key point to how to find the derivatives of scalar real-valued functions, such as the PDFs, with respect to a complex-valued scalar, vector, or matrix. One of the most useful book on this topic may be the one by Hj{\o}rungnes \cite{Hjorungnes11}, which is written in engineering-oriented manner. 
The theory of finding complex-valued derivatives in \cite{Hjorungnes11} is based on the complex differential of the objective function. 
Using the complex differential is much more easier to find a derivative than using the component-wise approach, such as the famous book by Magnus and Neudecker \cite{MagnusNeudecker07}, which mainly focuses on real-valued derivatives. 

It is worthy pointing out that the following fact is often used in deriving a  detector with simplified detection statistic or in a form whose statistical distribution is easy to be derived. Precisely, if a detector can be expressed as a monotonically increasing function of another one, then these two detectors are equivalent. We try to find a related reference. However, it is not found. Hence, we summarize the above fact in the following theorem.

\textbf{Theorem 1:}  Let $t_1$ and $t_2$ are two detectors, and
\begin{equation}
\label{ch02_ggt1t2}
t_2=g(t_1)
\end{equation}
monotonically increases with $t_1$. Then $t_1$ are $t_2$ have the same detection performance such that they have the identical probability of detection (PD) under the same PFA. 

\noindent
\textbf{Proof:} Let the PFAs of $t_1$ and $t_2$ be $\text{PFA}_1$ and $\text{PFA}_2$, respectively. Then
\begin{equation}
\label{ch02_PFA1}
\text{PFA}_1=\text{Pr}[t_1>\eta_1;\text{H}_0],
\end{equation}
\begin{equation}
\label{ch02_PFA2}
\text{PFA}_2=\text{Pr}[t_2>\eta_2;\text{H}_0],
\end{equation}
where $\eta_1$ and $\eta_2$ are detection thresholds of $t_1$ and $t_2$, respectively. According to \eqref{ch02_ggt1t2}, \eqref{ch02_PFA2} can be rewritten as
\begin{equation}
\label{ch02_PFA2a}
\text{PFA}_2=\text{Pr}[g(t_1)>\eta_2;\text{H}_0]=\text{Pr}[t_1>g^{-1}(\eta_2);\text{H}_0],
\end{equation}
where the second equality is owing to the fact that $g(t_1)$ is a momotonically increasing function of $t_1$, and $g^{-1}(\cdot)$ denotes the inverse function of $g(\cdot)$.
Comparing \eqref{ch02_PFA1} and \eqref{ch02_PFA2a}, and using $\text{PFA}_1=\text{PFA}_2$, we have
\begin{equation}
\label{ch02_eta1122}
\eta_1=g^{-1}(\eta_2).
\end{equation}
The PDs of $t_1$ and $t_2$ can be expressed as
\begin{equation}
\label{ch02_PD1}
\text{PD}_1=\text{Pr}[t_1>\eta_1;\text{H}_1]
\end{equation}
and
\begin{equation}
\label{ch02_PD2}
\text{PD}_2=\text{Pr}[t_2>\eta_2;\text{H}_1],
\end{equation}
respectively. Since $t_2=g(t_1)$ is a monotonically increasing function of $t_1$, \eqref{ch02_PD2} can be recast as
\begin{equation}
\label{ch02_PD2a}
\text{PD}_2=\text{Pr}[t_1>g^{-1}(\eta_2);\text{H}_1]=\text{Pr}[t_1>\eta_1;\text{H}_1]=\text{PD}_1,
\end{equation}
where the second equality is obtained according to \eqref{ch02_eta1122}. This completes the proof. $\hfill \blacksquare$ %

Adaptive detection is different from filtering-then-CFAR detection, which is widely adopted in most radar systems. Moreover, adaptive detection is highly related with adaptive filtering, although their purposes are different. In the following two subsections, we investigate the relationship between them.

\subsection{Relationship between adaptive detection and filtering-then-CFAR detection}
Nowadays, the mainly used detection scheme in most radar systems is the filtering-then-CFAR approach. Precisely, the test data are first filtered and then processed by the CFAR techniques. The CFAR processing is a technique which makes the detection threshold of a detector independent of noise covariance matrix. Or, equivalently, through CFAR processing, the statistical characteristics of the detector does not depend on the noise covariance matrix under the signal-absence hypothesis. There are many CFAR technologies, such as cell-averaging CFAR (CA-CFAR), 
greatest-of-selection CFAR (GO-CFAR), ordered statistic CFAR (OS-CFAR), and so on \cite{Richards10Book,Weinberg17Book}. 
It seems that the filtering-then-CFAR detection scheme is a natural approach for detecting a target in noise, since adaptive filtering can obtain high output SNR, which benefits the detection process.

The theoretical basis behind the filtering-then-CFAR detection scheme for multichannel data can be traced back to the classic paper \cite{BrennanReed73}.
Precisely, for airborne radar space-time two-dimensional signal processing, the test data, if containing the target signal, can be written as
\begin{equation}
\label{test}
\xbf=a\sbf+\nbf,
\end{equation}
where $\xbf$ is an $N_aN_p\times1$ test data vector,  $N_a$ is the number of the antennas, $N_p$ is the number of the pulses received by each antenna,  $\sbf=\sbf_{\text{p}}\otimes\sbf_{\text{a}} $ is an $N_aN_p\times1$ signal space-time steering vector, with $\sbf_{\text{p}}$ and $\sbf_{\text{a}}$ being an $N_p\times1$ time steering vector and an $N_a\times1$  space steering vector, respectively, $\otimes$ denotes the Kronecker product, and $\nbf$ is the noise, including clutter and thermal noise, distributed as circularly complex Gaussian distribution with covariance matrix $\Rbf$.

In \cite{BrennanReed73}, to detect the target in \eqref{test}, the test data vector $\xbf$ is first filtered by an $N_aN_p\times1$ weight vector $\wbf$. Hence, the output of the filter can be expressed as
\begin{equation}
\label{FltOutpt}
y=\wbf^H\xbf.
\end{equation}
For the filtered data $ y $, the optimum detector, in the Neyman-Pearson sense, is the likelihood ratio test, given by
\begin{equation}
\label{LRT}
t_{\text{LRT}}=\frac{f_1(y|\xbf=a\sbf+\nbf)} {f_0(y|\xbf=\nbf)},
\end{equation}
where $f_1(\cdot)$ and $f_0(\cdot)$ are the PDFs under signal-presence and signal-absence hypotheses, respectively. The optimum filter weight $\wbf$ can be obtained by maximizing \eqref{LRT}, written symbolically as
\begin{equation}
\label{LRTopt}
\wbf_\text{opt}=\mathop {\max }\limits_{\bf{w}} \frac{f_1(y|\xbf=a\sbf+\nbf)} {f_0(y|\xbf=\nbf)},
\end{equation}
which is shown to be equivalent to \cite{BrennanReed73}
\begin{equation}
\label{LRTopt2}
\wbf_\text{opt}=\mathop {\max }\limits_{\bf{w}} \frac{|\wbf^H\sbf|^2} {\wbf^H\Rbf\wbf},
\end{equation}
and the solution to \eqref{LRTopt2} is
\begin{equation}
\label{LRTopt3}
\wbf_\text{opt}=\mu\Rbf^{-1}\sbf,
\end{equation}
where $ \mu $ is an arbitrary non-zero constant.

A well-known equivalent solution to \eqref{LRTopt2} is the minimum variance distortionless response (MVDR), which is mathematically formed as \cite{LiStoica06book}
\begin{equation}
\label{Plm_mvdr}
\left\{ \begin{array}{l}
\mathop {\min }\limits_{\bf{w}} {\bf{w}^H\Rbf\wbf},\\
\text{s.t.}{\kern 4pt} {\kern 1pt} {\bf{w}^H\sbf} = 1,
\end{array} \right.
\end{equation}
and the corresponding solution is
\begin{equation}
\label{MVDR}
\wbf_\text{MVDR}=\frac{\Rbf^{-1}\sbf}{\sbf^H\Rbf^{-1}\sbf}.
\end{equation}
Taking \eqref{MVDR} into \eqref{FltOutpt} and performing the norm-squared operation leads to
\begin{equation}
\label{MF}
t_\text{MF}=\frac{|\sbf^H\Rbf^{-1}\xbf|^2}{(\sbf^H\Rbf^{-1}\sbf)^2}.
\end{equation}


Gathering the above results indicates that the optimum detection in \eqref{LRT} is equivalent to the optimum filtering in \eqref{LRTopt2}, and the optimum filter weight is given in \eqref{LRTopt3}. 
Based on the above results, a technique called space-time adaptive processing (STAP) came into being, which is regarded as one of most effective technology for airborne radar clutter cancellation, and numerous achievements have been obtained \cite{WangPeng00Egls,Klemm06,Guerci15}. 
Note that STAP is a filtering technique\footnote{Strictly speaking, the STAP technique is much less than its literal meaning. Precisely, STAP is a filtering technique to reject the clutter and jammer (if present) for airborne radar \cite{Ward94,Melvin04Overview}.
}, whose aim is to maximize the output SNR. To realize the final target detection, CFAR processing is needed. 

It is worth pointing out that the above equivalence between optimum detection and optimum filtering holds under \textit{certain processing flow} and \textit{certain assumptions}. The specific processing flow is filtering-then-detection. Precisely, the multichannel test data vector $\xbf$ is first filtered by the weight vector $\wbf$, resulting in the scalar-valued data $y$. Then, a detector is devised based on the filtered data $y$.
The assumption is that the noise $\nbf$ in the test data is Gaussian distributed, and its covariance matrix $ \Rbf $ is known in advance. Unfortunately, the above assumption is usually not satisfied in practice, since radar system works in varying environment.
When the noise covariance matrix $\Rbf$ is unknown, it is usually replaced by the sample covariance matrix (SCM), formed by using the training data received in the vicinity of the test data. Then the optimum filter in \eqref{LRTopt3} becomes the sub-optimum filter of the sample covariance inversion (SMI) \cite{ReedMallett74}.
To complete target detection, it also needs appropriate CFAR processing.

Note that the above filtering-then-CFAR detection scheme adopts adaptive filtering. However, there is another filtering-then-CFAR detection scheme, which performs non-adaptive filtering, such as moving target indication (MTI) and moving target detection (MTD) and pulse Doppler processing. The key point in the MTD and pulse Doppler processing is Doppler filtering using multiple pulses. However, the number of the pulses used in the MTD is much smaller than that used in the pulse Doppler processing. Moreover, the MTD is often used by ground-based radar, while the pulse Doppler processing is mainly used by airborne radar.
This non-adaptive filtering-then-CFAR detection scheme usually has lower complexity compared with the adaptive filtering-then-CFAR detection scheme, however, suffers from certain performance loss, since its filtering performance is limited.

For unknown noise, if the test and training data are directly utilized to devise multichannel adaptive detectors, then better detection performance can be obtained, compared with the above filtering-then-CFAR detection scheme. Adaptive detection is just a kind of this detection scheme. Precisely, for adaptive detection, the test and training data are jointly utilized to design an adaptive detector, and then it is compared with a detection threshold, set according to a pre-assigned PFA. If the value of a detector is greater than the threshold, a target is claimed. Otherwise, no target is claimed.

\begin{figure}
	\centering
		\includegraphics[width=0.5\textwidth]{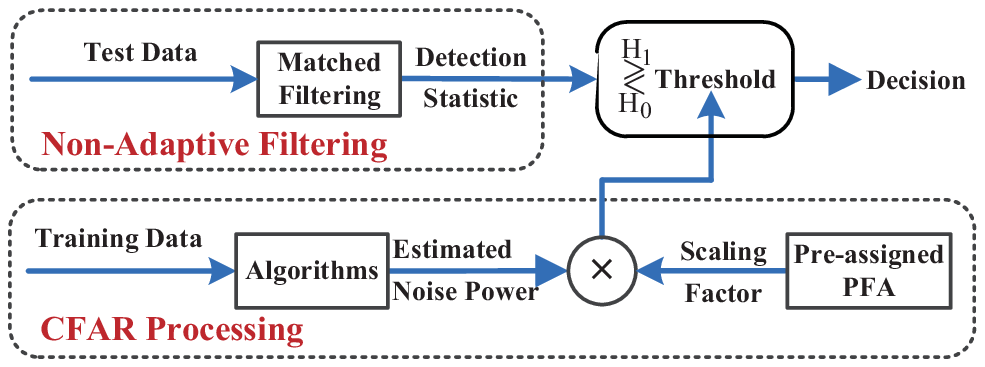}\\
{\footnotesize (a) Non-adaptive filtering-then-CFAR}\\
		\includegraphics[width=0.5\textwidth]{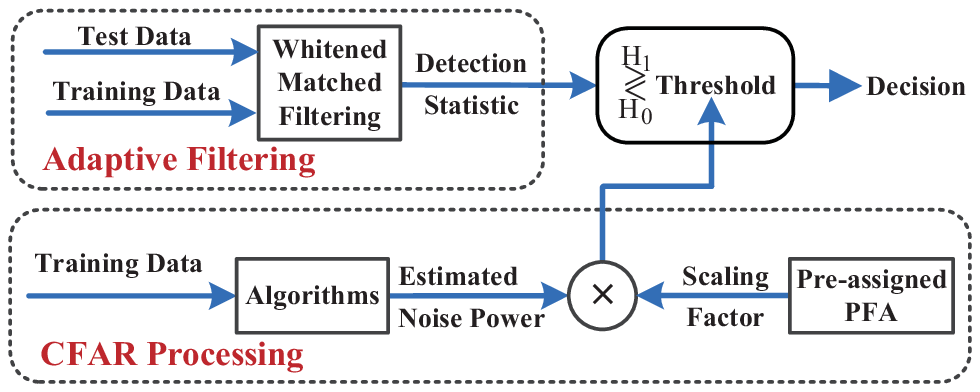}\\
{\footnotesize (b) Adaptive filtering-then-CFAR}\\
		\includegraphics[width=0.5\textwidth]{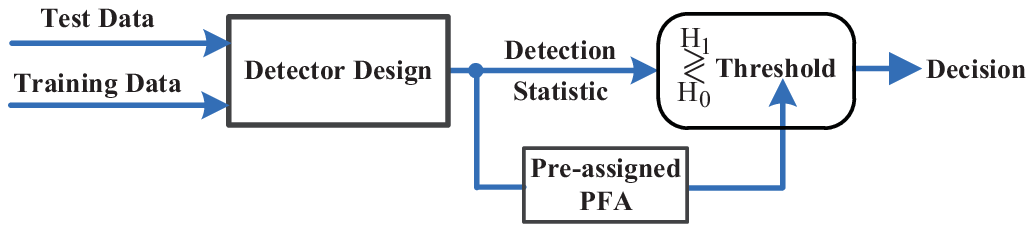}\\
{\footnotesize (c) Adaptive detection}
	\caption{Block diagrams for filtering-then-CFAR detection and adaptive detection}
	\label{BlockDiagram} 
\end{figure}

The block diagrams of filtering-then-CFAR detection and adaptive detection are summarized in Figure \ref{BlockDiagram}.
It can be concluded that the filtering-then-CFAR detection approach (adaptive or non-adaptive) needs two independent processing procedures, as its name indicates, i.e., filtering and CFAR processing. In contrast, independent filtering processing is not needed for adaptive detection, which achieves the function of filtering and CFAR processing simultaneously, both embedded in the detection statistic of the adaptive detection.

\subsection{Relationship between adaptive detectors and adaptive filters}
As explained above, adaptive filters  and adaptive detectors have different purposes, since the former tries to maximize the output SNR, while the latter tries to maximize the PD with a fixed PFA. However, adaptive filters and adaptive detectors have some common feature. They both adopt adaptivity. Precisely, they use training data to adaptively estimate the unknown noise covariance matrix. This is the essential point in adaptive processors.
Moreover, adaptive detectors have the function of adaptive filtering, which, however, is not achieved in an independent  procedure, as pointed above. 

\begin{figure}
	\centering		\includegraphics[width=0.6\textwidth]{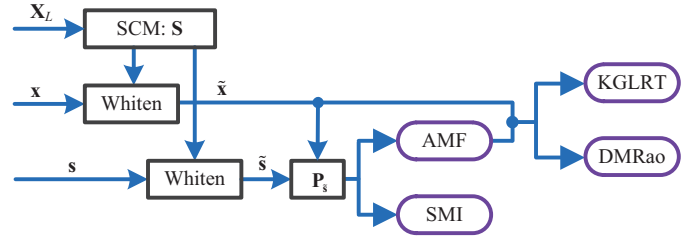}
		\caption{Block diagrams for one adaptive filter and three adaptive detectors}
	\label{BlockDiagramAdaptiveDetectionFiltering}
\end{figure}

As an example, Figure \ref{BlockDiagramAdaptiveDetectionFiltering} shows the block diagrams of one adaptive filter, namely, the SMI \cite{ReedMallett74}, and three adaptive detectors, namely, the KGLRT \cite{Kelly86}, adaptive matched filter (AMF) \cite{ChenReed91,RobeyFuhrmann92}, and De Maio's Rao (DMRao) \cite{DeMaio07} \footnote{The SMI is proposed based on the idea of filtering-then-CFAR detection. Mathematically, it can be written as $\left.\left[\mathop{\max}\limits_{\wbf} \frac{|\wbf^H\sbf|^2} {\wbf^H\Rbf\wbf}\right] \right|_{\Rbf=\frac{1}{L}\Sbf}$. The KGLRT, AMF and DMRao are proposed for the detection problem in \eqref{ch02_1} according to the criteria of GLRT, 2S-GLRT and Rao test, respectively. The AMF can also be obtained according to the Wald test.}.
The SMI can be obtained by replacing $\Rbf$ with the SCM $\Sbf$ in \eqref{MF}, resulting in
\begin{equation}
\label{SMI}
t_\text{SMI}=\frac{\tilde\xbf^H\Pbf_{\tilde\sbf}\tilde\xbf}{\tilde\sbf^H\tilde\sbf}.
\end{equation}
Moreover, the detection statistics of the KGLRT, AMF, and DMRao are
\begin{equation}
\label{KGLRT}
t_\text{KGLRT}=\frac{\tilde\xbf^H\Pbf_{\tilde\sbf}\tilde\xbf} {1+\tilde\xbf^H \tilde\xbf-\tilde\xbf^H\Pbf_{\tilde\sbf} \tilde\xbf},
\end{equation}
\begin{equation}
\label{AMF}
t_\text{AMF}={\tilde\xbf^H\Pbf_{\tilde\sbf}\tilde\xbf},
\end{equation}
and
\begin{equation}
\label{DMRao}
t_\text{DMRao}=
\frac{\tilde\xbf^H\Pbf_{\tilde\sbf}\tilde\xbf} {(1+\tilde\xbf^H\tilde\xbf)(1+\tilde\xbf^H \tilde\xbf-\tilde\xbf^H\Pbf_{\tilde\sbf} \tilde\xbf)},
\end{equation}
respectively, where $\tilde{\xbf}=\Sbf^{-\frac{1}{2}}\xbf$, $\tilde{\sbf}=\Sbf^{-\frac{1}{2}}\sbf$, $ \xbf $ is the test data vector, $ \sbf $ is the signal steering vector, $\Sbf= \Xbf_L \Xbf_L^H$ is the SCM\footnote{A more common SCM in adaptive filtering is defined as $\Sbf^\prime=\frac{1}{L} \Xbf_L \Xbf_L^H$. However,  for adaptive detection it is usually more convenient to use the SCM defined 
as $\Sbf= \Xbf_L \Xbf_L^H$.},  and $\Pbf_{\tilde{\sbf}}=\frac{\tilde{\sbf}\tilde{\sbf}^H}{\tilde{\sbf}^H\tilde{\sbf}}$ is the orthogonal projection matrix of ${\tilde{\sbf}}$.

The SMI and AMF can be taken as the outputs of certain adaptive filters, and then their corresponding weight vectors are\footnote{Note that the SMI weight in \eqref{weightSMI} satisfies the constraint $\wbf_\text{SMI}^H\sbf=1$.}
\begin{equation}
\label{weightSMI}
\wbf_\text{SMI}=\frac{\Sbf^{-1}\sbf}{\sbf^H\Sbf^{-1}\sbf}
\end{equation}
and
\begin{equation}
\label{weightAMF}
\wbf_\text{AMF}=\frac{\Sbf^{-1}\sbf}{\sqrt{\sbf^H\Sbf^{-1}\sbf}},
\end{equation}
respectively. However, the KGLRT and DMRao cannot be expressed as the output of a filter.


Two key functions of adaptive filtering are clutter rejection and signal integration. 
The former is achieved by the ``whiten'' model, accomplished by the matrix $\Sbf^{-\frac{1}{2}}$, while the latter is achieved by the orthogonal projection matrix ``$ \Pbf_{\tilde\sbf} $''.
It is seen from Figure \ref{BlockDiagramAdaptiveDetectionFiltering}, along with \eqref{SMI}-\eqref{DMRao}, that the SMI, KGLRT, AMF, and DMRao all have the function of adaptive filtering. Moreover, the AMF and SMI have the same filtering performance, since they have the same output SNR. This can be verified by substituting \eqref{weightSMI} and \eqref{weightAMF} into the quantity to be maximized in the right-hand side of \eqref{LRTopt2}. 
However, their detection performance is different, since the AMF has the CFAR property, whereas the SMI does not\footnote{The statistical performance analysis of the multi-band generalization of the SMI, called the modified SMI (MSMI), in \cite{WangCai90} indicates that the detection threshold of the SMI depends on the noise covariance matrix $\Rbf$.}. 

In summary, adaptive detectors use the test and training data to form specific structures, which are CFAR and have the function of filtering, embedded in the detection statistics. 



\section{Literature Review}

According to different criteria, the problem of adaptive detection can be sorted into different types. For example, according to the extension of a target, adaptive detection can be sorted into point target detection and distributed (spread) target detection; according to the fact that whether the signal is mismatched or not, adaptive detection can be sorted as detection in the absence of signal mismatch and detection in the presence of signal mismatch; according to statistical property of the noise, adaptive detection can be sorted into detection in Gaussian noise and detection in non-Gaussian noise; according to the characters of the test and training data, adaptive detection can be sorted into detection in HE and detection in non-homogeneous (heterogeneous) environment; etc.  However, the above classifications are too rough. 
Hence, we review the literature in the following {six} categories\footnote{We are sorry to any researcher whose work is overlooked or otherwise not discussed.}. For convenience, in each subsection we summarize the corresponding taxonomies in a table.
\subsection{Adaptive detection for point targets in the absence of signal mismatch}
\label{subsec:pnt-tgt-no-mis}
\begin{table}[htbp]
\footnotesize
\begin{center}
{\footnotesize
\caption{Related Taxonomy in Subsection \ref{subsec:pnt-tgt-no-mis}}
}
\begin{tabular}{ll}
\hline
Taxonomy &   Meaning  \\ \hline \hline 
HE  & A scenario that test and training data have the same noise covariance matrix.\\ \hline
PHE & \makecell[l]{ A scenario that test and training data have the same noise covariance matrix upon to unknown\\ scaling factor.} \\ \hline
Nonhomogeneity & \makecell[l]{A scenario that the data in the collection of test and training data do not have the same noise\\ covariance matrix.}  \\ \hline
Compound-Gaussian process &  \makecell[l]{A random process which is in the form of a product of two components.  One is the is the square \\root of a positive scalar random process (called texture, accounting for local power change),\\  while the other is a complex Gaussian process (called speckle, accounting for local scattering).} \\ \hline
Rank-one signal & {A kind of signal, modelled by the product of a known vector and an unknown scaling factor. } \\ \hline
Subspace signal & \makecell[l]{A kind of signal, modelled by the product of a known matrix and an unknown vector. That is\\ to say, a subspace signal lies in a known subspace but with unknown coordinates.} \\ \hline
\end{tabular}
\label{tab:taxonomy}
\end{center}
\end{table}

In the seminal paper \cite{Kelly86},  Kelly considered the detection problem for a point target in HE. Precisely, the point target has a known signal steering vector, embedded in Gaussian noise with unknown covariance matrix. To estimate the unknown noise covariance matrix, a set of IID training data was used, which is signal-free and shares the same noise covariance matrix with the test data. Then Kelly proposed the famous KGLRT. According to the 2S-GLRT, Chen \textit{et al.} \cite{ChenReed91} and Robey \textit{et al.} \cite{RobeyFuhrmann92} independently derived the well-known AMF, which has small complexity compared with the KGLRT. The corresponding Rao test was obtained by De Maio \cite{DeMaio07}, i.e., the DMRao, which has lower PD than the KGLRT and AMF. However, the DMRao has better performance in terms of rejecting mismatched signals. The corresponding Wald test was also derived by De Maio \cite{DeMaio04}, which coincides with the AMF. 
Noticeably, in 1994, Gerlach proposed the nonconcurrent mean level adaptive detector (N-MLAD) \cite{Gerlach94b} and concurrent mean level adaptive detector (C-MLAD) \cite{Gerlach94c}. The N-MLAD and C-MLAD are essentially the  AMF and DMRao, respectively; see also \cite{Gerlach95,Gerlach1995Convergence}. Moreover, the AMF was utilized in \cite{ReedGau98} 
for simultaneous detection and parameter estimation (i.e., target's Doppler and bearing).

The three detector KGLRT, DMRao, and AMF were all devised under the assumption of the HE.
However, the data may have different statistical properties, owing to rapidly changed environmental factors or instrumental factors, such as adaptation of conformal array, bistatic radar, or multisite radar.
Partially homogeneous environment (PHE) is a widely used nonhomogeneity model, which well characterizes the environment for airborne radars with low number of training data \cite{ConteDeMaio01} and also suitable for wireless communications with fades over multiple sources of interference \cite{KrautScharfButler05}. 
The GLRT for point target detection in PHE was derived by Kraut \textit{et al.}, denoted as the adaptive coherent estimator (ACE) \cite{KrautScharf99}. It was found in \cite{DeMaioIommelli08} that the Rao and Wald tests in PHE coincide with the ACE. In \cite{LiuLi15a}, a simple approach for the threshold setting of ACE, as well as the AMF, was provided. An invariance property of the ACE was given in \cite{ConteDeMaio03WSEAS}, and it was shown to be uniformly most powerful invariant (UMPI) in \cite{KrautScharfButler05}. More recently, it was shown in \cite{DeMaio19SPL} that the ACE using the fixed-point covariance estimate \cite{PascalChitour08} coincides with a maximal invariant component\footnote{It is observed that the upper-bound performance of the ACE is provide by the normalized matched filter (NMF), which was given in \cite{ConteLops95,Gini97a}. Moreover, the NMF was shown in \cite{DeMaioConte10b} to be the UMPI detector in spherically invariant random vector (SIRV) disturbance with a specific texture.}.
It is worth to pointing out that the ACE is effective in two kinds of non-homogeneous environment. One is spherically invariant noise \cite{ConteLops96} or compound-Gaussian noise \cite{ConteLops95}. The other is Bayesian heterogeneity. Precisely, the covariance matrix of the training data is subject to inverse complex Wishart distribution, and is proportional to the covariance matrix in the test data \cite{BidonBesson08b}. Moreover, the ACE is also called the adaptive normalized matched filter (ANMF) \cite{ConteLops96,ConteLops98} or normalized AMF (NAMF) \cite{Rangaswamy05}. In \cite{DeMaioFoglia05} the CFAR behavior using experimentally measured data was investigated for the KGLRT, AMF, and two variations of the ACE, namely, recursive ANMF (R-ANMF) \cite{ConteDeMaio02b} and persymmetric (RP-ANMF) \cite{ConteDeMaio04}. It was shown in \cite{DeMaioFoglia05} that all these detectors exhibit a false alarm rate higher than the preassigned value, and the RP-ANMF is most robust among them.
More recently, The problem of target separation detection (TSD) was considered in \cite{GaoAubry20AES}, where TSD tests were designed according to the GLRT. It was shown therein that the TSD tests can effectively monitor the event of target separation.

The above detectors are for rank-one signals, which have a known steering vector.
However, a signal may naturally lie in a subspace, but with unknown coordinates, such as polarimetric target detection \cite{DeMaioAlfano03,DeMaioAlfano04,AlfanoDeMaio04,LiuZhang12c,HaoGazor16AES}. 
This type of signal is called subspace signal, which can be mathematically expressed as the product of a full-column-rank matrix and a vector.
Under the background of polarimetric target detection, references \cite{ParkLiWang95} and \cite{DeMaioRicci01} generalized the KGLRT and AMF to the case of 2-dimensional subspace. Then, references \cite{RaghavanPulsone96,LombardoPastina01} generalized the KGLRT to the case of subspace with dimension greater than 2, and the detector can be named as the subspace-based GLRT (SGLRT). Similarly, the AMF was generalized to the case of subspace with dimension greater than 2 in \cite{LiuZhang12b}, and the detector was referred to as the subspace-based AMF (SAMF). The subspace versions of the DMRao and ACE were given in \cite{LiuXie14b} and \cite{KrautScharf01}, respectively, and the resulting detectors can be denoted as the subspace-based Rao (SRao) test and adaptive subspace detector (ASD), respectively. The statistical properties of the SGLRT was given in \cite{PastinaLombardo01,RaghavanPulsone96}, the statistical properties of the SAMF was given in \cite{LiuZhang12b}, the statistical properties of the ASD was given in \cite{LiuZhang11,LiuZhang12a}, and the statistical properties of the SRao was given in \cite{LiuWang16AES}.


\subsection{Adaptive detection for distributed targets in the absence of signal mismatch}
\label{subsec:dstr-tgt-no-mis}
\begin{table}[htbp]
\footnotesize
\begin{center}
{\footnotesize
\caption{Related Taxonomy in Subsection \ref{subsec:dstr-tgt-no-mis}}
}
\begin{tabular}{ll}
\hline
Taxonomy &   Meaning  \\ \hline\hline 
Distributed target  & A target which occupies more than one range bins for a radar system.\\ \hline
DD & \makecell[l]{A detection problem, for which the received echoes all come from the same direction. \\ However, the corresponding signal steering vector is only known to lie in a given subspace.} \\ \hline
GDD &  \makecell[l]{A detection problem, for which both the column and row components of a rank-one matrix-valued\\ signal are constrained to lie in known subspaces, but with unknown coordinates.} \\ \hline
DOS signal &  \makecell[l]{A kind of signal, which is matrix-valued and its row and column elements both lie in known subspaces\\ but with unknown coordinates.} \\ \hline
\end{tabular}
\label{tab:taxonomy}
\end{center}
\end{table}

For a high-resolution radar (HRR), a target may be spread in range, especially a big target, such as a large ship. It was shown in \cite{Hughes83} that a properly designed HRR can provide improved detection performance. This is mainly due to two factors. One is that increasing the capability of range resolution of the radar can reduce the amount of energy per range bin backscattered by the clutter. The other is that a distributed target is usually less fluctuated than an unresolved point target.

It was assumed in \cite{ConteDeMaio01} that the echoes reflected by the distributed target all came from the same direction, and the GLRT and 2S-GLRT for distributed target detection in HE and PHE were derived. The corresponding Rao and Wald tests in HE were derived in \cite{ShuaiKong12}, while the Rao and Wald test in PHE were given in \cite{HaoMa12}. The 2S-GLRT in HE in \cite{ConteDeMaio01} was known as the generalized AMF (GAMF). Similarly, we can name the GLRT in HE in \cite{ConteDeMaio01} as generalized KGLRT (GKGLRT), since it is a generalization of the KGLRT.
It is observed that the GLRT in HE proposed in \cite{ConteDeMaio01} shares the same form as the multiband GLR (MBGLR) in \cite{WangCai91}\footnote{The MBGLR was proposed for point target detection when a radar system has multiple frequency bands.}.

Reference \cite{ConteDeMaio03} investigated the problem of detecting a distributed target, whose signal steering vector was unknown. The GLRT, 2S-GLRT, modified 2S-GLRT (M2S-GLRT), and spectral norm test (SNT) were proposed. It was shown in \cite{LiuXie13b} that the 2S-GLRT and M2S-GLRT can be obtained according to the Wald test and Rao test, respectively. Some intuitive interpretations about the detectors were also given in \cite{LiuXie13b}. Recently, reference \cite{Raghavan20TSP} considered the case when the test data matrix was of rank two, and a generalization of ACE was proposed and its analytical performance was given.

In \cite{BessonScharf06b} it was assumed that the echoes backscattered by the distributed target all came from the same direction. However, the corresponding signal steering vector was only known to lie in a given subspace. This correspond detection problem was referred to as the direction detection (DD) therein, and the so-called generalized adaptive direction detector (GADD) was proposed according to the 2S-GLRT in PHE. From a mathematical point of view, for the problem of direction detection, the matrix-valued signal to be detected is of rank one, and its column components are constrained to a known subspace, while its row components are completely unknown.  A more general signal model was adapted in \cite{BoseSteinhardt96a}, where both the column and row components of a rank-one matrix-valued signal are constrained to lie in known subspaces, but with unknown coordinates. This kind of problem can be taken as a generalized direction detection (GDD). However, it did not use the training data in \cite{BoseSteinhardt96a}. Instead, it was assumed that the dimension of the test data satisfied certain constraint. Then a set of virtual training data can be obtained by using a unitary matrix transformation to the test data. As a consequence, the row structure of the signal was lost. Then the corresponding GLRT and 2S-GLRT were proposed therein. Essentially, the data model in \cite{BoseSteinhardt96a} was equivalent to that in \cite{BessonScharf06b}, but the environments were homogeneous.
The Wald test for the DD in HE was proposed in \cite{LiuXie14c}, and it was shown that there is no reasonable Rao test for the problem of direction detection. The problem of GDD in HE was exploited in \cite{LiuLiu2015b}, where the corresponding GLRT and 2S-GLRT were proposed. 
Moreover, the 2S-GLRT in PHE for GDD was given in \cite{Liu19SCIS}.

For the problem of detecting a distributed target, a systematic and comprehensive investigation was the report by Kelly and Forsythe in 1989 \cite{KellyForsythe89}, where the solid mathematical background for adaptive signal detection was given. In \cite{KellyForsythe89} the signal to be detected is matrix-valued and its row and column elements both lie in known subspaces but with unknown coordinates. This kind of signal model is referred to as the double subspace (DOS) signal in \cite{LiuXie14b,LiuXie14a}. The DOS signal model is very general and includes many types of point targets and distributed targets as the special cases. 
In \cite{KellyForsythe89}, no training data set was utilized. In contrast, a dimension constraint was posed on the test data. Then after a unitary matrix transformation on the test data, a set of virtual training data was obtained. Unfortunately, the row structure of the DOS signal is lost after the unitary matrix transformation.
The problem of detecting a DOS signal was generalized in \cite{LiuXie14b,LiuXie14a}, where true training data were assumed available, and many detectors were proposed and compared.

Compared with the detectors for point targets, the statistical performance of the detectors designed for distributed targets is difficult to be derived. In particular, the statistical performance 
of the GLRT and 2S-GLRT for distributed target in HE, proposed in \cite{ConteDeMaio01}, was given in \cite{WangCai91} and \cite{Raghavan13a}, respectively. 
Moreover, the result in \cite{WangCai91} was generalized in \cite{Raghavan13b} to the case of signal mismatch. 
Signal mismatch will be explained detailed in the next subsection.

\subsection{Adaptive detection in the presence of signal mismatch}
\label{subs:MisSig}
\begin{table}[htbp]
\footnotesize
\begin{center}
{\footnotesize
\caption{Related Taxonomy in Subsection \ref{subs:MisSig}}
}
\begin{tabular}{ll}
\hline
Taxonomy &   Meaning  \\ \hline\hline
Signal mismatch  & \makecell[l]{The phenomenon that the actual signal steering vector is not aligned with the nominal one adopted by \\ the radar system.}\\ \hline
Robustness & \makecell[l]{A property that the detection performance of a detector does not decrease severely with the increase of signal\\ mismatch.}  \\ \hline
Selectivity & A property that the detection performance of a detector decreases rapidly with the increase of signal mismatch.  \\ \hline
Directivity  & \makecell[l]{The property (including robustness and selectivity) of a detector when detecting a mismatch signal.} \\ \hline
Tunable detector & \makecell[l]{A kind of detector, which is parameterized by one or more positive scaling factors, called the tunable\\ parameters. By adjusting the tunable parameters, the directivity property of the detector can be changed.} \\ \hline
Cascaded detector & A kind of detector, formed by cascading a robust detector and a selective detector.\\ \hline
Weighted detector & A kind of detector, formed by weighting a robust detector and a selective detector.\\ \hline
\end{tabular}
\label{tab:taxonomy}
\end{center}
\end{table}
In practice, there often exists signal mismatch \cite{LiuLiu15SPL}. Precisely, the actual signal steering vector is not aligned with the nominal one adopted by the radar system. 
The statistical performance analysis for adaptive detectors in the presence of signal mismatch was first dealt with in \cite{Kelly89}, where it is shown that a key quantity controlling the detection performance of the KGLRT with mismatched signals is the generalized cosine-squared between the actual signal and the nominal signal in the whitened space. Based on the result in \cite{Kelly89}, the statistical performance of the AMF and ACE was given in \cite{Richmond00b}, while the performance of the DMRao was dealt with in \cite{DeMaio07}. The statistical performance of the subspace-based detectors was addressed in \cite{LiuLiu16SP} for the case of mismatched subspace signals, which is a generalization of the rank-one signal.

Signal mismatch can be caused by array error or target maneuvering.
Moreover, signal mismatch can also be caused by jamming signals coming from the radar sidelobe, due to electronic countermeasures (ECM).
For different sources of signal mismatch, different 
types of detectors are needed.
For the first case, a robust detector is preferred, which achieves satisfied detection performance when signal mismatch occurs. In contrast, for the second case, a selective detector is preferred, whose detection performance decreases rapidly with the increase of signal mismatch.

One method to design a robust detector for mismatched signals is adopting subspace signal model (for rank-one signals) \cite{RaghavanPulsone96} or enlarging the signal subspace (for subspace signals) \cite{ZeiraFriedlander97,ZeiraFriedlander98}.
Another method is constraining the actual angle or Doppler frequency lie in an compact interval \cite{DeMaioDeNicola10,LeeNguyen19}. Then, maximization of the concentrated likelihood function over the actual angle or Doppler can be formulated as a semidefinite programming (SDP) convex problem, and hence easily solved.
A third method is to assume that the actual signal lies in a convex cone, whose axes coincide with the nominal signal steering vector. Then a robust detector is designed by using second-order cone (SOC) programming \cite{DeMaio05,Besson06,Besson07a,DeMaioHuang11,HaoBandiera12}.
A fourth method is to adding a random component in the test data under the signal-presence hypothesis. This makes the hypothesis more plausible when signal mismatch happens \cite{ColucciaRicci19}.

To design a selective detector, one approach is to modify the original hypothesis test by adding a determinant unknown fictitious signal (or jammer) under the null hypothesis. The fictitious signal satisfies certain constraints. A useful constraint is that the fictitious signal is orthogonal to the nominal signal in the quasi-whitened space \cite{PulsoneRader01} or whitened space \cite{BandieraBesson07WABORT}. Then the resulting detector will be inclined to choose the null hypothesis when there is no target in the nominal direction but in other directions.
Under this idea, many selective detectors have been proposed, such as the adaptive beamformer orthogonal rejection test (ABORT) \cite{PulsoneRader01}, whitened ABORT (W-ABORT) \cite{BandieraBesson07WABORT}, their Bayesian variations \cite{HaoShang12}, and other modifications \cite{ColucciaRicci15b,LiuZhao17,LiuLiu2017AES}.
The proposed selective detectors in the aforementioned references were mainly under the assumption of the HE. In contrast, a selective detector was proposed in \cite{HaoYang12} for distributed target detection in PHE. However, the selectivity property of the proposed detector is limited. In \cite{LiuLiu18TSPPHE} a detector with improved selectivity 
was proposed for distributed target detection in PHE.

Another  approach to design selective detector is adding a random unknown fictitious signal under both the null and alternative hypotheses. An intuitive interpretation may be lack. However, it works in certain parameter setting, such as the double-normalized AMF (DN-AMF) \cite{OrlandoRicci10}. 

Note that the directivity (robustness or selectivity) of the above detectors cannot be adjusted. In other words, for a given detector, it either works as a robust detector or a selective detector, not both. This limits the flexibility of the detectors in detecting mismatched signals.
Tunable detectors, cascaded detectors, weighted detectors, as well as their combinations, can overcome the above limitation. 

Tunable detectors are mainly obtained by comparing the similarities in the detection statistics of two or more detectors with different directivity properties, and they, with specific tunable parameters, usually contain conventional detectors as their special cases. Directivity property of a tunable detector for mismatched signals can be smoothly changed by adjusting one or two parameters, called tunable parameters.
The first tunable detector was proposed by Kalson in 1992 \cite{Kalson92}, which contains the KGLRT and AMF as two special cases. However, the selectivity of this tunable detector cannot exceed the KGLRT.
Another tunable detector was proposed by Hao \textit{et al.} in \cite{HaoLiu10a}, termed as KRAO, which contains the KGLRT and DMRao as two special cases. The KRAO has enhanced selectivity but its robustness is limited.
In \cite{LiuXie14e} a tunable detector termed as KMABORT, was proposed, which contains the KGLRT, AMF, and ABORT as three special cases. The KMABORT is characterized by two tunable parameters, and hence it has more freedoms in detecting mismatched signals. 
However, its best robust property for mismatched signals is tantamount to that of the AMF. Fortunately, the AMF is very robust for mismatched signals, although it is not designed specially for robust detection of mismatched signals. 
A tunable detector, called KWA, was proposed in \cite{BandieraOrlando09c}, which contains the KGLRT, W-ABORT, and adaptive energy detector (AED)  \cite{RaghavanQiu95} as its special cases. The KWA can provide even more robust property than the AMF. As a special case of the KWA, 
the AED does not need the nominal signal steering vector, instead, it only tests whether there exists a signal with sufficient energy. In other words, it does not  differentiate between matched signals and mismatched signals. As a result, the AED is most robust. There are some other tunable detectors, such as the ones in \cite{LiuXie15b,LiuXie15a,LiuLiu16TSP2,LiuZhouLiu18TSP}.

A cascaded detector is forming by cascading a robust detector and a selective detector, and hence it has numerous pairs of detection thresholds. By changing the pair of detection thresholds, it can change the directivity property for mismatched signals. This type of cascaded detector is also called two-stage detector. A two-stage detector, referred to as 2SGLRT, cascading the KGLRT and AMF was proposed in \cite{PulsoneZatman00}. In \cite{Richmond00b}, a two-stage detector, called adaptive sidelobe blanker (ASB), was proposed, which cascades the AMF and ACE. In \cite{DeMaio07}, a two-stage detector, 
denoted as AMF-Rao, which cascades the AMF and DMRao. In \cite{BandieraBesson08a}, a two-stage detector, called WAS-ASB was proposed, which cascades the SGLRT and W-ABORT. In \cite{BandieraOrlando08}, a two-stage detector, called S-ASB was proposed, which cascades the SGLRT and ACE. In \cite{BandieraOrlando09c}, a two-stage detector called KWAS-ASB was proposed, which cascades the KWA and SGLRT. In \cite{HaoLiu10a}  two two-stage detectors were proposed, named as the KRAO-ASB and SKRAO-ASB. The former cascades the AMF and KRAO, while the latter cascades the SGLRT and KRAO.  In \cite{HaoLiu11}, a two-stage detector, called SD-RAO was proposed, which cascades the SGLRT and DMRao. The above two-stage detectors were all designed for rank-one signals. In contrast, a two-stage detector, named AESD, was proposed in \cite{DuanLiu17GRSL} for mismatched subspace signal by cascading the AED and ASD. The useful lecture \cite{BandieraOrlando09a} summarized the selective detectors ABORT and W-ABORT, the tunable detector KWA, the two-stage detectors ASB, AMF-Rao, S-ASB and WAS-ASB. Recently, a survey on the two-stage detector was given in \cite{DeMaioOrlando16AESm}.

A weighted detector is constructed by weighting a robust detector and a selective detector. By adjusting the weight, the directivity can be smoothly changed. A weighted detector, called SAMF-ASD, was proposed in \cite{LiuLi17_SP_MisSubs}.

All the tunable detectors, two-stage detectors, and weighted detectors above are designed for point target in HE. The ABORT was generalized in \cite{HaoYang12} for the distributed target detection both in HE and PHE. For distributed target detection, the W-ABORT was generalized in \cite{BandieraBesson07TSP_WABORT} and \cite{LiuLiu18TSPPHE} in HE and PHE, respectively. Moreover, a tunable detector for distributed target detection in PHE was proposed in \cite{LiuLiu18TSPPHE}, called tunable GLRT in PHE (T-GLRT-PHE).

Note that the capabilities of robustness or selectivity  of the two-stage detector and weighted detector cannot exceed their corresponding cascaded detectors and weighted detectors, respectively. In contrast, the tunable detector usually has much more freedoms to change the directivity for mismatched signals.

\subsection{Adaptive detection in interference}
\label{subs:Interference}
\begin{table}[htbp]
\footnotesize
\begin{center}
{\footnotesize
\caption{Related Taxonomy in Subsection \ref{subs:Interference}}
}
\begin{tabular}{ll}
\hline
Taxonomy &   Meaning  \\ \hline\hline
Noise interference  & A type of random interference, having the effect of thermal noise or clutter. \\ \hline
Coherent interference & \makecell[l]{A type of interference, having the effect of deceiving the radar system,\\ which only lies in a direction and occupies a Doppler bin.}  \\ \hline
Subspace interference  &  \makecell[l]{A type of coherent interference, which can be modelled by a subspace model.} \\ \hline
Orthogonal interference & {A type of coherent interference, which is orthogonal to the signal in some manner.}\\ \hline
\end{tabular}
\label{tab:taxonomy}
\end{center}
\end{table}
Most of the aforementioned detectors are designed without taking into account the presence of interference. In practice, however, there usually exists interference, besides noise and possible signal of interest. Interference can be caused by the intentional ECM or unintentional industrial production. 

Masking and deception are two main effects of interference on radar system.
Noise interference has the effect of masking the radar system, while coherent interference has the effect of deceiving the radar system.
Noise interference 
plays the role of thermal noise or clutter. Hence, it raises the level of the noise. As a result, in order to maintain CFAR property, the radar system has to raise the detection threshold, which reduces the radar sensitivity for target detection \cite[pp. 114-115]{Briggs04}.
Coherent interference usually imitates a real target, and hence it can deceive the radar system. This requires the interference works coherent to the radar system. Coherent interference can also be called false-target interference, including false-range interference, false-velocity interference, and false-direction interference. 

From the point of view of data model, coherent interference is usually constrained to lie in a known subspace, and hence is often referred to as subspace interference in the field of adaptive detection.
Much work was done by Scharf \emph{et al.} \cite{ScharfFriedlander94,BehrensScharf94,ScharfMcCloud02} for detecting a multichannel signal in subspace interference and thermal noise (or colored noise with known covariance matrix). 
Some other relative work in subspace interference and colored noise with known covariance matrix was given in \cite{BessonScharf05,BessonScharf06a,WangFang12,LiuZhang14}.

In practical applications, the noise covariance matrix is usually unknown, and needed to be estimated. For distributed target detection in subspace interference, it was assumed in \cite{BandieraDeMaio07a} that the noise covariance matrix was unknown. To estimate the noise covariance matrix, a set of sufficient training data was used. The GLRT and 2S-GLRT were derived both in HE and PHE therein. The PFA of the GLRT in HE was given in \cite{LiuLi19}. The corresponding Rao test and two-step Rao (2S-Rao) tests 
in HE and PHE were derived in \cite{LiuLiu2015d}. The Wald test and two-step Wald (2S-Wald) tests for point target detection in subspace interference were derived in \cite{LiuLiu19TAESWald}. 
Moreover, a modified Rao test was given in \cite{Wang20SP}, which took both the signal coordinate matrix and interference coordinate matrix as the relative parameter.
It is shown in \cite{LiuLiu19TAESWald} that in HE the 2S-GLRT, 2S-Rao, and Wald test (the other detectors all strongly related with these three detectors) whiten the noise (or equivalently reject the clutter) in the same manner. However, they reject the subspace interference in different manners.
Recently, the statistical performance of the GLRT for subspace interference was analyzed in \cite{LiuLi20} for the case that the signal was of rank one. Moreover, the statistical performance of the GLRT-based detectors for point target detection in subspace interference was analysed in \cite{LiuLiu18AES} for the case of signal mismatch, including the signal match as a special case. It was shown in \cite{LiuLiu18AES} that the coherent interference and signal mismatch affect the detection performance of the GLRT-based detectors through two generalized angles.
One is the angle between the whitened actual signal and the whitened interference subspace. The other is the angle of the actual signal and nominal signal matrix after they are both projected onto the interference-orthogonalized subspace.
Reference \cite{LiuLiu20a} investigated the detection problem in subspace interference when signal mismatch happens. Two selective detectors and a tunable detector were proposed, and their statistical performance was also given therein. The detection problem in subspace interference was addressed in \cite{AubryCarotenuto16SPL,DeMaioOrlando2016Invariance,CiuonzoDeMaio16a, CiuonzoDeMaio16b,CiuonzoDeMaio17TSP} in the framework of invariance principle. 
When the subspace interference lies in both the test and training data, it was pointed that in \cite{AubryDeMaio14b} that there is no effective GLRT, and a modified GLRT was proposed based on the method of sieves therein.

For the DD problem in the presence of subspace interference in HE, the GLRT and 2S-GLRT were developed in \cite{BandieraBesson07}, while the Wald test and 2S-Wald test were obtained in \cite{LiLiu18MSSP}. The corresponding 2S-GLRT and 2S-Wald tests 
in PHE were derived in \cite{DongLiu17SPL_DD_PHE}.

In the above references, sufficient information about the coherent interference is assumed available. However, this is not always the case in practice.
It was assumed in \cite{BandieraBesson13} that the interference subspace was unknown except for its dimension, and a GLRT-like detector was proposed therein.
In \cite{LiuWang16AES}, it was assumed that the coherent interference was unknown but it was orthogonal to the signal in the whitened space. This type of interference was called orthogonal interference therein\footnote{
The orthogonal interference satisfies the generalized eigenrelation (GER) defined in \cite{Richmond00c}, which can be approximately met in practice, especially for the out-of-mainbeam interference \cite{Richmond00a}. It is pointed out in \cite{Richmond00a} that using secondary data selection strategies, e.g., the power selected training \cite{RabideauSteinhardt99}, results in the orthogonality of the signal  and interference in the whitened space.}. Then three detectors were proposed, according to the criteria of GLRT, Rao test, and Wald test. Remarkably, the resulting three detectors share the same forms as the SGLRT, SRao, and SAMF, respectively. However, statistical performance analysis indicated that the orthogonal interference can degrade the detection performance \cite{LiuWang16AES}.
Moreover, it was assumed in \cite{BandieraDeMaio07b,DeMaioDeNicola09b,SvenssonJakobsson11} that there were uncertainties in signal and coherent interference. To account for these uncertainties, the signal and interference were constrained to certain proper cones. Then effective detectors were proposed by using convex optimization.

The adaptive detection in completely unknown coherent interference was dealt with in \cite{LiuLiu2015c}. At the stage of detector design, the unknown interference was assumed to lie in a subspace orthogonal to the signal. According to the GLRT and Wald test, two detectors were proposed, and the detector derived according to the GLRT was called adaptive orthogonal rejection detector (AORD). It was shown that the AORD has better detection performance than others in completely unknown interference. Another distinctive feature of the AORD is that it can even provide significantly performance improvement, compared with the KGLRT and AMF in the absence of interference. This was shown in \cite{LiuLiu_16SPL}, where the statistical performance of the AORD was also given.

The above references mainly deal with coherent interference.
It was assumed in \cite{Besson07b} that there was a completely unknown noise interference, and the corresponding GLRT for rank-one signals was shown to be equivalent to the ACE. The corresponding Rao test was given in \cite{OrlandoRicci10}, i.e., the DN-AMF, mainly adopted for mismatched signal detection, as explained in Subsection \ref{subs:MisSig}.
The above results were generalized in \cite{LiuHan17SCIS} when there existed additional coherent interference, and the GLRT, Rao test, and Wald test were derived for subspace signals.
In \cite{BessonOrlando07} the noise interference was constrained by the GER, and the GLRT was shown to be the same as the KGLRT. Moreover, the corresponding Rao and Wald tests were shown to be the DMRao and AMF, respectively \cite{HaoOrlando12a}. The results in \cite{BessonOrlando07,HaoOrlando12a} were generalized in \cite{ShangLiu18GER} for subspace signals.
It was assumed in \cite{TangWang20} that the noise interference lies in a subspace orthogonal to the signal subspace, and a detector was proposed according to the 2S-Rao test, named as two-step
orthogonal SAMF (2S-OSAMF). Numerical examples shew that the 2S-OSAMF has better detection performance than its competitors even the noise interference is completely unknown.

In \cite{Orlando17} the authors considered the problem of determining whether the test data contained a noise interference or not. This problem was solved by formulating the problem as a binary hypothesis test, and a detector was designed according to GLRT criterion.
In \cite{AddabboBesson19} the authors considered the problem of detecting a signal in the presence of noise interference, which only occupied parts of training data. Two GLRT-related detectors were proposed, which were shown to have better performance than the existing detectors.
In \cite{YanAddabbo20} the authors considered two scenarios for the signal detection problem in interference. One was that only noise interference existed, and the other is that both noise interference and coherent interference existed. For the first scenario, an effective estimate for the interference covariance was proposed and then utilized in the AMF, which can mitigate the deleterious effects of the noise interference. 
For the second scenario, a compressive sensing-based GLRT was proposed.
Some other detection problems involved in noise interference were given in
\cite{Raghavan19,Besson19,Besson20SPL}.

\subsection{Adaptive detection with limited training data}
\label{subs:LmtTrnDat}
\begin{table}[htbp]
\footnotesize
\begin{center}
{\footnotesize
\caption{Related Taxonomy in Subsection \ref{subs:LmtTrnDat}}
}
\begin{tabular}{ll}
\hline
Taxonomy &   Meaning  \\ \hline\hline
Low-rank structure & \makecell[l]{Noise covariance matrix is a sum of a scaled identity matrix and a low-rank matrix, with eigenvalues\\ much greater than unity.}  \\ \hline
Persymmetry & Noise covariance matrix is persymmetric about its cross diagonal and Hermitian about its diagonal. \\ \hline
Spectral symmetry &  Ground clutter has a symmetric PSD centred around the zero-Doppler frequency. \\ \hline
\end{tabular}
\label{tab:taxonomy}
\end{center}
\end{table}
For adaptive processing, e.g., adaptive detection or adaptive filtering, it usually needs sufficient training data to estimate the unknown noise covariance matrix. 
In particular, it was shown in \cite{ReedMallett74} that the adaptive filter SMI needs at least $2N-3$ IID training data to maintain 3 dB SNR loss, compared with the optimum filter (with known noise covariance matrix), with $N$ being the dimension of the test data. 
This is known as the Reed-Mallett-Brenann (RMB) rule \cite{ReedMallett74}\footnote{Recently, a simple proof of the RMB rule has been given in \cite{LiuLiu2017RMB}. It is worth pointing out for adaptive detection, more than $2N-3$ IID training data are required to maintain 3 dB SNR loss, compared with the optimum detector, as shown in Figure 3 in the following.}.
However, this requirement may not be always satisfied in practice.
Taking an example of the STAP filtering for airborne radar, if the number of antenna elements is 30, the number of pulses is 40, and the system bandwidth is $10\ \text{MHz} $, in order to meet the requirement of the RMB rule, each filter needs received data within a range of roughly $36 \ \text{km}$. The IID assumption usually cannot be guaranteed in such a wide range\footnote{In other words, in many applications only a few number of data are IID. There are many approaches to choose qualified data, such as reiterative censored fast maximum likelihood (CFML) \cite{Gerlach02}, generalized inner product (GIP) \cite{RangaswamyMichels04}, approximate maximum likelihood (AML) \cite{HanDeMaio19AES}, etc.}.

\textit{A priori} information-based method and dimension reduction are two main kinds of approach to alleviate the requirement of sufficient IID training data.
\subsubsection{\textit{A priori} information-based methods}
\textit{A priori} information-based method 
includes several sub-kinds, namely, Bayesian methods, parametric methods,  special structure-based methods, etc. 

For Bayesian methods\footnote{The Bayesian methods were also used to model the detection problem in non-homogeneous environment, e.g., \cite{BessonTourneret07,BidonBesson08a}.}, the noise covariance matrix is ruled by a certain statistical distribution \cite{LiuHan18}, and the distribution parameters can be obtained by using limited training data. In \cite{DeMaioFarina10} the noise covariance matrix was assumed to be subject to a given inverse Wishart distribution, and the Bayesian one-step GLRT (B1S-GLRT) and Bayesian 2S-GLRT (B2S-GLRT) were proposed. It was shown by simulated and experiment data that these two Bayesian detectors can provide better detection performance than the conventional ones with low sample support. Notice that the B1S--GLRT and B2S--GLRT can be taken as the Bayesian generalizations of the KGLRT and AMF, respectively. The Bayesian version of the ACE was derived in \cite{WangSahinoglu11,ZhouZhang12}. The Bayesian method was also adopted in \cite{HaoShang12,FrancescoBesson15} to devise selective detectors with limited training data.
Noticeably, the Bayesian method can be used even no training data are available \cite{BandieraBesson11}.

Parametric (or model-based) method approximates the interference spectrum with a low-order multichannel autoregressive (AR) model \cite{RomanRangaswamy00}. In other words, the noise covariance matrix can be well characterized by using only a few parameters. Hence, this method largely reduces the required training data. At the same time, it also reduces the computational complexity. In \cite{RomanRangaswamy00}, the parametric AMF (PAMF) was proposed. The PAMF was shown to be equivalent to the parametric Rao test in \cite{SohnLi07b}, where the asymptotic (in the case of large sample case) statistical distribution was also derived. The corresponding parametric GLRT was obtained in \cite{SohnLi07a}, which was shown to have better detection performance than the PAMF. 
In \cite{LiMichels06}, the nonstationary PAMF (NS-PAMF) and nonstationary normalized PAMF (NS-PAMF) were proposed for adaptive signal detection in hyperspectral imaging.
There are many other 
parametric detectors, e.g., \cite{MichelsHimed00,AlfanoRSN04,SohnLi08,AbramovichJohnson08a,AbramovichJohnson08b,AbramovichSpencer10, WangLi10a,WangLi10b,AbramovichRangaswamy11,JiangLi12b,JianHe13,WangWang14AR,ShiHao15,Mennad2017AR, GaoLi18TSP,YanHao20a}.

The ``structure'' for special structure-based methods is for the noise covariance matrix, which may have different kinds of special structures for different antenna configurations or different radar operating environments. The special structures for the noise covariance matrix include low-rank structure, Toeplitz \cite{Fuhrmann91}, Kronecker \cite{Raghavan17Kronecker,WangXia17Kronecker}, persymmetry, spectral symmetry, etc.

For the low-rank structure, which is data-dependent, the noise covariance matrix is a sum of a scaled identity matrix (corresponding to weak thermal noise) and a low-rank matrix (corresponding to strong clutter), with eigenvalues much greater than unity.
Then, with limited training data, 
the principal component approximation of the SCM is usually a better estimation for the noise covariance matrix than the SCM itself \cite{HaimovichBarNess91}. 
Under this guideline, many reduced-rank approaches have been developed.
Precisely, the reduced-rank versions of the KGLRT, AMF, and ACE were exploited in \cite{WangLiu14} for the problem of space-time adaptive detection (STAD) in  airborne radar with the data received by multiple sensors under different pulses. 
There are many other well-known reduced-rank detectors or filters, such as the principal component analysis (PCA) \cite{HaimovichBarNess91}, cross-spectral metric (CSM) \cite{GoldsteinReed1997}, multistage Wiener filter (MWF) \cite{GoldsteinReed98,GoldsteinReed99}, auxiliary-vector filter (AVF) \cite{PadosKarystinos2001}, joint iterative optimization (JIO) \cite{FaDeLamare11},  conjugate gradient (CG)-based AMF (CG-AMF) \cite{ChenLi15}, \footnote{It is worthy pointing out that the MWF, AVF, and CG are equivalent to each other \cite{ChenMitra02,ScharfChong08}, and they all belong to the Krylov subspace technique, which was originally used in numerical calculation \cite{BroydenVespucci04} and have been recently successfully used in signal processing \cite{Dietl07Krylov}. Remarkably, the Krylov subspace technique needs neither matrix inversion nor eigenvalue decomposition (EVD), and it can provide better performance than the EVD-based methods \cite{LiuXie14d,LiuXie15Krylov}. } and some others \cite{GauReed98,ReedGau99a,ReedGau99b,LiuXie14f,LiSongLiu18RSN}. 
Moreover, the diagonally loaded versions of the KGLRT, AMF and ACE were investigated in \cite{AbramovichSpencer07a,Liu11DLSTAD}. Diagonal loading can be often taken as a kind of reduced-rank method, since it uses the low-rank structure information of the noise covariance matrix.

Persymmetry is another useful structure for the covariance matrix estimation with low sample support. For the persymmetric covariance matrix, it is persymmetric about its cross diagonal and Hermitian about its diagonal. This structure exists when symmetrically spaced linear arrays and/or pulse trains  are used. In addition, persymmetry can be found in other situations, e.g., standard rectangular arrays, uniform cylindrical arrays (with an even number of elements), and some standard exagonal arrays \cite{DeMaioOrlando16TSP}. 
In \cite{Nitzberg80} and \cite{DeMaio03}, the maximum likelihood estimates of persymmetric covariance matrices were provided in the absence and presence of white noise, respectively. It has been proven in \cite{LiuLiu16TSP1,LiuOrlando19} that the exploitation of persymmetry is tantamount to doubling the number of training data in adaptive processing. For target detection by exploiting the persymmetry, references \cite{PaillouxForster11,Hao15TAESPersymmetric,LiuCui15AES,DeMaioOrlando15,DeMaioOrlando16b} considered the case of point target with a single observation in HE, while references  \cite{CaiWang91PersMSMI,CaiWang92,LiuLi19TSP,LiuLiu19TSP1} considered the case of distributed target or point target with multiple observations/multi-bands in HE. Persymmetric detectors in HE with improved rejection capabilities were given in \cite{HaoOrlando14a}.
In the PHE, several persymmetric detection algorithms were designed in \cite{CasilloDeMaio07,HaoOrlando12b,GaoLiao14,HaoOrlando14b,WangLi16SP,CiuonzoOrlando6SPL}. 
The above references exploiting persymmetry mainly focus on rank-one signal detection. In contrast, persymmetric detection of subspace signals was considered in \cite{LiuLiu18TSPPsmtrc,LiuSun19,MaoGao19,LiuLiu20b,LiuJian20,LiuLiu20AES}. Moreover, persymmetry can also be used in non-Gaussian noise \cite{ConteDeMaio03a,ConteDeMaio04,PaillouxForster11,GaoLiao13,GuoTao17,LiuLiu18SP} or multiple-input multiple-output (MIMO) radars \cite{LiuLi15,LiuLiu18b,LiuHan19,LiuHan19a}.

Spectral symmetry exists in ground clutter, when it is observed by a stationary monostatic radar system.  Precisely, the ground clutter has a symmetric power spectral density (PSD) centred around the zero-Doppler frequency. This special structure is confirmed by real data in \cite{BillingsleyFarina99,ConteDeMaio05}. 
Other situations where spectral symmetry exists were discussed in details in \cite{DeMaioOrlando16TSP}.
Exploiting the spectral symmetry, adaptive detectors were proposed for the HE \cite{DeMaioOrlando16TSP,YanMassaro17} and PHE \cite{FogliaHao17AES}. Simulation results indicate that utilizing the spectral symmetry is equivalent to doubling the number of the training data.

The above special structures can be combined together to further improve the performance, for instance, Bayesian method plus parametric method \cite{WangLi11a}, parametric method plus persymmetry \cite{WangSahinoglu12,GaoLiao15}, low-rank structure plus persymmetry \cite{GinolhacForster12}, persymmetry plus spectral symmetry \cite{HaoOrlando16SPL,FogliaHao17DSP_PHE,CarotenutoDeMaio19}.

\subsubsection{Dimension reduction methods}
The method of utilizing \textit{a priori} information may suffer from significantly performance loss, if the prior information greatly departs from the actual one. 
Another approach to alleviate the requirement of sufficient IID training data is dimension reduction, which is data-independent. To this end, a reduced-dimension transformation is applied to the test and training data before adaptive processing. This has the effect of projecting the noise covariance matrix onto a low-dimension subspace. As a result, the required number of IID training data can be considerably reduced, and the computational complexity is reduced as well. Various reduced-dimension approaches have been proposed, 
such as auxiliary channel receiver (ACR) \cite{Klemm87}, extended factor approach (EFA) \cite{DiPietro92}, space-time multiple-beam (STMB) \cite{WangChen03}, sum-difference STAP ($\Sigma\Delta$--STAP) \cite{BrownSchneible00}, best channel method (BCM) \cite{ZhangHe14},  alternating low-rank decomposition (ALRD) \cite{CaiWu18AES}, among others \cite{YangWang19SP}.

The aforementioned approaches were proposed for filtering. In contrast, the joint-domain localized GLR (JDL-GLR) detector was proposed in \cite{WangCai94} for airborne radar target detection. The JDL-GLR first transforms the test and training data into a reduced-dimension space, and then uses the KGLRT structure to form the final detector.
Another similar reduced-dimension GLRT was proposed  in \cite{AyoubHaimovich00}. 
In \cite{ReedGau98} two reduced-dimension detectors were proposed, which adopted the AMF structure.
In \cite{JinFriedlander05b}, a reduced-dimension detector was proposed by using subarray processing.
Recently, a random matrix-based reduced-dimension detector was given in \cite{Besson20SP}. The detector also uses the KGLRT structure. However, the reduced-dimension matrix is chosen in a different manner. Precisely, one column of the reduced-dimension matrix is aligned with signal steering vector, while the other columns are chosen randomly in the subspace orthogonal to the signal steering vector.

In \cite{WangZhao19a,Wang20b} the test and training data were first projected on the one-dimensional signal subspace, resulting in scalar data. 
Then using the resultant scalar data, two reduced-dimension detectors were designed.
The above reduced-dimension detectors are mainly for rank-one signals. In contrast, a reduced-dimension detector for subspace signal detection was proposed in \cite{LiuLiu19JFI}, referred to as subspace transformation-based detector (STBD). It is shown in \cite{LiuLiu19JFI} that the STBD, which can also serve as a filter, can provide improved detection and filtering performance even in some sample-abundant scenarios, besides the case of limited training data.

Besides the \textit{A priori} information-based method and dimension reduction method, there may be some other technologies to alleviate (or even not need) the requirement of training data. For example, in \cite{Wang20SPb} the authors considered the problem of detecting a multichannel spatial signal in unknown noise without training data. To estimate the unknown noise covariance matrix, a number of echo signals reflected by the test data were utilized.

\subsection{Adaptive detection for MIMO radar}
\label{subs:MIMO}
\begin{table}[htbp]
\footnotesize
\begin{center}
{\footnotesize
\caption{Related Taxonomy in Subsection \ref{subs:MIMO}}
}
\begin{tabular}{ll}
\hline
Taxonomy &   Meaning  \\ \hline\hline
Distributed MIMO radar & MIMO radar with widely separate antennas.  \\ \hline
Colocated MIMO radar & MIMO radar with closely spaced antennas.  \\ \hline
Spatial diversity  & \makecell[l]{The transmit antennas are far enough from each other, and hence the target radar cross sections\\ can be taken as independent random variables for different transmit-receive paths. With a spatially \\diverse set of ``looks'', each set of received data carries independent information about the target.}   \\ \hline
\end{tabular}
\label{tab:taxonomy}
\end{center}
\end{table}
A MIMO radar adopts multiple elements at both transmit and receive antennas. The transmitted waveforms are linearly independent or orthogonal \cite{LiStoica09book}. 
According to the antenna configuration, there are two basic categories for MIMO radar. One is distributed MIMO radar, whose antennas are far from each other \cite{HaimovichBlum08SPM}, while the other is colocated MIMO radar, whose antennas are closely spaced \cite{LiStoica07SPM}.

Strictly speaking, the review of MIMO radar target detection can also be carried out from above five aspects, or be included in the above five aspects. 
However, as an emerging research area, MIMO radar has received considerable attention. 
Hence, we would like to review MIMO radar target detection in an independent subsection from the following three aspects: adaptive detection for distributed MIMO radar, adaptive detection for colocated MIMO radar, and adaptive detection for other types of MIMO radar.

\subsubsection{Adaptive detection for distributed MIMO radar}
For the distributed MIMO radar detection, it was shown in \cite{FishlerHaimovich06TSP} that the distributed MIMO radar can provide better detection performance than traditional phased-array radar in high SNR regions. This improvement is due to the fact that spatial diversity can alleviate the impact of target scintillation, and spatial diversity gain is higher than the coherent processing gain of phased-array radar. Based on the results in \cite{FishlerHaimovich06TSP}, the expressions for the PD of the GLRT was derived in \cite{DuThompson08} when the target consists of a finite number of small scatterers. Reference \cite{TajerJajamovich10} considered the problem of joint target detection and parameter estimation, and it was shown that distributed MIMO radars provide significant improvement over phased-array radars for distributed targets. 
Reference \cite{AkcakayaNehorai10} dealt with the MIMO radar detection problem when phase synchronization mismatch arose between the transmit and receive antennas. The phase error was modelled as the von Mises distribution, and the corresponding GLRT was derived. Polarimetric MIMO radar detection in Gaussian noise was investigated in \cite{GogineniNehorai10TSP}, and it was shown that optimal design of the antenna polarizations leads to better detection performance than MIMO radars transmitting fixed polarized waveforms over all antennas.

In the above references, the target's movement feature was not taken into consideration. When the transmitted waveform was orthogonal and Doppler processing was adopted, the GLRT and 2S-GLRT were derived in \cite{SheikhiZamani08} for moving target detection in Gaussian background, and the expression for the PFA of the GLRT was given in \cite{LiuZhang13SP}. It was assumed in \cite{LiuLi15} that the noise covariance matrix had the persymmetric property, then the GLRT, as well as its statistical property, was derived for distributed MIMO radar which transmitted orthogonal waveforms and adopted Doppler processing. Under the same antenna configuration, as well as adopting the Doppler processing, the 2S-GLRT was given in \cite{LiCui14} for compound-Gaussian clutter, while the corresponding 2S-Rao and 2S-Wald tests were derived in \cite{LiCui15SP}. When the distributed MIMO radar transmitted orthogonal waveforms and adopted Doppler processing, reference \cite{LiCui15c} derived the GLRT for polarimetric moving target detection in the Gaussian noise. The detection problem in \cite{SheikhiZamani08} was generalized in \cite{HeLehmann10AES} by assuming that the target velocity was unknown, and it was shown that distributed MIMO radar has better detection performance than the phased-array radar when detecting a target with small radial velocities and environment is homogeneous. The distributed MIMO radar detection in non-homogeneous clutter was considered in \cite{WangLi11b}, where the corresponding GLRT was derived and analytically evaluated. It was also shown that the GLRT in \cite{WangLi11b} has better detection performance than the detector in \cite{HeLehmann10AES}, as well as the corresponding phased-array detector.

Note that in the above references orthogonal waveforms are adopted. Under the assumption of white Gaussian noise, it is shown in \cite{AkcakayaNehorai11} that a detector will suffer from certain detection performance loss if the orthogonality property of the waveforms transmitted by different antennas is not satisfied. 
However, the above result may not suitable for colored noise. Reference \cite{DeMaioLops08} derived the GLRT for distributed MIMO radar with arbitrary transmitted waveforms and arbitrary time-correlation of the noise, and it was shown that there is an inherent trade-off between diversity and integration, and that no uniformly optimum waveform design strategy exists.  In \cite{DeMaioLops07}, the GLRT was derived for distributed MIMO radar, with arbitrary transmit waveform and adopting Doppler processing. It was assumed that all transmit-receive pairs share the same known covariance matrix, then the expressions for the PD of the GLRT was given \cite{DeMaioLops07}, according to which the optimum transmit waveform was given. Reference \cite{NaghshModarresHashemi12} generalized the data model in \cite{DeMaioLops07} to the case that different transmit-receive pairs have different but known covariance matrices. Then the statistical performance of the corresponding GLRT was given for Swerling I target.

The signal model in \cite{DeMaioLops07} was also adopted in \cite{LiCui12,ZhangCui13SPL,LiYang19SP,CuiKong12d,CuiKong12c}, however, the noise was assumed to be compound-Gaussian. Precisely, the 2S-Rao and 2S-Wald tests for distributed MIMO radar were given in \cite{LiCui12}, while several Bayesian 2S-GLRTs were derived in \cite{ZhangCui13SPL,LiYang19SP}. The 2S-Rao and 2S-Wald tests in \cite{LiCui12} was generalized to polarimetric distributed MIMO radar detection in \cite{KongCui11} for point targets and in \cite{CuiKong12} for distributed targets. Moreover, reference \cite{CuiKong12d} derived the 2S-GLRT for point target detection with polarimetric distributed MIMO radar in compound-Gaussian clutter, and it was generalized in \cite{CuiKong12c} for the case of distributed target detection.

\subsubsection{Adaptive detection for colocated MIMO radar}
For the colocated MIMO radar detection, reference \cite{BekkermanTabrikian06} derived the GLRT and its asymptotic statistical distribution for colocated MIMO radar after beamforming in white Gaussian noise. Reference \cite{CuiKong12b} proposed three 2S-GLRTs for colocated MIMO radar with randomly distributed arrays in compound-Gaussian clutter, and it was shown that the configuration of randomly distributed arrays achieve detection performance improvement at the directions with strong clutter. 

Remarkably, it was shown in \cite{LiXu08TSP} that colocated MIMO radars make it possible that detecting a target or estimating its parameters does not need training data or even range compression. Without the training data, the problem of parameter estimation for colocated MIMO radar was addressed in \cite{XuLi08AES}, where the GLRT was derived to suppress the false peak induced by strong jammer. The corresponding Rao and Wald tests, as well as their statistical properties, were given in \cite{LiuWang15MIMO}. When signal mismatch occurs, a tunable MIMO radar detector was proposed in \cite{LiuZhouLiu18TSP}, which includes the Rao and Wald tests in \cite{LiuWang15MIMO} as special cases. The proposed tunable detector in \cite{LiuZhouLiu18TSP} has flexibility in controlling the direction property, selectivity or robustness, for mismatched signals. Two robust detectors were proposed in \cite{LiuLi19b} for mismatched signals by assuming the actual signal lying in certain subspaces. The GLRT in \cite{XuLi08AES} was generalized in \cite{LiuLiu18b} when the persymmetry of noise covariance matrix was exploitation. It was shown by simulation and experimental data that by utilization the persymmetry, the proposed detector in \cite{LiuLiu18b} can achieve better detection performance. The correspond persymmetric Rao test Wald test were given in \cite{LiuHan19a} and \cite{LiuHan19}, and in \cite{LiuHan19a} a  two-stage detector was also given for mismatched signal detection by cascading the above persymmetric Rao and Wald tests.

More recently, for a colocated MIMO radar, a robust Wald-type test were proposed in \cite{FortunatiSanguinetti20}. Performance analysis showed that there always exists a sufficient number of (virtual) antennas such that the required performance are satisfied, without prior knowledge of the noise statistical property. This type of MIMO radar was referred to as the massive MIMO radar therein. Moreover, in \cite{LanMarino20early} three adaptive GLRTs were proposed for colocated MIMO radar equipped with frequency diverse array (FDA).
\subsubsection{Adaptive detection for other types of MIMO radar}
There are several types of variations of the distributed MIMO radar and colocated MIMO radar, such as phased MIMO (Phased-MIMO) radar \cite{HassanienVorobyov10}, hybrid MIMO phased array radar (HMPAR) \cite{FuhrmannBrowning10}, transmit subaperturing MIMO (TS-MIMO) radar \cite{LiHimed10}, and multi-site radar system MIMO (MSRS-MIMO) \cite{XuDai11}. The MSRS-MIMO radar has multiple widely separate sub-arrays, and each sub-arrays has multiple colocated antennas.
According to the waveforms, the MSRS-MIMO radar can be classified as two kinds. One is that the waveforms are different or orthogonal in different transmit antennas \cite{ChenZheng17}. The other is that the waveforms transmitted by the antennas are scaled versions of a single waveform\cite{ChaoChen12AEU}. For convenience, the first type of MSRS-MIMO radar is referred to as the distributed-colocated MIMO radar, while the latter one is referred to as the distributed-phased MIMO radar\footnote{It is worth pointing out that the data model of the distributed-phased MIMO radar is the same as the conventional distributed MIMO radar which adopts coherent pulse processing in each sub-array, such as \cite{WangLi11b,WangLi13TSP,LiWang15}.}.

When the waveform are orthogonal, the GLRT for distributed-colocated MIMO radar was obtained in non-Gaussian environment in \cite{ChongPascal10JSTSP}, and the expression for the PFA was given therein under the constraint that the product of the number of transmit elements and receive elements is the same for each pair of transmit-receive sub-array. The results for the PFA in \cite{ChongPascal10JSTSP} was generalized in \cite{ZhangLiu15} by eliminating the above constraint. For distributed-colocated MIMO radar with non-orthogonal waveform, the GLRT in Gaussian noise was derived in \cite{XuLi07TSP}, while the two-step Rao and Wald tests were given in \cite{KongCui10}. In non-Gaussian background, the 2S-GLRT, Rao test, and Wald test were exploited in \cite{CuiKong10RSN} for distributed-colocated MIMO radar with non-orthogonal waveform. Moreover, reference \cite{WangJiang11} considered the problem of detecting a mismatched signal in distributed-phased MIMO radar, and proposed three selective detectors.

Before closing this section, we summarize important progress in Table I.

\begin{table}[htbp]
\footnotesize
\begin{center}
{\small
\caption{Important Progress in Multichannel Adaptive Signal Detection}
}
\begin{tabular}{llll}
\hline
Year & \qquad\qquad Important Progress & Author(s) & Ref.  \\ \hline 
{1986} & \makecell[l]{first paper on adaptive detection} & \makecell[l]{Kelly} & \makecell[l]{\cite{Kelly86}} \\ \hline
{1989} & \makecell[l]{solid mathematical background for\\ adaptive signal detection} & {Kelly and Forsythe} &  {\cite{KellyForsythe89}} \\ \hline
\makecell[l]{1991\\ 1992} & \makecell[l]{well-known detector for point\\ targets: AMF} & \makecell[l]{Chen, Fobey, \emph{et al.}} & \makecell[l]{\cite{ChenReed91}\\\cite{RobeyFuhrmann92}} \\ \hline
1992 & \makecell[l]{tunable detector for mismatched\\ signals} & \makecell[l]{Kalson} & \makecell[l]{\cite{Kalson92}}\\ \hline
1992 & \makecell[l]{persymmetry structure based detect-\\or with limited training data} & \makecell[l]{Cai and Wang} & \makecell[l]{\cite{CaiWang92}} \\ \hline
\makecell[l]{1995/\\1999} & \makecell[l]{well-known detector for point\\ targets: ACE} & \makecell[l]{Conte, Lops,\\ Kraut, Scharf, \emph{et al.}} & \makecell[l]{\cite{ConteLops95}\\ \cite{KrautScharf99}} \\ \hline 1996 & \makecell[l]{subspace-based signal detection} & \makecell[l]{Raghavan, Pulsone,\\ \emph{et al.}} & \makecell[l]{\cite{RaghavanPulsone96}}\\ \hline
{1996} & \makecell[l]{direction detection} & \makecell[l]{Bose and Steinhardt} & \makecell[l]{\cite{BoseSteinhardt96a}}\\ \hline
1997 & \makecell[l]{distributed target detection} &  \makecell[l]{Gerlach, Steiner, \\ \emph{et al.}} & \makecell[l]{\cite{GerlachSteiner97}}\\ \hline
2000 & \makecell[l]{two-stage detector for mismatched\\ signals} & \makecell[l]{Pulsone and Zatman} & \makecell[l]{\cite{PulsoneZatman00}} \\ \hline
2000 & \makecell[l]{low-rank structure based detector\\ with limited training data} & \makecell[l]{Ayoub and\\ Haimovich} & \makecell[l]{\cite{AyoubHaimovich00}} \\ \hline
\makecell[l]{2000/\\2006} & \makecell[l]{parametric detector  with\\ limited training data} & \makecell[l]{Roman, Rangaswamy,\\ Li, Michels,  \emph{et al.}} & \makecell[l]{\cite{RomanRangaswamy00}\\\cite{LiMichels06}}\\ \hline
2001 & \makecell[l]{selective detector for mismatched\\ signals} & \makecell[l]{Pulsone and Rader} & \makecell[l]{\cite{PulsoneRader01}}\\ \hline
2003 & 
\makecell[l]{adatpive detectors based on Rao\\ and Wald tests} & \makecell[l]{Conte and De Maio} & \makecell[l]{\cite{ConteDeMaio03a}}\\ \hline
2004 & \makecell[l]{multiple target detection} & \makecell[l]{Gini, Bordoni, \\ \emph{et al.} } & \makecell[l]{\cite{GiniBordoni04}}\\ \hline
2005 & \makecell[l]{adaptive detection based on\\ convex optimization} & \makecell[l]{De Maio} & \makecell[l]{\cite{DeMaio05}} \\ \hline
2007 & \makecell[l]{adaptive detection in subspace\\ interference} & \makecell[l]{Bandiera, De Maio,\\ \emph{et al.}} & \makecell[l]{\cite{BandieraDeMaio07a}} \\ \hline
2007 & \makecell[l]{ Bayesian detector in heterogeneous\\ environment} & \makecell[l]{Besson, Tourneret,\\ \emph{et al.}} & \makecell[l]{\cite{BessonTourneret07}}\\ \hline
2008 & \makecell[l]{MIMO radar detection in unknown\\ noise} & \makecell[l]{Xu, Li, \emph{et al.}} & \makecell[l]{\cite{XuLi08AES}}\\ \hline
2010 & \makecell[l]{spectral symmetry based detector\\ with limited training data} & \makecell[l]{De Maio, Orlando, \\ \emph{et al.}} & \makecell[l]{\cite{DeMaioOrlando16TSP}}\\ \hline
2014 & \makecell[l]{Rao and Wald tests for complex-\\ valued signals with circularly\\ symmetric random parameters} & \makecell[l]{Liu, Wang, \emph{et al.}} & \makecell[l]{\cite{LiuWang14}}\\ \hline
2014 & \makecell[l]{double subspace signal detection} & \makecell[l]{Liu, Xie, \emph{et al.}} & \makecell[l]{\cite{LiuXie14b,LiuXie14a}}\\ \hline
2016 & \makecell[l]{Rao test for complex-valued signals\\ with circularly or non-circularly\\ symmetric random parameters} & \makecell[l]{Kay and Zhu} & \makecell[l]{\cite{KayZhu16TSP}} \\ \hline
2016 & \makecell[l]{adaptive detection based on \\ covariance structure classification} & \makecell[l]{Carotenuto, De Maio,\\ \emph{et al.}} & \makecell[l]{\cite{CarotenutoDeMaio17}} \\ \hline
\end{tabular}
\label{ch06_tab1}
\end{center}
\end{table}

\section{Typical adaptive detectors for different detection problems}
\label{Sec:Well-knownDetectors}
Multichannel adaptive signal detection was first investigated for a point target in 1986 by Kelly \cite{Kelly86}. Based on Kelly's work, all kinds of problems were dealt with, and numerous detectors were proposed.
In this section, we first summarize the statistical properties of many well-known detectors for point targets, since the statistical properties are the primary tool to evaluate the detection performance of the detectors. Then, we generalize the case of point target detection to distributed target detection and signal detection in the presence of interference\footnote{We choose the above three cases because they are representative in the field of adaptive detection and extensively studied in the literature.}.

\subsection{Adaptive detectors for point targets and their statistical distributions}
\label{subsec:8detectors}
Note that subspace signal model is more general than the rank-one signal model adopted in \eqref{test}. It is pointed out in \cite{ScharfFriedlander94} that the matched subspace detector is the general building block of signal processing, and it contains the rank-one matched filter or detector as a special case. Hence, in this subsection the detectors for point targets are all based on subspace signal model.

For the detection problem in \eqref{ch02_1}, if the signal $\sbf$ lies in a known subspace spanned by an $N\times p$ full-column-rank matrix $\Hbf$, then we have $\sbf=\Hbf\thetabf$, with $\thetabf$ being a $p\times1$ unknown coordinate vector. 
In the HE, the noise covariance matrix in the test data $\xbf$ is the same as that in the training data $\xbf_{\text{e},l}$. Then, for the detection problem in \eqref{ch02_1} with $\sbf$ being replaced by $\Hbf\thetabf$, the GLRT\cite{PastinaLombardo01}, Rao test\cite{LiuXie14b}, and Wald test\cite{LiuXie14b} are
\begin{equation}
\label{SGLRT}
t_\text{SGLRT}=
\frac{\tilde\xbf^H\Pbf_{\tilde\Hbf}\tilde\xbf}{1+\tilde\xbf^H\tilde\xbf-\tilde\xbf^H\Pbf_{\tilde\Hbf}\tilde\xbf},
\end{equation}
\begin{equation}
\label{SRao}
t_\text{SRao}=
\frac{\tilde\xbf^H\Pbf_{\tilde\Hbf}\tilde\xbf}{(1+\tilde\xbf^H\tilde\xbf)(1+\tilde\xbf^H\tilde\xbf-\tilde\xbf^H\Pbf_{\tilde\Hbf}\tilde\xbf)},
\end{equation}
and
\begin{equation}
\label{SAMF}
t_\text{SAMF}=
{\tilde\xbf^H\Pbf_{\tilde\Hbf}\tilde\xbf},
\end{equation}
respectively, where $\tilde\Hbf=\Sbf^{-\frac{1}{2}}\Hbf$ and $\Pbf_{\tilde\Hbf}=\tilde\Hbf(\tilde\Hbf^H\tilde\Hbf)^{-1}\tilde\Hbf^H$. The detectors in \eqref{SGLRT}-\eqref{SAMF} are referred as the SGLRT, SRao, and SAMF, respectively.

In the PHE, 
the noise covariance matrices in the test and training data can be modified as $\Rbf_{\text{e}}=\sigma^2\Rbf$ and $\Rbf$, respectively, with $\sigma^2$ being an unknown positive scaling factor, standing for the power mismatch between the test and training data.
In the PHE, the GLRT, Rao test, and Wald test coincide with each other, and are found to be \cite{LiuXie14a}
\begin{equation}
\label{ASD}
t_\text{ASD}= \frac{\tilde\xbf^H\Pbf_{\tilde\Hbf}\tilde\xbf}{\tilde\xbf^H\tilde\xbf},
\end{equation}
which is named as ASD in \cite{KrautScharf01}. Note that the SGLRT, SRao, SAMF, and ASD are the subspace generalizations of the KGLRT, DMRao, AMF, and ACE, respectively.

The above four detectors are designed without taking into account the possibility of signal mismatch.
On the one hand, signal mismatch may be caused by antenna error, mutual coupling, or target maneuvering.
On the other hand, signal mismatch can also be caused by a strong target or a jamming signal located in the radar sidelobe, generated by the ECM.
For different sources of signal mismatch, different directivity properties (the capability of selectivity or robustness to signal mismatch) of the detector are preferred.
For the first case, a robust detector is needed, which maintains good detection performance in the presence of signal mismatch. In contrast, for the second case, a selective detector is preferred, whose detection performance decreases rapidly  with the increase of signal mismatch.

To design selective detectors in the case of signal mismatch, an effective approach is adding artificially determinant factitious jammer under hypothesis $\text{H}_0$ \cite{PulsoneRader01}. Then, the detection problem in \eqref{ch02_1} can be modified to be
\begin{equation}
\label{ch02_selective}
\left\{ {\begin{array}{l}
	\text{H}_0 :{ {{ \xbf}}}={ { \nbf}} +\qbf , ~\xbf_{\text{e},l}=\nbf_{\text{e},l},~ l=1,2,\cdots,L,\\
	\text{H}_1 :{ {{\xbf}}}= {\Hbf\thetabf}+{ { \nbf}} , ~\xbf_{\text{e},l}=\nbf_{\text{e},l},~ l=1,2,\cdots,L,\\ 
	\end{array}} \right.
\end{equation}
where the $N\times1$ unknown vector $\qbf$ denotes the artificially injected determinant factitious jammer. Remarkably, the injection of the factitious jammer $\qbf$ makes the resulting detector tend to choose hypothesis $\text{H}_0$ if signal mismatch occurs.
When $\qbf$ is constrained to be orthogonal to the signal subspace in the quasi-whitened space, i.e.,
\begin{equation}
\label{HiSq0}
\Hbf^H\Sbf^{-1}\qbf=\mathbf{0}_{p\times1},
\end{equation}
the GLRT for the detection problem in \eqref{ch02_selective} is 
\begin{equation}
\label{SABORT}
t_\text{SABORT}=\frac{1+\tilde\xbf^H\Pbf_{\tilde\Hbf}\tilde\xbf}{1+\tilde\xbf^H\tilde\xbf-\tilde\xbf^H\Pbf_{\tilde\Hbf}\tilde\xbf},
\end{equation}
which is a special case of the adaptive direction detector with mismatched signal rejection of type 2 (ADD-MSR2) in \cite{LiuLiu2017AES}, and a subspace generalization of the ABORT, proposed in \cite{PulsoneRader01}. Hence, for convenience, the detector in \eqref{SABORT} is referred to as the subspace-based ABORT (SABORT).

If the determinant factitious jammer in \eqref{HiSq0} is modified as
\begin{equation}
\label{HiRq0}
\Hbf^H\Rbf^{-1}\qbf=\mathbf{0}_{p\times1}.
\end{equation}
That is to say, the factitious jammer is orthogonal to the signal subspace in the truely whitened space. Then the
GLRT for the detection problem in \eqref{ch02_selective} becomes 
\begin{equation}
\label{WSABORT}
t_\text{W-SABORT}=\frac{1+\tilde\xbf^H\tilde\xbf}{(1+\tilde\xbf^H\tilde\xbf-\tilde\xbf^H\Pbf_{\tilde\Hbf}\tilde\xbf)^2},
\end{equation}
which is a special case of the adaptive direction detector with mismatched signal rejection of type 1 (ADD-MSR1) in \cite{LiuLiu2017AES}, and a subspace generalization of the W-ABORT in \cite{BandieraBesson07WABORT}. The detector in \eqref{WSABORT} is denoted as the whitened SABORT (W-SABORT) for convenience.

Moreover, another approach to devise a selective detector is injecting an unknown, rank-one, noise-like, fictitious jammer $\vbf$ under both hypotheses. As a results, the noise covariance matrix in the test data becomes $\Rbf=\Rbf_{\text{e}}+\vbf\vbf^H$. Then, the selective detector derived according to the Rao test is
\begin{equation}
\label{DNSAMF}
t_\text{DN-SAMF}=\frac{\tilde\xbf^H\Pbf_{\tilde\Hbf}\tilde\xbf}{\tilde\xbf^H\tilde\xbf(1+\tilde\xbf^H\tilde\xbf-\tilde\xbf^H\Pbf_{\tilde\Hbf}\tilde\xbf)},
\end{equation}
which is a special case of the Rao test in \cite{LiuHan17SCIS} 
and a subspace generalization of the DN-AMF in \cite{OrlandoRicci10}. For convenience, the detector in \eqref{DNSAMF} is denoted as the doubly normalized SAMF (DN-SAMF).

Different from the above devised selective detectors, a robust detector to the signal mismatch may be preferred in many applications. A robust detector can be designed by assuming the desired signal to be detected is completely unknown. In other words, the signal $\sbf$ in \eqref{ch02_1} is unknown, or equivalently, the dimension of the signal matrix $\Hbf$ is $N\times N$. Then the corresponding GLRT is given by \cite{RaghavanQiu95}
\begin{equation}
\label{AED}
t_\text{AED}= \tilde\xbf^H \tilde\xbf,
\end{equation}
which can be denoted as AED. It is shown in \cite{LiuXie14b} that the Rao test and Wald test are both equivalent to the GLRT, i.e., the AED in \eqref{AED}.


Since the case of signal match can be taken as a special case of signal mismatch (i.e., the mismatched angle is zero), we only summarize the statistical properties of the above detectors in the presence of signal mismatch.
As mentioned above, the statical performance of the detectors in the presence of signal mismatch was first dealt with by Kelly in \cite{Kelly89} for the KGLRT in the case of rank-one signal. Based on this result, the statistical performance of the SGLRT, SAMF, and ASD was given in \cite{LiuLiu16SP}. In the following, we summarize the statical properties of the above eight detectors, some of which were not found in the open literature. 

To obtain the statical distributions of the detectors, it is convenient to introduce the following quantity
\begin{equation}
\label{betaHE}
\beta=\frac{1}{1+\tilde{\xbf}^H\tilde\xbf-\tilde{\xbf}^H\Pbf_{\tilde\Hbf}\tilde\xbf},
\end{equation}
which can be taken as a loss factor.

If signal mismatch happens, the actual signal, denoted as $\sbf_0$, may not completely lie in the signal subspace spanned by the columns of $\Hbf$.
Then, it is shown in \cite{LiuLiu16SP} that the statistical distribution of the SGLRT in \eqref{SGLRT}, with $\beta$ given, under hypothesis $\text{H}_1$ is complex noncentral F-distribution, with $p$ and $L-N+1$ DOF and a noncentrality parameter $\beta\rho~\text{cos}^2\phi$, written symbolically as
\begin{equation}
\label{ch03_SGLRT_distribution22}
t_{\text{SGLRT}}|[\beta,\text{H}_1]\sim{\cal{CF}}_{p,L-N+1}\left(\beta\rho_{\text{pnt}} ~\text{cos}^2 \phi \right),
\end{equation}
where $\rho_{\text{pnt}}$ is the output SNR, defined as
\begin{equation}
\label{ch03_SNR_rho}
\rho_\text{pnt}=\sbf_0^H\Rbf^{-1 }\sbf_0,
\end{equation}
\begin{equation}
\label{ch03_cos2phi}
\text{cos}^2\phi=\frac{\sbf_0^H\Rbf^{-1}\Hbf(\Hbf^H\Rbf^{-1}\Hbf)^{-1} \Hbf^H\Rbf^{-1}\sbf_0}  {\sbf_0^H\Rbf^{-1}\sbf_0},
\end{equation}
and the notation ``$|[\beta,\text{H}_1]$'' denotes the fact that the above statistical distribution holds under hypothesis $\text{H}_1$ on the condition that $\beta$ is given. Equation \eqref{ch03_cos2phi} can be rewritten as
\begin{equation}
\label{ch03_cos2phi2}
\text{cos}^2\phi=\frac{\bar\sbf_0^H\Pbf_{\bar\Hbf}\bar\sbf_0}  {\bar\sbf_0^H\bar\sbf_0},
\end{equation}
where $\sbf_0=\Rbf^{-\frac{1}{2}}\sbf_0$, $\bar\Hbf=\Rbf^{-\frac{1}{2}}\Hbf$,  $\Pbf_{\bar\Hbf}=\bar\Hbf(\bar\Hbf^H\bar\Hbf)^{-1}\bar\Hbf^H$.
It follows from \eqref{ch03_cos2phi2} that the quantity $\text{cos}^2\phi$ measures cosine-squared of the angle between the whitened actual signal $\bar\sbf_0$ and the whitened nominal signal subspace spanned by the columns of $\bar\Hbf$. $\text{cos}^2\phi$ plays a key role in controlling the detection performance of a detector in the presence of signal mismatch. This is numerically shown in the next section.

Moreover, it is shown in \cite{LiuLiu16SP} that the statistical distribution of the loss factor $\beta$ in \eqref{betaHE} under hypothesis $\text{H}_1$ is a complex noncentral Beta distribution, with $L-N+p+1$ and $N-p$ DOFs and a noncentrality parameter $\delta^2$, written symbolically as
\begin{equation}
\label{ch03_sBeta_distribution}
\beta|\text{H}_1\sim{\cal C}{\cal B}_{L-N+p+1,N-p}(\delta^2),
\end{equation}
where
\begin{equation}
\label{ch03_delta33}
\delta^2=\rho_{\text{pnt}}~\text{sin}^2\phi,
\end{equation}
and $\text{sin}^2\phi=1-\text{cos}^2\phi$.

In contrast, under hypothesis $\text{H}_0$, the statistical distributions of the SGLRT in \eqref{SGLRT} and the loss factor $\beta$ in \eqref{betaHE} become 
\begin{equation}
\label{ch03_SGLRT_distribution00}
t_{\text{SGLRT}}|[\beta,\text{H}_0]\sim{\cal{CF}}_{p,L-N+1},
\end{equation}
and
\begin{equation}
\label{ch03_betas_distribution00}
\beta|\text{H}_0\sim{\cal C}{\cal B}_{L-N+p+1,N-p},
\end{equation}
respectively.

The analytical expressions for the PDF and cumulative distribution function (CDF) of the complex noncentral F-distribution and complex noncentral Bea distribution were exploited in detail in Kelly and  Forsythe's classic report \cite{KellyForsythe89}, also summarized in \cite{Richmond00b,BandieraOrlando09a}. One can use these CDFs and PDFs to derive the expressions for the PDs and PFAs of the above detectors.

It is straightforward to verify that the following seven equations hold
\begin{equation}
\label{ch03_SAMF_SGLRT}
t_\text{SAMF}=\frac{t_\text{SGLRT}}{\beta},
\end{equation}
\begin{equation}
\label{ch03_ASD_SGLRT}
t_\text{ASD}=\frac{t_\text{SGLRT}}{1-\beta},
\end{equation}
\begin{equation}
\label{ch03_SRao_SGLRT}
t_\text{SRao}=\frac{\beta t_\text{SGLRT}}{1+t_\text{SGLRT}},
\end{equation}
\begin{equation}
\label{ch03_SABORT_SGLRT}
t_\text{SABORT}= \beta + t_\text{SGLRT},
\end{equation}
\begin{equation}
\label{ch03_WABORT_SGLRT}
t_\text{W--SABORT}= (1+t_\text{SGLRT})\beta,
\end{equation}
\begin{equation}
\label{ch03_DNSAMF_SGLRT}
t_\text{DN--SAMF}=\frac{\beta t_\text{SGLRT}}{(1-\beta)(1-\beta+t_\text{SGLRT})},
\end{equation}
\begin{equation}
\label{ch03_AED_SGLRT}
t_\text{AED}=\frac{1-\beta+t_\text{SGLRT}}{\beta}.
\end{equation}
Based on the conditional distribution of the SGLRT in \eqref{ch03_SGLRT_distribution22} and the statistical distribution of the loss factor $\beta$ in \eqref{ch03_sBeta_distribution}, 
along with the statistical dependences in \eqref{ch03_SAMF_SGLRT}-\eqref{ch03_AED_SGLRT}, one can readily obtain analytical expressions for the PDs and PFAs of the detectors. Interested readers can refer to \cite{LiuLiu16SP} for examples.
Note that one can obtain the expressions for the PD and PFA of the AED in a more direct manner by deriving the statistical distribution of the AED \cite{LiuWang14b}. Precisely, according to Theorem 3.2.13 in \cite[p.98]{Muirhead05} or Theorem 5.2.2 in \cite[p.176]{Anderson03}, the statistical distribution of the AED in \eqref{AED} under hypotheses $\text{H}_1$ and $\text{H}_0$ are
\begin{equation}
\label{ch02_SD_AED1}
t_{\text{AED}}|{\text{H}_1}\sim{\cal CF}_{N,L-N+1}(\rho_{\text{pnt}})
\end{equation}
and
\begin{equation}
\label{ch02_SD_AED0}
t_{\text{AED}}|{\text{H}_0}\sim{\cal CF}_{N,L-N+1},
\end{equation}
respectively.

In order to evaluate the detection performance of the detectors under different numbers of training data, we consider the detector with known noise covariance matrix. Precisely, when $\Rbf$ is known, the GLRT for the detection problem in \eqref{ch02_1} with $\sbf$ being replaced by $\Hbf\thetabf$ is 
\begin{equation}
\label{ch03_SMF}
t_{\text{SMF}}
={\xbf^H\Rbf^{-1}\Hbf(\Hbf^H\Rbf^{-1}\Hbf)^{-1}\Hbf^H\Rbf^{-1}\xbf}, 
\end{equation}
which is referred to as the subspace-based matched filter (SMF). It 
 can also be obtained by the criteria of GLRT, Rao and Wald tests. The statistical distribution of the SMF in \eqref{ch03_SMF} under hypothesis $\text{H}_1$ is a complex noncentral Chi-square distribution with $p$ DOFs and a noncentrality parameter $\rho$ \cite{LiuWang16AES}, written symbolically as
\begin{equation}
\label{ch03_SD_SMF1}
{t_\text{SMF}}|\text{{H}}_1\sim{\cal C}\chi _p^2({\rho_{\text{pnt}}}).
\end{equation}
Under hypothesis  $\text{H}_0$, the above distribution becomes central, i.e.,
\begin{equation}
\label{ch03_SD_SMF0}
{t_\text{SMF}}|\text{{H}}_0\sim{\cal C}\chi _p^2
\end{equation}

%

\subsection{Numerical examples}

In this subsection we compare the detection performance of the detectors with numerical examples, and only focus on the case of HE. Two cases are considered, namely, the case of no signal mismatch and the case of signal mismatch. 
The PD curves of all detectors are obtained by using the theoretical results, and confirmed by Monte Carlo simulations, which are not shown for a clear display.

\begin{figure}
	\centering
	\includegraphics[width=0.6\textwidth]{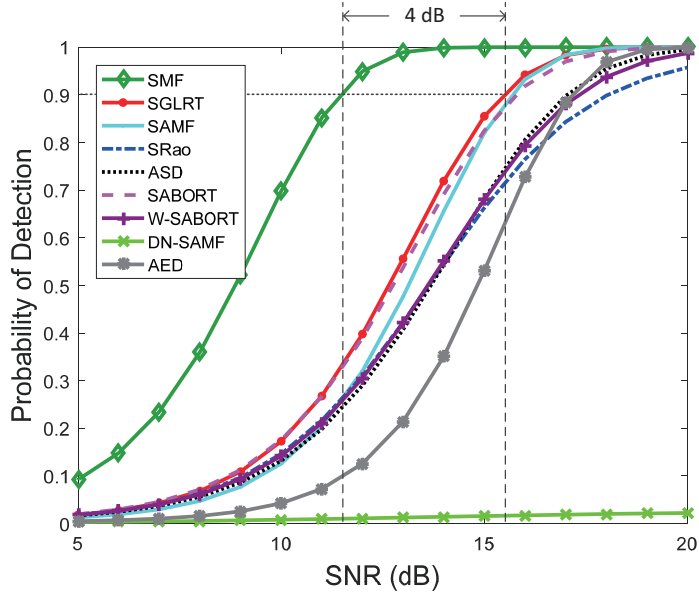}
	\caption{PD versus SNR. $N=12$, $p=2$, $L=2N$, and $\text{PFA}=10^{-3}$.}
	\label{fig_Pd_vs_SNR} 
\end{figure}

Figure \ref{fig_Pd_vs_SNR} compares the detection performance of the adaptive detectors under different SNRs in the absence of signal mismatch. For comparison purpose, the result for the SMF is also reported.
The results indicate that, for the chosen parameters, the SGLRT, among the eight adaptive detectors, has the highest PD and slightly better than the SAMF and SABORT, the DN-SAMF has the lowest PD, and the PDs of the ASD, W-SABORT, SRao, and AED are in between. Moreover, the detection performance loss of the SGLRT in terms of SNR is roughly 4 dB when $\text{PD}=0.9$, compared with the SMF.
This is quite different from adaptive filtering, since it is well-known from the RMB rule \cite{ReedMallett74} that $2N$ independent identically distributed (IID) training data can maintain 3 dB SNR loss, compared with the optimum filter.
The above detection loss is owing to two factors \cite{Kelly86}. One is the effective SNR loss factor (similar to adaptive filtering), and the other is the CFAR loss of the adaptive detectors. The effective SNR loss factor depends roughly on the ratio of $L$ to $N$, while the CFAR loss depends solely on $L$, whose increase results in the decrease of the CFAR loss.



\begin{figure}[tbp]
	\centering
	\includegraphics[width=0.6\textwidth] {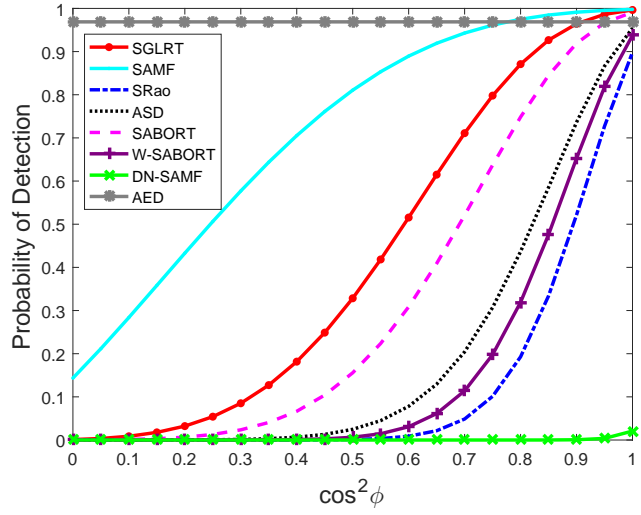}\\
	\caption{PD versus $\cos^2\phi$. $N=12$, $p=2$, $L=2N$, $\text{SNR}=18~\text{dB}$, and $\text{PFA}=10^{-3}$. }
	\label{fig_Pd_vs_cos2_N12p2L2N}
\end{figure}

Figure \ref{fig_Pd_vs_cos2_N12p2L2N} shows the detection performance of the adaptive detectors under different amount of signal mismatch. As expected,  the AED is the most robust and its PD does not vary with the change of $\cos^2\phi$. However, its PD cannot attain unity for the chosen parameters. The robustness of the SAMF, SGLRT, SABORT, ASD, W-SABORT, SRao, and DN-SAMF reduces in sequence.

\begin{figure}[tbp]
	\centering
	\includegraphics[width=0.35\textwidth] {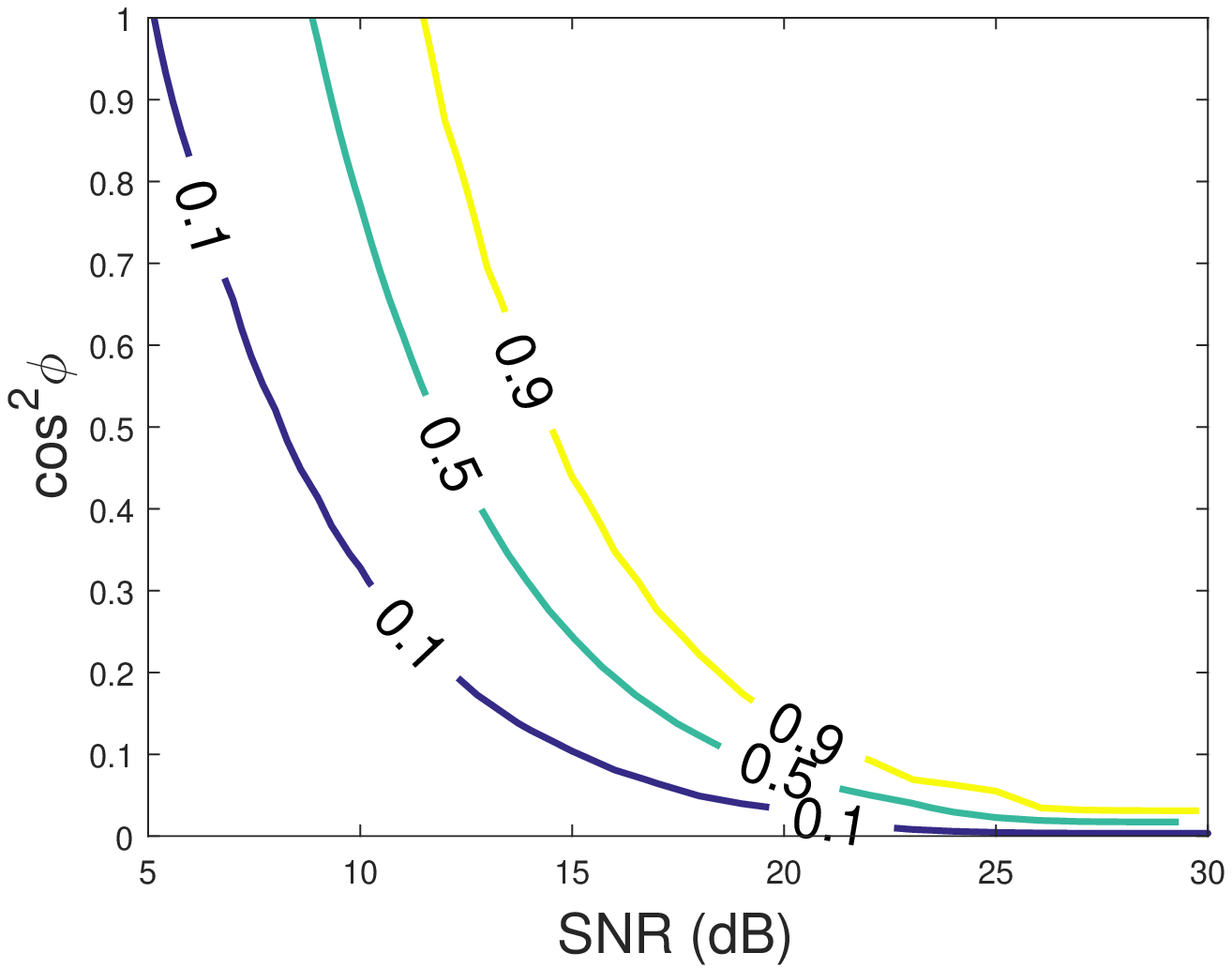}
	\includegraphics[width=0.35\textwidth] {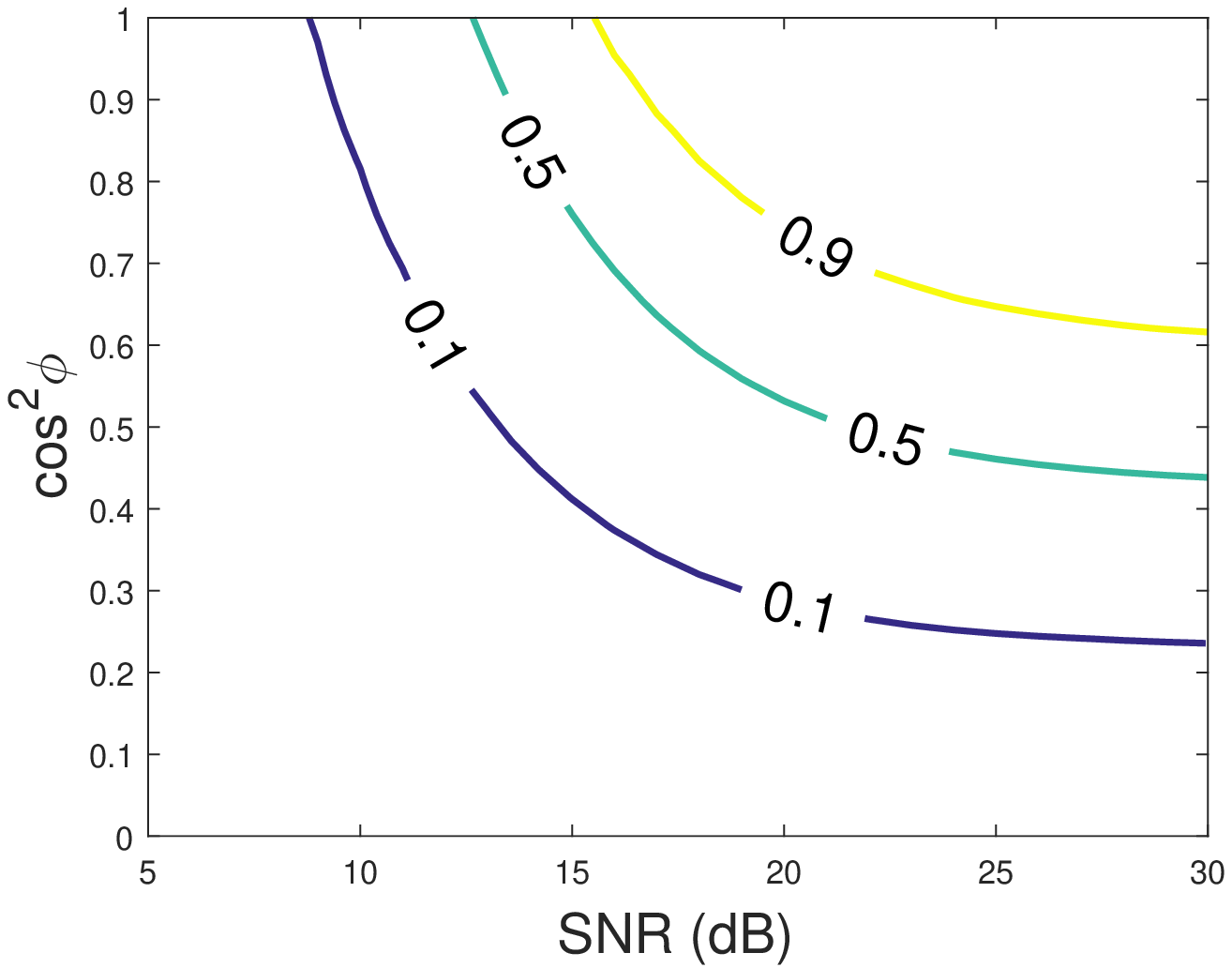}\\
	 {\scriptsize (a) SMF \qquad \qquad \qquad  \qquad \qquad \qquad  (b) SGLRT}\\
	\includegraphics[width=0.35\textwidth] {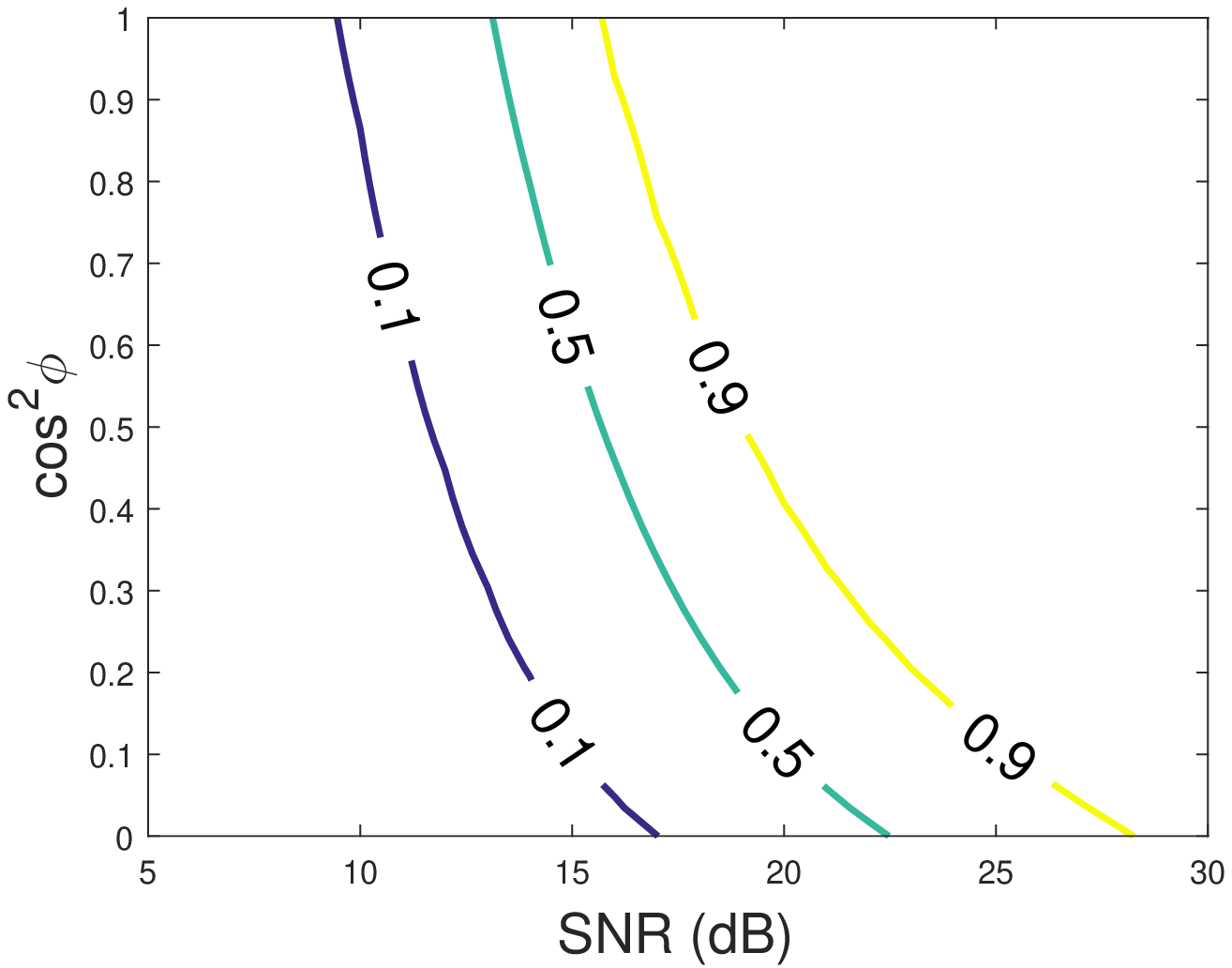}
	\includegraphics[width=0.35\textwidth] {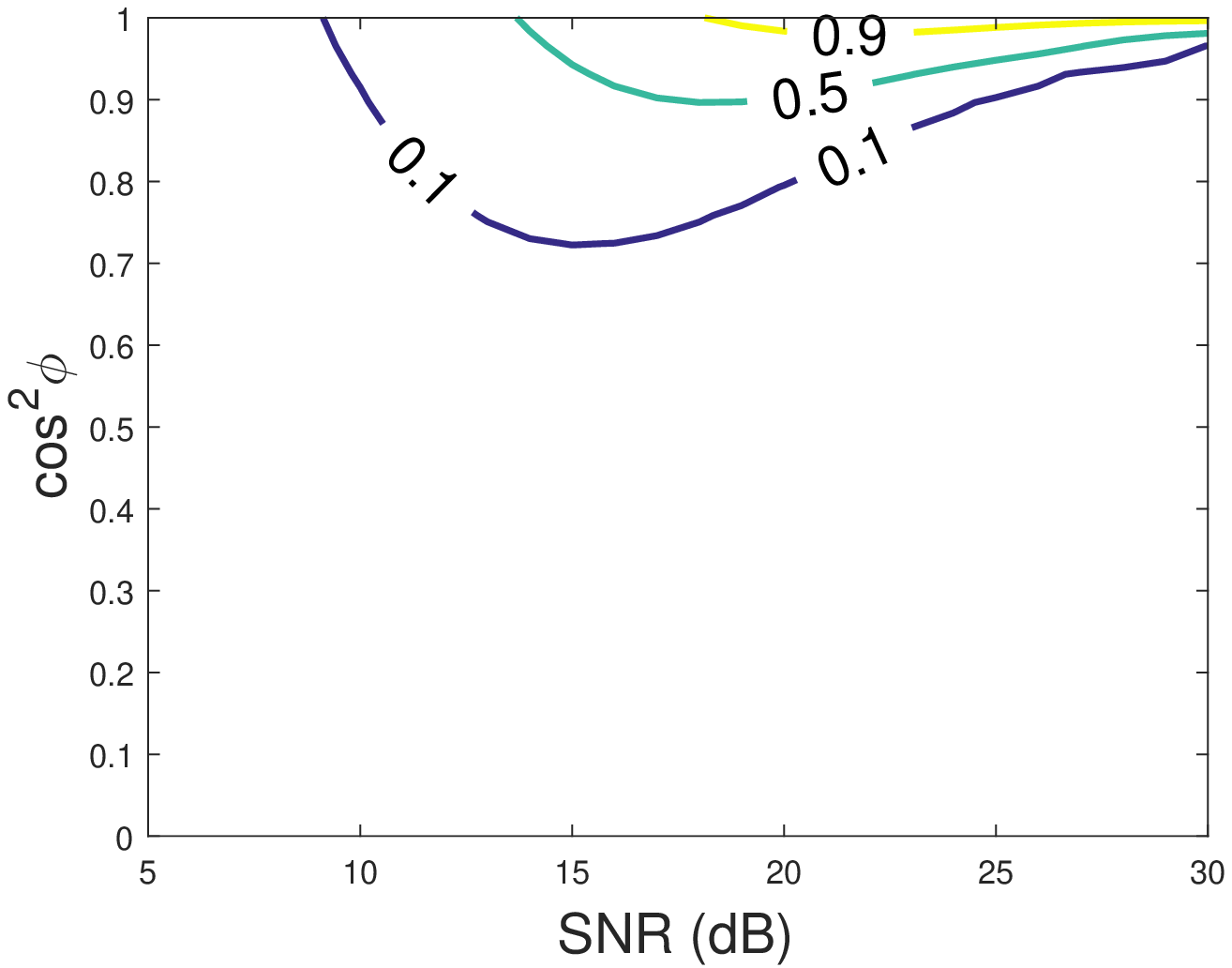}\\
	 {\scriptsize (c) SAMF \qquad \qquad \qquad  \qquad \qquad \qquad (d) SRao}\\
	\includegraphics[width=0.35\textwidth] {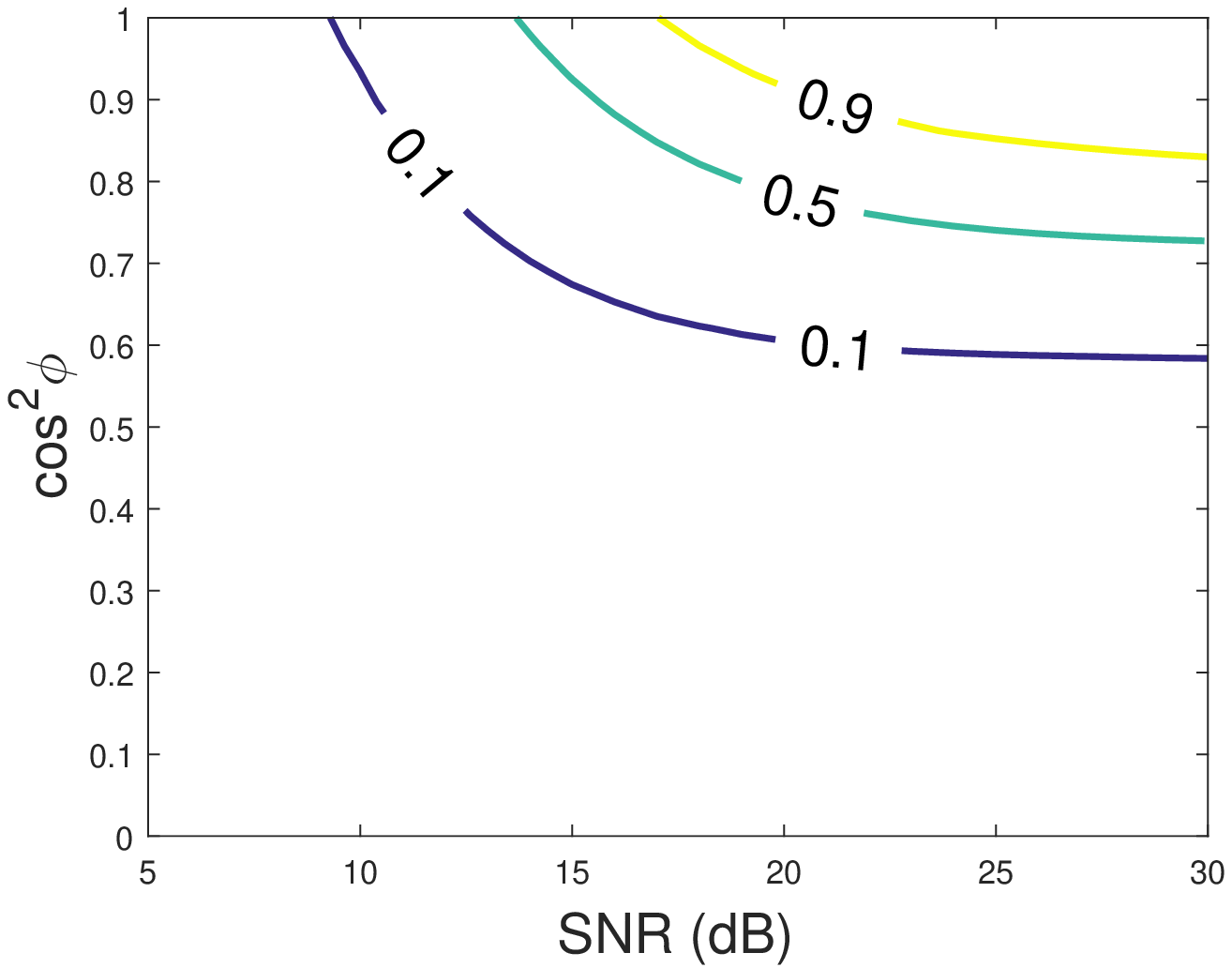}
	\includegraphics[width=0.35\textwidth] {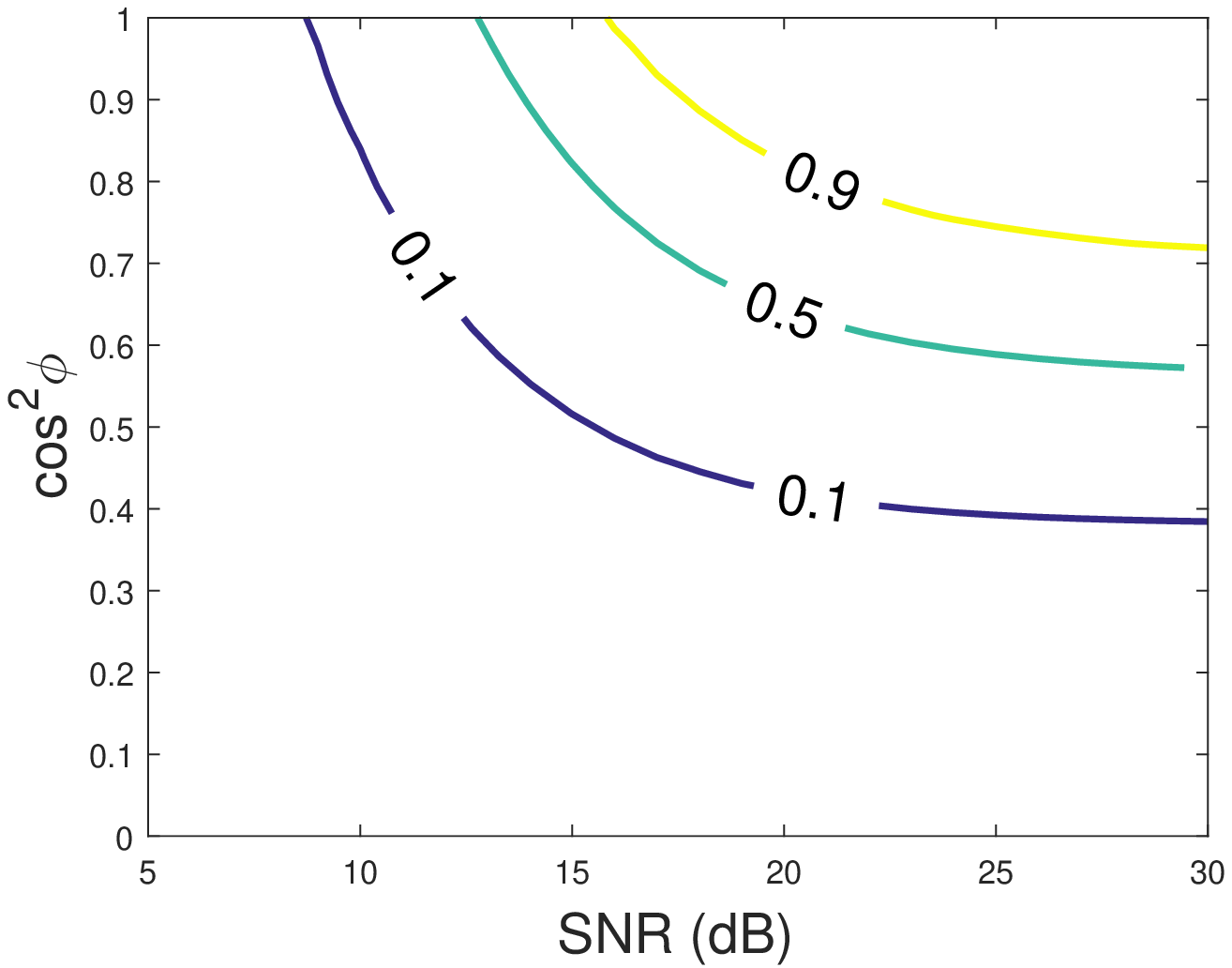}\\
	 {\scriptsize (e) ASD \qquad \qquad \qquad  \qquad \qquad \qquad  (f) SABORT}\\
	\includegraphics[width=0.35\textwidth] {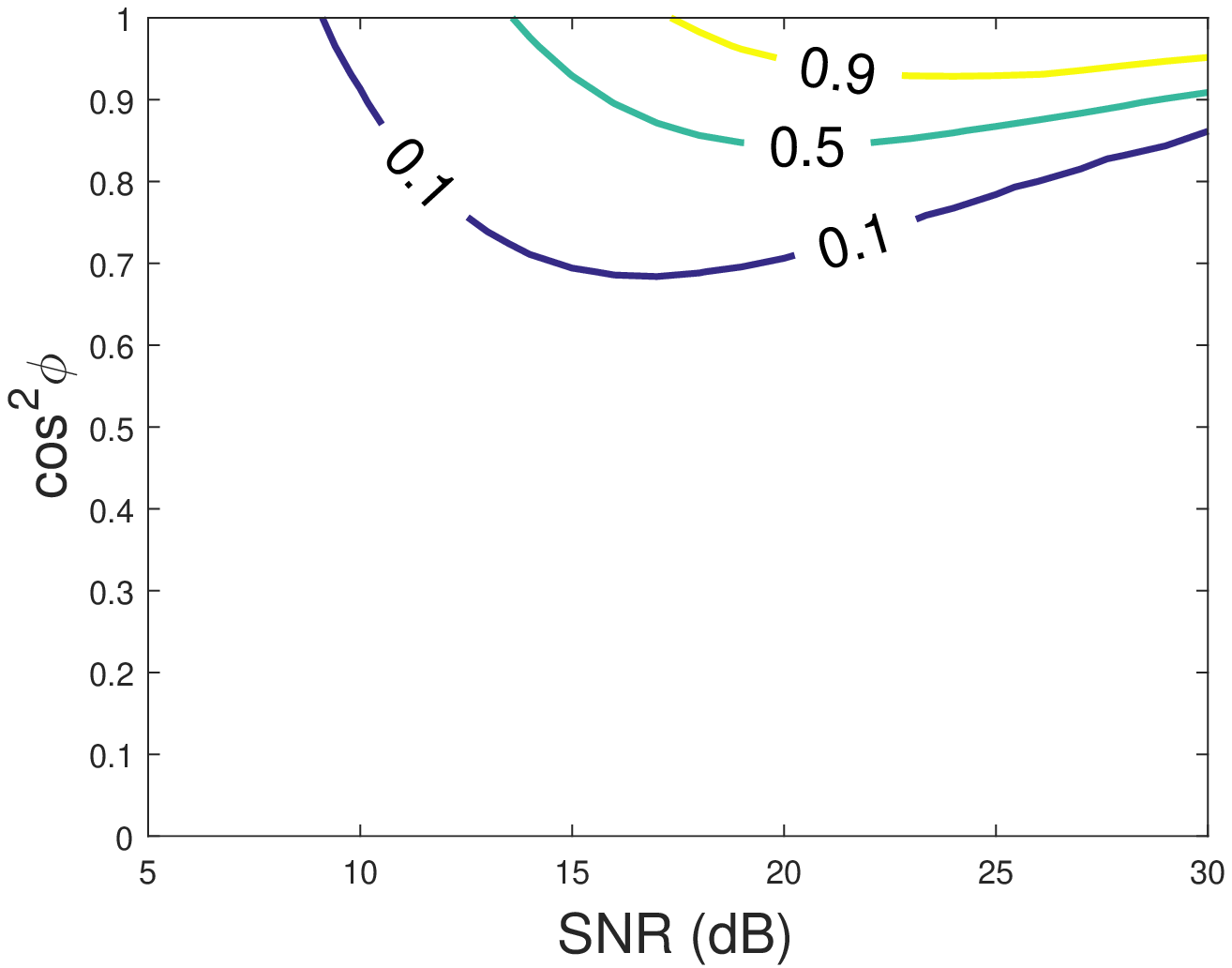}
	\includegraphics[width=0.4\textwidth] {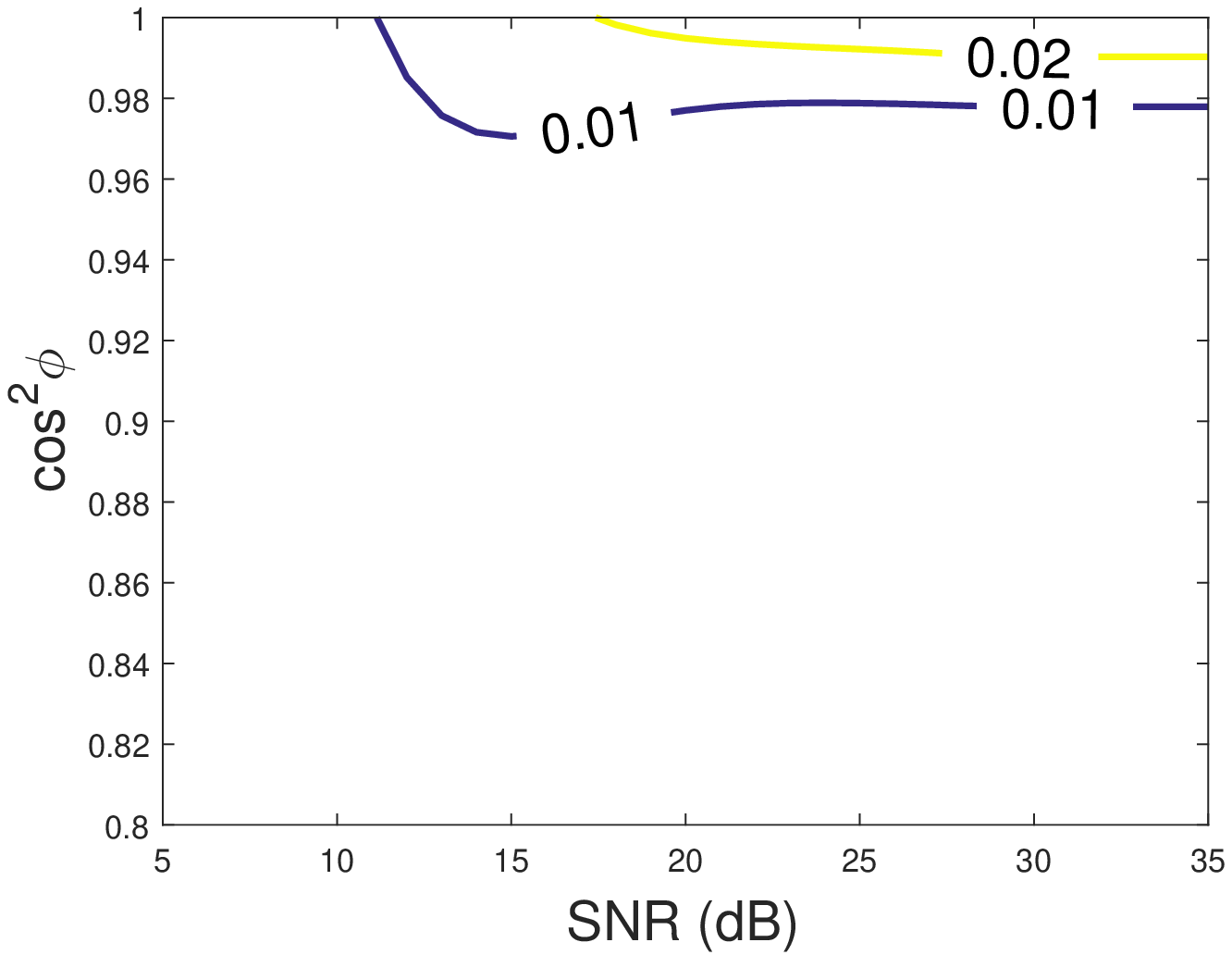}\\
   {\scriptsize (g) W-SABORT \qquad \qquad \qquad \qquad \qquad \qquad (h) DN-SAMF}\\
	\includegraphics[width=0.4\textwidth] {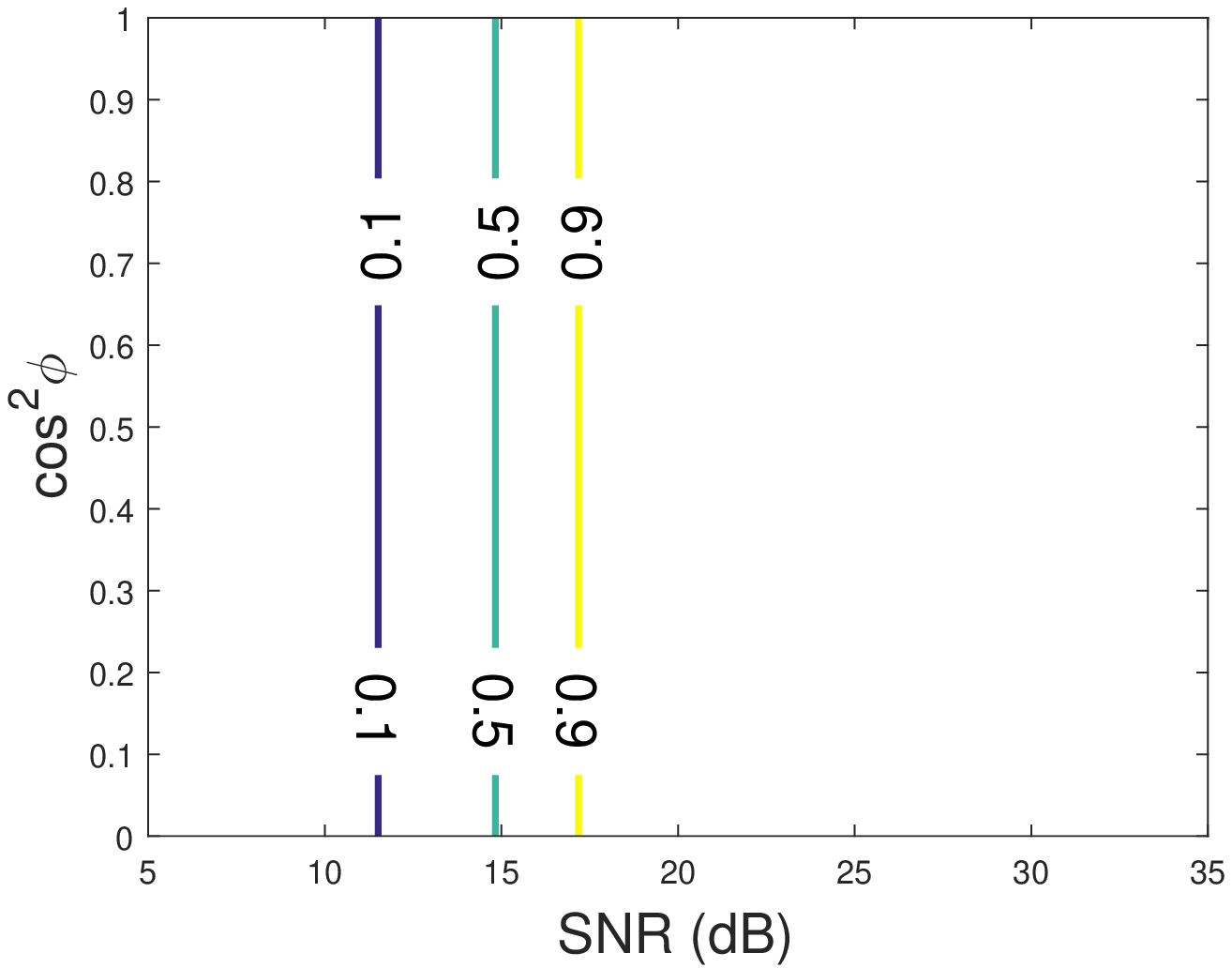}\\
   {\scriptsize (i) AED}\\
\caption{Contours of the PDs versus SNR and $\cos^2\phi$. $N=12$, $p=2$, $L=2N$, and $\text{PFA}=10^{-3}$. }
	\label{fig_Contour_N12p2L2N}
\end{figure}

Another method to illustrate the detection performance for mismatched signals is showing the contours of PDs as functions of SNR and $\cos^2\phi$, first introduced in \cite{PulsoneRader01} and named as mesa plot. This is displayed in Figure \ref{fig_Contour_N12p2L2N} for the above detectors. 
The directivities of the detectors are the same as those in Figure \ref{fig_Pd_vs_cos2_N12p2L2N}. However, more information can be inferred from Figure \ref{fig_Contour_N12p2L2N}.
Taking the SAMF for example, it is very robust to signal mismatch. It can provide a PD as high as 0.9 as long as the SNR is high enough, even the whitened actual signal is orthogonal to the whitened nominal signal subspace, i.e., the case of $\cos^2\phi=0$.
In contrast, for a selective detector, such as the SABORT, it does not achieve a PD higher than 0.5 when $\cos^2\phi<0.55$, no matter how high the SNR is.
It is worth pointing out for the chosen parameters, the SAMF and SABORT have comparable PDs for matched signals as shown in Figure \ref{fig_Pd_vs_SNR}.
Hence, if a selective detector is needed, the SABORT is a better candidate than the SAMF.

Before closing this section, we would like to give the following three remarks. First, it is known from \eqref{ch03_SGLRT_distribution00}, \eqref{ch03_betas_distribution00}, and \eqref{ch03_SAMF_SGLRT}-\eqref{ch03_AED_SGLRT} that all the adaptive detectors exploited above have the CFAR property with respect to the noise covariance matrix $\Rbf$.
Second, only the ASD, among the above eight adaptive detectors, possesses the CFAR property in PHE, 
although the ASD has lower PD than some other detectors in HE.
Third, the DN-SAMF can behave quite well when the number of system dimension $N$ is large enough, as shown in \cite{OrlandoRicci10}.

\subsection{Generations of point-target-based adaptive detectors}
The detection problem in \eqref{ch02_1} has been generalized in many aspects, and hence, many other adaptive detectors have been proposed besides the ones shown in the above subsection.
Distributed target detection (without interference) and signal detection in interference are two import generations, which will be shown below.

\subsubsection{Adaptive detectors for distributed targets}
A large target usually occupies multiple range bins, especially for high-resolution radar system \cite{LongLiang19}. In this case, the detection problem in \eqref{ch02_1} should be modified as
\begin{equation}
\label{test_distributedTargt}
\left\{ {\begin{array}{l}
 \text{H}_0 :{ {{ \Xbf}}}={ { \Nbf}} , ~\xbf_{\text{e},l}=\nbf_{\text{e},l},~ l=1,2,\cdots,L,\\
 \text{H}_1 :{ {{\Xbf}}}= {\sbf\abf^H}+{ { \Nbf}} , ~\xbf_{\text{e},l}=\nbf_{\text{e},l},~ l=1,2,\cdots,L,\\ 
\end{array}} \right.
\end{equation}
where $\Xbf$ is an $N\times K$ matrix denoting the test data, with $N$ being the number of system channels and $K$ being the number of range bins occupied by the distributed target, $\Nbf$ is the noise in the test data, $\sbf$ is the signal steering vector, $\abf$ is the coordinate vector of the signal, $\xbf_{\text{e},l}$ is the $l$th training data vector, and $\nbf_{\text{e},l}$ is the noise in $\xbf_{\text{e},l}$.
The columns of $\Nbf$ are IID, having the noise covariance matrix $\Rbf_t$. Denote the noise covariance matrix of $\nbf_{\text{e},l}$ as $\Rbf$. Then, in HE, $\Rbf_t=\Rbf$, while in PHE $\Rbf_t=\sigma^2\Rbf$, with $\sigma^2$ being the unknown power mismatch between the test data and training data.

For the detection problem in \eqref{test_distributedTargt}, the GLRT and its two-step variation for the HE and PHE were all proposed in \cite{ConteDeMaio01}. Precisely, for the HE, the GLRT and 2S-GLRT are
\begin{equation}
\label{ch04_GKGLRT2}
t_{\text{GKGLRT}}=\frac{\tilde\sbf^H\tilde\Xbf(\Ibf_K+\tilde\Xbf^H \tilde\Xbf)^{-1} \tilde\Xbf^H\tilde\sbf} {\tilde\sbf^H\tilde\sbf- \tilde\sbf^H\tilde\Xbf(\Ibf_K+\tilde\Xbf^H\tilde\Xbf)^{-1}\tilde\Xbf^H\tilde\sbf}
\end{equation}
and
\begin{equation}
\label{ch04_GAMF}
t_{\text{GAMF}}=\frac{\tilde\sbf^H\tilde\Xbf\tilde\Xbf^H\tilde\sbf} {\tilde\sbf^H\tilde\sbf},
\end{equation}
respectively.
Moreover, for the PHE, the GLRT and 2S-GLRT are
\begin{equation}
\label{ch04_glrt_PHE}
t_{\text{GLRT--PHE}}=\frac{(\hat\sigma_0^2)^{\frac{NK}{L+K}} \left|\Ibf_K+\dfrac{1}{\hat\sigma_0^2}{\tilde\Xbf^H\tilde\Xbf}\right|} {(\hat\sigma_1^2)^{\frac{NK}{L+K}}\left|\Ibf_K+\dfrac{1}{\hat\sigma_1^2} {\tilde\Xbf^H\Pbf_{\tilde\sbf}^\bot\tilde\Xbf}\right|}
\end{equation}
and
\begin{equation}
\label{ch04_GASD}
t_{\text{GASD}}=\frac{\tilde\sbf^H\tilde\Xbf\tilde\Xbf^H\tilde\sbf} {\tilde\sbf^H\tilde\sbf~\text{tr}(\tilde\Xbf^H\tilde\Xbf)},
\end{equation}
respectively. In \eqref{ch04_glrt_PHE}, $\hat\sigma_0^2$ and $\hat\sigma_1^2$ are the sole solutions of
\begin{equation}
\label{ch04_solution_sigma0}
\sum_{k_0=1}^{r_0}\frac{\lambda_{k_0}}{\lambda_{k_0}+\sigma^2}=\frac{NK}{L+K}
\end{equation}
and
\begin{equation}
\label{ch04_solution_sigma1}
\sum_{k_1=1}^{r_1}\frac{\xi_{k_1}}{\xi_{k_1}+\sigma^2}=\frac{NK}{L+K},
\end{equation}
respectively, where $r_0=\min(N,K)$, $r_1=\min(N-1,K)$, $\lambda_{k_0}$ is the ${k_0}$th non-zero eigenvalue of
$\tilde\Xbf^H\tilde\Xbf$, 
$k_0=1,2,\cdots,r_0$, and $\xi_{k_1}$ is the $k_1$th non-zero eigenvalue of
$\tilde\Xbf^H\Pbf_{\tilde\sbf}^{\bot}\tilde\Xbf$, 
$ k_1=1,2,\cdots,r_1$.

The detectors in \eqref{ch04_GAMF} and \eqref{ch04_GASD} were referred to as generalized AMF (GAMF) and generalized adaptive subspace detector (GASD), respectively in \cite{ConteDeMaio01}. For convenience, the detector in \eqref{ch04_GKGLRT2} is denoted as GKGLRT in this paper.

Moreover, for the detection problem in \eqref{test_distributedTargt} in HE, the Wald test is the same as the GAMF, while the Rao test was proposed in \cite{ShuaiKong12}, described as\footnote{Using matrix inversion lemma, it is easy to show that equation \eqref{ch04_GRao} can be recast as $t_{\text{Rao-HE}}=\frac{\tilde\sbf^H\tilde{\Xbf} (\Ibf_K+\tilde\Xbf^H \Pbf_{\tilde\sbf}^\bot\tilde\Xbf)^{-1} (\Ibf_K+\tilde\Xbf^H\tilde\Xbf)^{-1}\tilde\Xbf^H\tilde\sbf}  {\tilde\sbf^H\tilde\sbf}$. }
\begin{equation}
\label{ch04_GRao}
t_{\text{Rao-HE}}=\frac{\sbf^H(\Sbf+\Xbf\Xbf^H)^{-1}\Xbf\Xbf^H(\Sbf+\Xbf\Xbf^H)^{-1}\sbf}  {\sbf^H(\Sbf+\Xbf\Xbf^H)^{-1}\sbf}.
\end{equation}
For the detection problem in \eqref{test_distributedTargt} in PHE, the  Rao test and Wald test were proposed in \cite{HaoMa12}, given by
\begin{equation}
\label{ch04_Rao_PHE}
\begin{aligned}
t_\text{Rao--PHE}=~\frac{1}{\hat\sigma_0^2}\text{tr}\left[ \Xbf^H\hat\Rbf_0^{-1}\Hbf(\Hbf^H\hat\Rbf_0^{-1}\Hbf)^{-1} \Hbf^H\hat\Rbf_0^{-1}\Xbf\right]
\end{aligned}
\end{equation}
and
\begin{equation}
\label{ch04_Wald_PHE}
t_{\text{Wald--PHE}}=\frac{1}{\hat\sigma_1^2}\text{tr} \left[\Xbf^H\hat\Rbf_1^{-1}\Hbf(\Hbf^H\hat\Rbf_1^{-1}\Hbf) ^{-1}\Hbf^H\hat\Rbf_1^{-1}\Xbf\right],
\end{equation}
respectively, where $\hat\sigma_0^2$ and $\hat\sigma_1^2$ are the sole solutions of \eqref{ch04_solution_sigma0} and \eqref{ch04_solution_sigma1}, respectively,
\begin{equation}
\label{ch04_R_MLE0}
\hat\Rbf_0=\frac{1}{L+K}\left(\Sbf+\dfrac{1}{\hat\sigma^2} {\Xbf\Xbf^H}\right)
\end{equation}
and
\begin{equation}
\label{ch04_R_MLE_PHE}
\hat\Rbf_1=\frac{1}{L+K}\Sbf^{\frac{1}{2}}\left( \Ibf_N+\dfrac{1}{\hat\sigma_1^2} {  \Pbf_{\tilde\sbf}^\bot\tilde\Xbf\tilde\Xbf^H \Pbf_{\tilde\sbf}^\bot }\right)\Sbf^{\frac{1}{2}}.
\end{equation}

Different from the case of point target, it is more difficult to derive the statistical performance for the distributed-target-based detectors. At presence, only the statistical performance of the GKGLRT and GAMF is known. The statistical distribution of the GKGLRT was first proposed in \cite{WangCai91} for the case of no signal mismatch, and then generalized to the case of signal mismatch in  \cite{Raghavan13b}. The statistical distribution of the GAMF was given in \cite{Raghavan13a}.
Precisely, under hypothesis $\text{H}_1$, the conditional distribution of the GKGLRT in \eqref{ch04_GKGLRT2} is
\begin{equation}
\label{ch04_GLRT_HE_p1_cSD}
t_{\text{GKGLRT}} |\text{H}_1\sim{\cal C}{\cal F}_{K,L-N+1}\left(\rho_{\text{dstr}}\cos^2\phi_\text{rk1}\beta_\text{GKGLRT}\right),
\end{equation}
where
\begin{equation}
\label{ch04_SNR_p1}
\rho_{\text{dstr}}=\abf^H\abf\cdot\sbf_0^H\Rbf^{-1}\sbf_0,
\end{equation}
can be taken as the output SNR, with $\sbf_0$ being the actual signal steering vector,
\begin{equation}
\label{}
\cos^2\phi_\text{rk1}=\frac{|\sbf_0^H\Rbf^{-1}\sbf|^2}{\sbf_0^H\Rbf^{-1}\sbf_0~ \sbf^H\Rbf^{-1}\sbf}
\end{equation}
is generalized cosine-squared between the actual signal $\sbf_0$ and the nominal signal $\sbf$ in the whitened space, and $\beta_\text{GKGLRT}$ is a loss factor for the GKGLRT, haveing the statistical distribution
\begin{equation}
\label{ch04_CB00}
\beta_\text{GKGLRT}|\text{H}_1\sim {\cal C}{\cal B}_{L+K-N+1,N-1}(\rho\sin^2\phi_\text{rk1}),
\end{equation}
with $\sin^2\phi_\text{rk1}=1-\cos^2\phi_\text{rk1}$.
Under hypothesis $\text{H}_0$, equations \eqref{ch04_GLRT_HE_p1_cSD} and \eqref{ch04_CB00} become
\begin{equation}
\label{ch04_GLRT_HE_p0_cSD}
t_{\text{GKGLRT}}|\text{H}_0\sim{\cal C}{\cal F}_{K,L-N+1}
\end{equation}
and
\begin{equation}
\label{ch04_CB0x0}
\beta_\text{GKGLRT}|\text{H}_0\sim {\cal C}{\cal B}_{L+K-N+1,N-1},
\end{equation}
respectively.
Moreover, under hypothesis $\text{H}_1$ the conditional distribution of the GAMF in \eqref{ch04_GAMF} is
\begin{equation}
\label{ch04_GAMF_con_SD}
{\beta_{\text{GAMF}}}~t_\text{GAMF}|\text{H}_1\sim{\cal CF}_{K,L-N+1}(\beta_{\text{GAMF}}\rho_\text{dstr}),
\end{equation}
where ${\beta_{\text{GAMF}}}$ is a loss factor for the GAMF, with the statistical distribution
\begin{equation}
\label{ch04_beta_GAMF_SD}
\beta_{\text{GAMF}}|[\text{H}_1~\text{and}~\text{H}_0]\sim{\cal CB}_{L-N+2,N-1}.
\end{equation}
Under hypothesis $\text{H}_0$, equation \eqref{ch04_GAMF_con_SD} turns to be
\begin{equation}
\label{ch04_GAMF_con_SD0}
{\beta_{\text{GAMF}}}~t_\text{GAMF}|\text{H}_0\sim{\cal CF}_{K,L-N+1}.
\end{equation}

There are two kinds of further generalizations of the detection problem in \eqref{test_distributedTargt}. One is that the signal steering vector $\sbf$ lies in a given subspace spanned by an $N\times p$ full-column matrix $\Hbf$. Hence, $\sbf$ can be expressed as $\sbf=\Hbf\thetabf$, with $\thetabf$ being $p\times1$ unknown coordinates. It follows that \eqref{test_distributedTargt} becomes
\begin{equation}
\label{test_DD}
\left\{ {\begin{array}{l}
 \text{H}_0 :{ {{ \Xbf}}}={ { \Nbf}} , ~\xbf_{\text{e},l}=\nbf_{\text{e},l},~ l=1,2,\cdots,L,\\
 \text{H}_1 :{ {{\Xbf}}}= {\Hbf\thetabf\abf^H}+{ { \Nbf}} , ~\xbf_{\text{e},l}=\nbf_{\text{e},l},~ l=1,2,\cdots,L,\\ 
\end{array}} \right.
\end{equation}
The GLRT and 2S-GLRT in HE were proposed in \cite{BoseSteinhardt96a}, described as
\begin{equation}
\label{GLRDD}
t_{\text{GLRDD}}=\lambda_{\max}\left[\tilde\Xbf^H\Pbf_{\tilde\Hbf} \tilde\Xbf(\Ibf_K+\tilde\Xbf^H\tilde\Xbf)^{-1} \right]
\end{equation}
and
\begin{equation}
\label{AMDD}
t_{\text{AMDD}}=\lambda_{\max}\left(\tilde\Xbf^H\Pbf_{\tilde\Hbf} \tilde\Xbf \right),
\end{equation}
respectively, where $\lambda_{\max}(\cdot)$ denotes the maximum eigenvalue of the matrix argument. It was shown in \cite{LiuXie14c} that there is no reasonable Rao test for the detection problem in \eqref{test_DD}, the 2S-Wald test is the same as the detector in \eqref{AMDD}, and the Wald test is given by
\begin{equation}
\label{SNRDD}
t_{\text{SNRDD}}=\frac{\thetabf_{\max}^H\tilde\Hbf^H\tilde\Xbf \tilde\Xbf^H\tilde\Hbf\thetabf_{\max}}{\thetabf_{\max}^H\tilde\Hbf^H \tilde\Hbf\thetabf_{\max}},
\end{equation}
where $\thetabf_{\max}$ is a principal eigenvector (the eigenvector corresponding to the maximum eigenvalue) of the matrix $(\tilde\Hbf^H\tilde\Hbf)^{-1}\tilde\Hbf^H\tilde\Xbf(\Ibf_K+\tilde\Xbf^H\tilde\Xbf) ^{-1}\tilde\Xbf^H\tilde\Hbf$. The detectors in \eqref{GLRDD}, \eqref{AMDD}, and \eqref{SNRDD} are referred to as GLR-based direction detector (GLRDD), adaptive matched direction detector (AMDD), SNR-based direction detector (SNRDD) in \cite{LiuXie14c}.
The 2S-GLRT for the detection problem in \eqref{test_DD} in PHE was proposed in \cite{BessonScharf06b}, given by\footnote{The GADD in \eqref{GADD} can be also derived according to 2S-Wald test.}
\begin{equation}
\label{GADD}
t_{\text{GADD}}=\frac{\lambda_{\max}\left(\tilde\Xbf^H\Pbf_{\tilde\Hbf} \tilde\Xbf \right)}{\text{tr}\left(\tilde\Xbf^H \tilde\Xbf \right)},
\end{equation}
which was denoted as GADD therein.

It follows from \eqref{GLRDD}-\eqref{GADD} that the detectors choose a direction among the subspace spanned by the columns of $\Hbf$. In other words, the detection problem in \eqref{test_DD} tantamount to finding a direction with the largest possibility in a gvien subspace, and hence it is called direction detection in \cite{BessonScharf06b}.

The problem of direction detection can be further generalized when both the column component and row component of the signal to be detected lie in given subspaces. To be precise, the test data under hypothesis $\text{H}_1$ becomes ${ {{\Xbf}}}= {\Hbf\thetabf\alphabf^H\Cbf}+{ { \Nbf}}$, with $\Cbf$ a given $M\times K$ full-row-rank matrix and $\alphabf$ an $M\times1$ vector. This kind of problem is denoted as generalized direction detection in \cite{LiuLiu2015b}, where the GLRT and 2S-GLRT in HE were proposed in therein. Moreover, the Wald test in HE was given in \cite{Liu20SCIS}, and the 2S-GLRT in PHE was derived in \cite{Liu19SCIS}.

Different from \eqref{test_DD}, another generalization of \eqref{test_distributedTargt} is the case that each column of the test data $\Xbf$ has a slightly different steering vector in the sense that these steering vectors are different but all come from the same subspace. Hence, the detection model in \eqref{test_distributedTargt} can be modified as
\begin{equation}
\label{test_HP}
\left\{ {\begin{array}{l}
 \text{H}_0 :{ {{ \Xbf}}}={ { \Nbf}} , ~\xbf_{\text{e},l}=\nbf_{\text{e},l},~ l=1,2,\cdots,L,\\
 \text{H}_1 :{ {{\Xbf}}}= {\Hbf\Phibf}+{ { \Nbf}} , ~\xbf_{\text{e},l}=\nbf_{\text{e},l},~ l=1,2,\cdots,L,\\ 
\end{array}} \right.
\end{equation}
where $\Hbf$ is an $N\times p$ full-column-rank matrix, and $\Phibf$ is a $p\times K$ matrix standing for the coordinates.
The GLRT in HE was proposed in Kelly and Forsythe’s classic report \cite{KellyForsythe89}, while the Rao tes and Wald test in HE can be obtained according to the results in \cite{LiuXie14b}. Precisely, the GLRT, Rao test, and Wald test are given by
\begin{equation}
\label{2}
{{t}_\text{GLRT}}=\frac{\left| {{\Ibf}_{K}}+{{\Xbf}^H}{{\Sbf}^{-1}}\Xbf \right|}{\left| {{\Ibf}_{K}}+{{\Xbf}^H}{{\Sbf}^{-1}}\Xbf-{{\Xbf}^H}{{\Sbf}^{-1}}\Hbf{{({{\Hbf}^H}{{\Sbf}^{-1}}\Hbf)}^{-1}}{{\Hbf}^H}{{\Sbf}^{-1}}\Xbf \right|},
\end{equation}
\begin{equation}
\label{3}
{{t}_\text{Rao}}=\text{tr}\left\{ {{\Xbf}^H}{{(\Sbf+\Xbf{{\Xbf}^H})}^{-1}}\Hbf{{\left[{{\Hbf}^H}{{(\Sbf+\Xbf{{\Xbf}^H})}^{-1}}\Hbf\right]}^{-1}}{{\Hbf}^H}{{(\Sbf+\Xbf{{\Xbf}^H})}^{-1}}\Xbf \right\},
\end{equation}
and
\begin{equation}
\label{4}
{{t}_\text{Wald}}=\text{tr}\left[ {{\Xbf}^H}{{\Sbf}^{-1}}\Hbf{{({{\Hbf}^H}{{\Sbf}^{-1}}\Hbf)}^{-1}}{{\Hbf}^H}{{\Sbf}^{-1}}\Xbf \right],
\end{equation}
respectively.

Note that when $p=N$ in the detection problem in \eqref{test_HP}, the signal steering vectors lie in the whole observation space. Or, equivalently, the steering vectors are completely unknown. The correspond GLRT, 2S-GLRT, and a modified 2S-GLRT (M2S-GLRT) in HE were proposed in \cite{ConteDeMaio03}. It was also shown in \cite{LiuXie13b} that the M2S-GLRT is essentially the corresponding Rao test, while the 2S-GLRT can also be derived according to the Wald test.
Moreover, in \cite{LiuXie14b,LiuXie14a} the test data in \eqref{test_HP} were generalized to the case ${ {{\Xbf}}}= {\Hbf\Phibf\Cbf}+{ { \Nbf}} $, with $\Phibf$ being a $p\times M$ unknown matrix, $\Cbf$ being an $M\times K$ known full-row-rank matrix. The corresponding signal model was called double subspace (DOS) model\footnote{The DOS signal model was first introduced in \cite{KellyForsythe89}. However, it was assumed in \cite{KellyForsythe89} that no training data were available. Instead, it was assumed $K\ge M+N$. This constraint ensures the existence of a set of virtual training data, generated by a certain unitary matrix to the test data.} in \cite{LiuXie14b,LiuXie14a}, where many adaptive detectors were proposed.

\subsubsection{Adaptive detectors in the presence of interference}
Most of the above detectors are designed without taking into account the possibility of interference, which usually exists in practice. Interference can be caused intentionally (jamming due to the ECM) or unintentionally (communication signals or radar signals transmitted by other radar systems). In this case, the detection problem in \eqref{ch02_1} can be modified as
\begin{equation}
\label{test_jamming}
\left\{ {\begin{array}{l}
\text{H}_0 :{ {{ \xbf}}}={ { \nbf}} , ~\xbf_{\text{e},l}=\nbf_{\text{e},l},~ l=1,2,\cdots,L,\\
\text{H}_1 :{ {{\xbf}}}= {\sbf}+\jbf+{ { \nbf}} , ~\xbf_{\text{e},l}=\nbf_{\text{e},l},~ l=1,2,\cdots,L,\\ 
\end{array}} \right.
\end{equation}
where $\jbf$ stands for the interference.
Roughly speaking, there are two main kinds of interference. One is coherent interference, while the other is noise interference. For the former, it works like a real target, which usually lies in a certain spatially direction and/or occupies a Doppler bin. Hence, the coherent interference can be modelled by a subspace model.
For the latter, it works like thermal noise or clutter. As a result, the noise interference changes the noise covariance matrix of the test data.

Based on the above analysis, the coherent interference can be modelled as $\jbf=\Jbf\phibf$, where the $N\times q$ full-column-rank matrix spans the subspace where the interference lies, and the $q\times1$ vector $\phibf$ denotes the unknown coordinates. For coherent interference and subspace signals (i.e., the signal in \eqref{test_jamming} can be expressed as $\sbf=\Hbf\thetabf$), the GLRT and 2S-GLRT in HE and PHE for the detection problem in \eqref{test_jamming} were all proposed in \cite{BandieraDeMaio07a}, and the GLRT and 2S-GLRT in PHE coincide with each other. Precisely, the GLRT and 2S-GLRT in HE are
\begin{equation}
\label{GLRT-HE-I}
t_{\text{GLRT-HE-I}}=\dfrac{\tilde{{\mathbf {x}}}^H\mathbf{P}_{\mathbf {P}_{\tilde{\mathbf {J}}}^\bot \tilde{\mathbf {H}}} \tilde{{\mathbf {x}}}}{1+\tilde{{\mathbf {x}}}^H\mathbf{P}_{\tilde{\mathbf {J}}}^\bot \tilde{{\mathbf {x}}}-\tilde{{\mathbf {x}}}^H\mathbf{P}_{\mathbf {P}_{\tilde{\mathbf {J}}}^\bot \tilde{\mathbf {H}}} \tilde{{\mathbf {x}}}}
\end{equation}
and
\begin{equation}
\label{2S-GLRT-HE-I}
t_{\text{2S-GLRT-HE-I}}=\tilde{{\mathbf {x}}}^H\mathbf{P}_{\mathbf {P}_{\tilde{\mathbf {J}}}^\bot \tilde{\mathbf {H}}} \tilde{{\mathbf {x}}},
\end{equation}
respectively, while the GLRT in PHE is
\begin{equation}
\label{GLRT-PHE-I}
t_{\text{GLRT-PHE-I}}=\dfrac{\tilde{{\mathbf {x}}}^H\mathbf{P}_{\mathbf {P}_{\tilde{\mathbf {J}}}^\bot \tilde{\mathbf {H}}} \tilde{{\mathbf {x}}}} {\tilde{{\mathbf {x}}}^H\mathbf{P}_{\tilde{\mathbf {J}}}^\bot \tilde{{\mathbf {x}}}},
\end{equation}
where $\Pbf_{\Pbf_{\tilde\Jbf}^\bot\tilde\Hbf}=\Pbf_{\tilde\Jbf}^\bot\tilde\Hbf (\tilde\Hbf^H\Pbf_{\tilde\Jbf}^\bot\tilde\Hbf)^{-1}\tilde\Hbf^H\Pbf_{\tilde\Jbf}^\bot$, $\Pbf_{\tilde\Jbf}^\bot=\Ibf_N-\Pbf_{\tilde\Jbf}$, and $\Pbf_{\tilde\Jbf}=\tilde\Jbf (\tilde\Jbf^H\tilde\Jbf)^{-1}\tilde\Jbf^H$.
For convenience, the detectors in \eqref{GLRT-HE-I}, \eqref{2S-GLRT-HE-I}, and \eqref{GLRT-PHE-I} are referred to the GLRT in HE with interference rejection (GLRT-HE-I), 2S-GLRT in HE with interference rejection (2S-GLRT-HE-I), and GLRT in PHE with interference rejection (GLRT-PHE-I), respectively.

For coherent interference, the Rao test and 2S-Rao test were proposed in \cite{LiuLiu2015d}, while the Wald test and 2S-Wald test were derived in \cite{LiuLiu19TAESWald}. Precisely, in HE the Rao test and 2S-Rao test are
\begin{equation}
\label{Rao-HE-I}
t_{\text{Rao-HE-I}}=\dfrac{\tilde{{\mathbf {x}}}^H\mathbf{P}_{\tilde{\mathbf {J}}}^\bot \mathbf {P}_{\tilde{\mathbf {H}}} \mathbf {P}_{\tilde{\mathbf {J}}}^\bot \tilde{{\mathbf {x}}}}{(1+\tilde{{\mathbf {x}}}^H\mathbf{P}_{\tilde{\mathbf {J}}}^\bot \tilde{{\mathbf {x}}})(1+\tilde{{\mathbf {x}}}^H\mathbf{P}_{\tilde{\mathbf {J}}}^\bot \mathbf {P}_{\tilde{\mathbf {H}}}^\bot \mathbf {P}_{\tilde{\mathbf {J}}}^\bot \tilde{{\mathbf {x}}})}
\end{equation}
and
\begin{equation}
\label{2S-Rao-HE-I}
t_{\text{2S-Rao-HE-I}}=\tilde{{\mathbf {x}}}^H\mathbf{P}_{\tilde{\mathbf {J}}}^\bot \mathbf {P}_{\tilde{\mathbf {H}}} \mathbf {P}_{\tilde{\mathbf {J}}}^\bot \tilde{{\mathbf {x}}},
\end{equation}
respectively, while in PHE the Rao test is the same as the 2S-Rao test, given by
\begin{equation}
\label{Rao-PHE-I}
t_{\text{Rao-PHE-I}}= \dfrac{\tilde{{\mathbf {x}}}^H\mathbf{P}_{\tilde{\mathbf {J}}}^\bot \mathbf {P}_{\tilde{\mathbf {H}}} \mathbf {P}_{\tilde{\mathbf {J}}}^\bot \tilde{{\mathbf {x}}}}  {\tilde{{\mathbf {x}}}^H\mathbf{P}_{\tilde{\mathbf {J}}}^\bot \tilde{{\mathbf {x}}}}.
\end{equation}
The Wald test is the same as the 2S-Wald test both in HE and PHE, given by
\begin{equation}
\label{Wald-HE-I}
t_{\text{Wald-HE-I}}=\tilde{{\mathbf {x}}}^H\mathbf{P}_{\tilde{\mathbf {H}}\vert \tilde{\mathbf {J}}}^H \mathbf {P}_{\tilde{\mathbf {H}}\vert \tilde{\mathbf {J}}} \tilde{{\mathbf {x}}}
\end{equation}
and
\begin{equation}
\label{Wald-PHE-I}
t_{\text{Wald-PHE-I}}=\dfrac{\tilde{{\mathbf {x}}}^H\mathbf{P}_{\tilde{\mathbf {H}}\vert \tilde{\mathbf {J}}}^H \mathbf {P}_{\tilde{\mathbf {H}}\vert \tilde{\mathbf {J}}} \tilde{{\mathbf {x}}}}  {\tilde{{\mathbf {x}}}^H\mathbf{P}_{\tilde{\mathbf {B}}}^\bot \tilde{{\mathbf {x}}}},
\end{equation}
respectively, where $\mathbf {P}_{\tilde {\mathbf H} \vert \tilde {\mathbf J} } =\tilde {\mathbf H} (\mathbf{\tilde H} ^H\mathbf{P}_{\tilde {\mathbf J} }^\bot \tilde {\mathbf H} )^{-1}\tilde {\mathbf H} ^H\mathbf{P}_{\mathbf{\tilde J}}^\bot$ is the oblique projection matrix onto the subspace spanned by $\tilde\Hbf$ along the subspace spanned by $\tilde\Jbf$.  
Detailed analysis and comparison of the above detectors can be found in \cite{LiuLiu19TAESWald}.

At present, only the GLRT-HE-I, 2S-GLRT-HE-I, and GLRT-PHE-I have known statistical properties, given in \cite{LiuLiu18AES}. Precisely, the conditional distribution of the GLRT-HE-I in \eqref{GLRT-HE-I} with a fixed $\beta _\text{I}$ under hypothesis $\text{H}_1$, is
\begin{equation}
\label{StDist_GLRT_HE_Mis}
t_\text{GLRT--HE--I}|[\beta_\text{I},\text{H}_1]\sim{\cal C}{{\cal F}_{p,L - N + q + 1}}({\rho_{\text{eff}}} {\beta _\text{I}}), 
\end{equation}
where
\begin{equation}
\label{rho_HE_eff_org}
\rho_{\text{eff}}={\bar \sbf_0}^H \Pbf_{\bar\Jbf}^\bot \Pbf_{\Pbf_{\tilde\Jbf}^\bot\tilde\Hbf} \Pbf_{\bar\Jbf}^\bot\bar \sbf_0
\end{equation}
is defined as the effective SNR (eSNR), and ${\beta_{\text{I}}}$ is loss factor defined as
\begin{equation}
\label{Beta_HE-I}
{\beta_{\text{I}}} = \frac{1}{1 + {\tilde \xbf^H}\Pbf_{\tilde \Jbf}^ \bot \tilde \xbf - {\tilde \xbf^H}{\Pbf_{\Pbf_{\tilde \Jbf}^ \bot \tilde \Hbf}}\tilde \xbf}.
\end{equation}
The statistical distribution of ${\beta_{\text{I}}}$ under hypothesis $\text{H}_1$ is
\begin{equation}
\label{Beta_HE-I_H1SD}
\beta _{\text{I}}|\text{H}_1 \sim{\cal C}{{\cal B} _{L - N + p + q + 1,N - p - q}} (\delta^2_{\text{I}}),
\end{equation}
where
\begin{equation}
\label{delta2_HE}
\delta^2_{\text{I}}=
{\bar \sbf_0^H \Pbf_{\bar\Jbf}^\bot \Pbf_{\Pbf_{\bar\Jbf}^\bot\bar\Hbf}^\bot  \Pbf_{\bar\Jbf}^\bot   {\bar\sbf_0}},
\end{equation}
with $\Pbf_{\Pbf_{\tilde\Jbf}^\bot\tilde\Hbf} ^\bot= \Ibf_N-\Pbf_{\Pbf_{\tilde\Jbf}^\bot\tilde\Hbf}$.
Under hypothesis $\text{H}_0$, \eqref{StDist_GLRT_HE_Mis} and \eqref{Beta_HE-I_H1SD} reduce to
\begin{equation}
\label{StDist_GLRT_HE-I_0}
t_\text{GLRT--HE--I}|[\beta_\text{I},\text{H}_0]\sim{\cal C}{{\cal F}_{p,L - N + q + 1}} 
\end{equation}
and
\begin{equation}
\label{Beta_HE-I_H0SD}
\beta _\text{I}|\text{H}_0 \sim{\cal C}{{\cal B} _{L - N + p + q + 1,N - p - q}},
\end{equation}
respectively.

More geometric interpretation about the eSNR in \eqref{rho_HE_eff_org} can be found in \cite{LiuLiu18AES}.
Moreover, the following two equations can be easily verified
\begin{equation}
\label{2SGLRTHE51}
t_\text{2S--GLRT--HE--I}  = \frac{t_\text{GLRT--HE--I}}  {\beta _\text{I}},
\end{equation}
\begin{equation}
\label{GLRT-PHE-I22}
t_\text{GLRT--PHE--I}  = \frac{t_\text{GLRT--HE--I}}  {1-\beta _\text{I}}.
\end{equation}
Using \eqref{2SGLRTHE51} and \eqref{GLRT-PHE-I22}, along with \eqref{StDist_GLRT_HE_Mis}, \eqref{Beta_HE-I_H1SD}, \eqref{StDist_GLRT_HE-I_0} and \eqref{Beta_HE-I_H0SD}, we can obtain the analytical expressions for the PDs and PFAs of the 2S-GLRT-HE-I and GLRT-PHE-I.

For completely unknown noise interference, it was shown in \cite{Besson07b} that the GLRT for rank-one signals is equivalent to the ACE. In  The corresponding Rao test was derived in \cite{OrlandoRicci10}, i.e., the DN-AMF, originally adopted for mismatched signal detection. The results in \cite{Besson07b,OrlandoRicci10} were generalized in \cite{LiuHan17SCIS} when additional coherent interference existed.
In \cite{BessonOrlando07} the noise interference was assumed to be orthogonal to the signal of interest in the whitened space, and it was shown that the GLRT coincides with the KGLRT. Moreover, it was shown in \cite{HaoOrlando12a} that the corresponding Rao and Wald tests are the same as the DMRao and AMF, respectively. The results in \cite{BessonOrlando07,HaoOrlando12a} were generalized in \cite{ShangLiu18GER} for the case of subspace signals.
Some other generalizations for noise interference can be found in  \cite{AddabboBesson19,Raghavan19,Besson19,YanAddabbo20,Besson20SPL,TangWang20}.

\section{Conclusions}
In this paper, we investigated the detector design criteria for adaptive detection, analysed the relationship between adaptive detection and the filtering-then-CFAR detection approach, as well as the relationship between adaptive detectors and adaptive filters, gave a comprehensive review, summarized and compared typical adaptive detectors.
Adaptive detection jointly uses the test and training data to form an adaptive detector. Compared with the filtering-then-CFAR detection approach (adaptive or non-adaptive), adaptive detection has many distinct features. It achieves the function of filtering and CFAR processing simultaneously, and hence, it has simple detection procedure. Moreover, it can provide better detection performance. 

We hope that this paper will stimulate new researches on adaptive detection. Some
possible further research tracks are listed below.
1) The statistical performance of many adaptive detectors are needed to be studies, such as the Rao and Wald tests in subspace interference \cite{LiuLiu2015d,LiuLiu19TAESWald}, the 2S-GLRT in HE in the presence of signal mismatch \cite{ConteDeMaio01}. Obtaining these results can reveal how the signal mismatch and/or interference affect the detection performance.
2) Nowadays, multichannel signal detection has been combined with compressive sensing or sparse representation, which is an emerging signal processing technique for efficiently acquiring and reconstructing a compressible signal, by using much fewer samples. Several compressive sensing-based detectors were proposed, such as \cite{LiSong12,WangLiu15,RazaviValkama16,WimalajeewaVarshney17,WimalajeewaVarshney18, MaGan19,ZhangSward19} and the references therein. However, most proposed detectors based on compressive sensing are for known noise or white Gaussian noise with unknown variance. Much more challenging task is for colored noise with unknown covariance matrix.
3) Most existing adaptive detectors were designed under specific assumptions on the noise, 
either homogeneous, partially homogeneous, compound-Gaussian, or structure nonhomogeneity.
However, the actual noise may be different from the assumed
one, due to system and environment uncertainties. As a consequence, the designed detectors may suffer from significant performance loss. Therefore, it is necessary to devise fully adaptive detection approaches which can adjust the detection strategy to accommodate the changing environments.
Recently, some preliminary analysis on classification of noise covariance structure in Gaussian background was proposed in \cite{CarotenutoDeMaio17,CarotenutoDeMaio19,CarotenutoOrlando19,LiuBiondi19}.
4) Recently, some preliminary results of machine learning were utilized in adaptive detection \cite{ColucciaFascista20TSP,ColucciaFascista20SP,ZaimbashiLi20}. However, it was not fully addressed the fundamental problem that why and how the detection performance can be improved by using machine learning technologies.


In this paper, we mainly focused on the Gaussian background. In practice, the environment may exhibit non-Gaussian character \cite{ConteLops95,Gerlach99,GiniGreco02a,PascalChitour08,SangstonGini12}. Interesting readers can refer to a recently overview paper \cite{SangstonFarina16} on compound-Gaussian clutter, for the case that the relevant properties of the clutter are assumed to be known in advance.

\Acknowledgements{This work was supported in part by  National Natural Science Foundation of China (Grant Nos. 62071482 and 61871469), in part by the National Natural Science Foundation of China and Civil Aviation Administration of China under Grant U1733116, in part by the Youth Innovation Promotion Association CAS (Grant CX2100060053), in part by the National Key Research and Development Program of China (Grant 2018YFB1801105), and by China Postdoctoral Science Foundation (Grant 2020T130493).}


%
%

%
%

{\small
\bibliographystyle{IEEEtran}
\bibliography{D:/LaTexReference/Detection}

\begin{thebibliography}{100}
\providecommand{\url}[1]{#1}
\csname url@samestyle\endcsname
\providecommand{\newblock}{\relax}
\providecommand{\bibinfo}[2]{#2}
\providecommand{\BIBentrySTDinterwordspacing}{\spaceskip=0pt\relax}
\providecommand{\BIBentryALTinterwordstretchfactor}{4}
\providecommand{\BIBentryALTinterwordspacing}{\spaceskip=\fontdimen2\font plus
\BIBentryALTinterwordstretchfactor\fontdimen3\font minus
  \fontdimen4\font\relax}
\providecommand{\BIBforeignlanguage}[2]{{%
\expandafter\ifx\csname l@#1\endcsname\relax
\typeout{** WARNING: IEEEtran.bst: No hyphenation pattern has been}%
\typeout{** loaded for the language `#1'. Using the pattern for}%
\typeout{** the default language instead.}%
\else
\language=\csname l@#1\endcsname
\fi
#2}}
\providecommand{\BIBdecl}{\relax}
\BIBdecl

\bibitem{GiniGreco99b}
F.~Gini and M.~V. Greco, ``Suboptimum approach to adaptive coherent radar
  detection in compound-{Gaussian} clutter,'' \emph{IEEE Transactions on
  Aerospace and Electronic Systems}, vol.~35, no.~3, pp. 1095--1104, 1999.

\bibitem{ChongPascal10JSTSP}
C.~Y. Chong, F.~Pascal, J.-P. Ovarlez, and M.~Lesturgie, ``{MIMO} radar
  detection in non-{Gaussian} and heterogeneous clutter,'' \emph{IEEE Journal
  of Selected Topics in Signal Processing}, vol.~4, no.~1, pp. 115--126, 2010.

\bibitem{GinolhacForster14}
G.~Ginolhac, P.~Forster, F.~Pascal, and J.-P. Ovarlez, ``Exploiting persymmetry
  for low-rank space time adaptive processing,'' \emph{Signal Processing},
  vol.~97, pp. 242--251, 2014.

\bibitem{Kelly86}
E.~J. Kelly, ``An adaptive detection algorithm,'' \emph{IEEE Transactions on
  Aerospace and Electronic Systems}, vol.~22, no.~1, pp. 115--127, 1986.

\bibitem{DeMaioGreco16Book}
A.~De~Maio and S.~G. Greco, \emph{Modern Radar Detection Theory}.\hskip 1em
  plus 0.5em minus 0.4em\relax SciTech Publishing, 2016.

\bibitem{GiniFarina02b}
F.~Gini and A.~Farina, ``Vector subspace detection in compound-{Gaussian}
  clutter part {I}: Survey and new results,'' \emph{IEEE Transactions on
  Aerospace and Electronic Systems}, vol.~38, no.~4, pp. 1295--1311, 2002.

\bibitem{SangstonFarina16}
K.~J. {Sangston} and A.~{Farina}, ``Coherent radar detection in
  compound-{Gaussian} clutter: Clairvoyant detectors,'' \emph{IEEE Aerospace
  and Electronic Systems Magazine}, vol.~31, no.~11, pp. 42--63, November 2016.

\bibitem{Lemonte16}
A.~J. Lemonte, \emph{The Gradient Test: Another Likelihood-Based Test}.\hskip
  1em plus 0.5em minus 0.4em\relax Cambridge University Press, 2016.

\bibitem{Durbin70}
J.~Durbin, ``Testing for serial correlation in least-squares regression when
  some of the regressors are lagged dependent variables,'' \emph{Econometrica},
  vol.~38, no.~3, pp. 410--421, 1970.

\bibitem{Scharf91}
L.~L. Scharf, \emph{Statistical Signal Processing: Detection, Estimation, and
  Times Series Analysis}.\hskip 1em plus 0.5em minus 0.4em\relax New York:
  Addison-Wesley Publishing Company, 1991.

\bibitem{Kay05}
S.~M. Kay, ``The multifamily likelihood ratio test for multiple signal model
  detection,'' \emph{IEEE Geoscience and Remote Sensing Letters}, vol.~12,
  no.~5, pp. 369--371, 2005.

\bibitem{AbramovichSpencer07a}
Y.~I. Abramovich, N.~K. Spencer, and A.~Y. Gorokhov, ``Modified {GLRT} and
  {AMF} framework for adaptive detectors,'' \emph{IEEE Transactions on
  Aerospace and Electronic Systems}, vol.~43, no.~3, pp. 1017--1051, 2007.

\bibitem{CarotenutoDeMaio15GRS}
V.~{Carotenuto}, A.~{De Maio}, C.~{Clemente}, and J.~J. {Soraghan}, ``Invariant
  rules for multipolarization {SAR} change detection,'' \emph{IEEE Transactions
  on Geoscience and Remote Sensing}, vol.~53, no.~6, pp. 3294--3311, 2015.

\bibitem{CarotenutoDeMaio15GRSL}
V.~{Carotenuto}, A.~{De Maio}, C.~{Clemente}, and J.~{Soraghan}, ``Unstructured
  versus structured {GLRT} for multipolarization {SAR} change detection,''
  \emph{IEEE Geoscience and Remote Sensing Letters}, vol.~12, no.~8, pp.
  1665--1669, 2015.

\bibitem{CarotenutoDeMaio16GRS}
V.~{Carotenuto}, A.~{De Maio}, C.~{Clemente}, J.~J. {Soraghan}, and
  G.~{Alfano}, ``Forcing scale invariance in multipolarization {SAR} change
  detection,'' \emph{IEEE Transactions on Geoscience and Remote Sensing},
  vol.~54, no.~1, pp. 36--50, 2016.

\bibitem{CiuonzoCarotenuto17}
D.~{Ciuonzo}, V.~{Carotenuto}, and A.~{De Maio}, ``On multiple covariance
  equality testing with application to {SAR} change detection,'' \emph{IEEE
  Transactions on Signal Processing}, vol.~65, no.~19, pp. 5078--5091, Oct
  2017.

\bibitem{DeMaioDeNicola09a}
A.~De~Maio, S.~De~Nicola, and A.~Farina, ``{GLRT} versus {MFLRT} for adaptive
  {CFAR} radar detection with conic uncertainty,'' \emph{IEEE Signal Processing
  Letters}, vol.~16, no.~8, pp. 707--710, 2009.

\bibitem{DeMaioOrlando17GRS}
A.~{De Maio}, D.~{Orlando}, L.~{Pallotta}, and C.~{Clemente}, ``A multifamily
  {GLRT} for oil spill detection,'' \emph{IEEE Transactions on Geoscience and
  Remote Sensing}, vol.~55, no.~1, pp. 63--79, Jan 2017.

\bibitem{DeMaioHan18}
A.~{De Maio}, S.~Han, and D.~Orlando, ``Adaptive radar detectors based on the
  observed {FIM},'' \emph{IEEE Transactions on Signal Processing}, vol.~66,
  no.~14, pp. 3838--3847, 2018.

\bibitem{GerlachSteiner99}
K.~Gerlach and M.~J. Steiner, ``Adaptive detection of range distributed
  targets,'' \emph{IEEE Transactions on Signal Processing}, vol.~47, no.~7, pp.
  1844--1851, 1999.

\bibitem{DeMaio02b}
A.~De~Maio, ``Polarimetric adaptive detection of range-distributed targets,''
  \emph{IEEE Transactions on Signal Processing}, vol.~50, no.~9, pp.
  2152--2159, 2002.

\bibitem{DeMaioFarina07}
A.~De~Maio, A.~Farina, and K.~Gerlach, ``Adaptive detection of range spread
  targets with orthogonal rejection,'' \emph{IEEE Transactions on Aerospace and
  Electronic Systems}, vol.~43, no.~2, pp. 738--752, 2007.

\bibitem{AubryDeMaio14b}
A.~Aubry, A.~De~Maio, D.~Orlando, and M.~Piezzo, ``Adaptive detection of
  point-like targets in the presence of homogeneous clutter and subspace
  interference,'' \emph{IEEE Signal Processing Letters}, vol.~21, no.~7, pp.
  848--852, 2014.

\bibitem{AubryDeMaio15TSP}
A.~Aubry, A.~{De Maio}, G.~Foglia, and D.~Orlando, ``Multipath exploitation for
  adaptive radar detection,'' \emph{IEEE Transactions on Signal Processing},
  vol.~63, no.~5, pp. 1268--1280, 2015.

\bibitem{RongAubry20TSP}
Y.~{Rong}, A.~{Aubry}, A.~{De Maio}, and M.~{Tang}, ``Diffuse multipath
  exploitation for adaptive detection of range distributed targets,''
  \emph{IEEE Transactions on Signal Processing}, vol.~68, pp. 1197--1212, 2020.

\bibitem{Rao05ScoreTest}
C.~R. Rao, ``Score test: Historical review and recent developments,'' in
  \emph{Advances in Ranking and Selection, Multiple Comparisons, and
  Reliability: Methodology and Applications}, N.~Balakrishnan, H.~N. Nagaraja,
  and N.~Kannan, Eds.\hskip 1em plus 0.5em minus 0.4em\relax Boston, MA:
  Birkh{\"a}user Boston, 2005, pp. 3--20.

\bibitem{Kay98}
S.~M. Kay, \emph{Fundamentals of Statistical Signal Processing: Detection
  Theory}.\hskip 1em plus 0.5em minus 0.4em\relax Englewood Cliffs, NJ:
  Prentice-Hall, 1998.

\bibitem{PagadaraiWyglinski11}
S.~Pagadarai, A.~Wyglinski, and C.~Anderson, ``An evaluation of the {Bayesian}
  {CRLB} for time-varying {MIMO} channel estimation using complex-valued
  differentials,'' in \emph{IEEE Pacific Rim Conference on Communications,
  Computers and Signal Processing}.\hskip 1em plus 0.5em minus 0.4em\relax
  IEEE, 2011, pp. 818--823.

\bibitem{LiuWang14}
W.~Liu, Y.~Wang, and W.~Xie, ``Fisher information matrix, {Rao} test, and
  {Wald} test for complex-valued signals and their applications,'' \emph{Signal
  Processing}, vol.~94, pp. 1--5, 2014.

\bibitem{KayZhu16TSP}
S.~Kay and Z.~Zhu, ``The complex parameter {Rao} test,'' \emph{IEEE
  Transactions on Signal Processing}, vol.~64, no.~4, pp. 6580--6588, 2016.

\bibitem{Hjorungnes11}
A.~Hj{\o}rungnes, \emph{Complex-Valued Matrix Derivatives: With Applications in
  Signal Processing and Communications}.\hskip 1em plus 0.5em minus 0.4em\relax
  New York: Cambridge University Press, 2011.

\bibitem{MagnusNeudecker07}
J.~R. Magnus and H.~Neudecker, \emph{Matrix Differential Calculus with
  Applications in Statistics and Econometrics}, 3rd~ed.\hskip 1em plus 0.5em
  minus 0.4em\relax New York: Wiley, 2007.

\bibitem{Richards10Book}
M.~A. Richards, J.~A. Scheer, and W.~A. Holm, \emph{Principles of Modern Radar,
  Volume I - Basic Principles}.\hskip 1em plus 0.5em minus 0.4em\relax Raleigh:
  SciTech Publishing, 2010.

\bibitem{Weinberg17Book}
G.~Weinberg, \emph{Radar Detection Theory of Sliding Window Processes}.\hskip
  1em plus 0.5em minus 0.4em\relax Boca Raton: CRC Press, 2017.

\bibitem{BrennanReed73}
L.~E. Brennan and L.~S. Reed, ``Theory of adaptive radar,'' \emph{IEEE
  Transactions on Aerospace and Electronic Systems}, vol.~9, no.~2, pp.
  237--252, 1973.

\bibitem{LiStoica06book}
J.~Li and P.~Stoica, \emph{Robust Adaptive Beamforming}.\hskip 1em plus 0.5em
  minus 0.4em\relax Hoboken: Wiley, 2006.

\bibitem{WangPeng00Egls}
Y.~Wang and Y.~Peng, \emph{Space-Time Adaptive Processing}.\hskip 1em plus
  0.5em minus 0.4em\relax Beijing: Tsinghua University Press, 2000.

\bibitem{Klemm06}
R.~Klemm, \emph{Principles of Space-Time Adaptive Processing}, 3rd~ed.\hskip
  1em plus 0.5em minus 0.4em\relax London: The Institution of Electrical
  Engineers, 2006.

\bibitem{Guerci15}
J.~R. Guerci, \emph{Space-Time Adaptive Processing for Radar}, 2nd~ed.\hskip
  1em plus 0.5em minus 0.4em\relax Boston: Artech House, 2015.

\bibitem{Ward94}
J.~Ward, ``Space-time adaptive processing for airborne radar,'' MIT Lincoln
  Laboratory, Lexington, Technical Report, 1994.

\bibitem{Melvin04Overview}
W.~L. Melvin, ``A {STAP} overview,'' \emph{IEEE Aerospace and Electronic
  Systems Magazine}, vol.~19, no.~1, pp. 19--35, 2004.

\bibitem{ReedMallett74}
I.~S. Reed, J.~D. Mallett, and L.~E. Brennan, ``Rapid convergence rate in
  adaptive arrays,'' \emph{IEEE Transactions on Aerospace and Electronic
  Systems}, vol.~10, no.~6, pp. 853--863, 1974.

\bibitem{ChenReed91}
W.-S. Chen and I.~S. Reed, ``A new {CFAR} detection test for radar,''
  \emph{Digital Signal Processing}, vol.~1, no.~4, pp. 198--214, 1991.

\bibitem{RobeyFuhrmann92}
F.~C. Robey, D.~R. Fuhrmann, E.~J. Kelly, and R.~Nitzberg, ``A {CFAR} adaptive
  matched filter detector,'' \emph{IEEE Transactions on Aerospace and
  Electronic Systems}, vol.~28, no.~1, pp. 208--216, 1992.

\bibitem{DeMaio07}
A.~De~Maio, ``{Rao} test for adaptive detection in {Gaussian} interference with
  unknown covariance matrix,'' \emph{IEEE Transactions on Signal Processing},
  vol.~55, no.~7, pp. 3577--3584, 2007.

\bibitem{WangCai90}
H.~Wang and L.~Cai, ``On adaptive multiband signal detection with the {SMI}
  algorithm,'' \emph{IEEE Transactions on Aerospace and Electronic Systems},
  vol.~26, no.~5, pp. 768--773, 1990.

\bibitem{DeMaio04}
A.~De~Maio, ``A new derivation of the adaptive matched filter,'' \emph{IEEE
  Signal Processing Letters}, vol.~11, no.~10, pp. 792--793, 2004.

\bibitem{Gerlach94b}
K.~Gerlach, ``A mean level adaptive detector using nonconcurrent data,''
  \emph{IEEE Transactions on Aerospace and Electronic Systems}, vol.~30, no.~1,
  pp. 258 -- 265, 1994.

\bibitem{Gerlach94c}
------, ``A comparison of two adaptive detection schemes,'' \emph{IEEE
  Transactions on Aerospace and Electronic Systems}, vol.~30, no.~1, pp.
  30--40, 1994.

\bibitem{Gerlach95}
------, ``The effects of signal contamination on two adaptive detectors,''
  \emph{IEEE Transactions on Aerospace and Electronic Systems}, vol.~31, no.~1,
  pp. 297--309, 1995.

\bibitem{Gerlach1995Convergence}
K.~Gerlach and F.~C. Lin, ``Convergence performance of binary adaptive
  detectors,'' \emph{IEEE transactions on aerospace and electronic systems},
  vol.~31, no.~1, pp. 329--340, 1995.

\bibitem{ReedGau98}
I.~S. Reed, Y.-L. Gau, and T.~K. Truong, ``{CFAR} detection and estimation for
  {STAP} radar,'' \emph{IEEE Transactions on Signal Processing}, vol.~34,
  no.~3, pp. 722--735, 1998.

\bibitem{ConteDeMaio01}
E.~Conte, A.~De~Maio, and G.~Ricci, ``{GLRT}-based adaptive detection
  algorithms for range-spread targets,'' \emph{IEEE Transactions on Signal
  Processing}, vol.~49, no.~7, pp. 1336--1348, 2001.

\bibitem{KrautScharfButler05}
S.~Kraut, L.~L. Scharf, and R.~W. Butler, ``The adaptive coherence estimator: A
  uniformly most-powerful-invariant adaptive detection statistic,'' \emph{IEEE
  Transactions on Signal Processing}, vol.~53, no.~2, pp. 427--438, 2005.

\bibitem{KrautScharf99}
S.~Kraut and L.~L. Scharf, ``The {CFAR} adaptive subspace detector is a
  scale-invariant {GLRT},'' \emph{IEEE Transactions on Signal Processing},
  vol.~47, no.~9, pp. 2538--2541, 1999.

\bibitem{DeMaioIommelli08}
A.~De~Maio and S.~Iommelli, ``Coincidence of the {Rao} test, {Wald} test, and
  {GLRT} in partially homogeneous environment,'' \emph{IEEE Signal Processing
  Letters}, vol.~15, pp. 385--388, 2008.

\bibitem{LiuLi15a}
J.~Liu, H.~Li, and B.~Himed, ``Threshold setting for adaptive matched filter
  and adaptive coherence estimator,'' \emph{IEEE Signal Processing Letters},
  vol.~22, no.~1, pp. 11--15, 2015.

\bibitem{ConteDeMaio03WSEAS}
E.~Conte and A.~{De Maio}, ``An invariant framework for adaptive detection in
  partially homogeneous environment,'' \emph{WSEAS Transactions on Circuits},
  vol.~2, no.~1, pp. 282--287, 2003.

\bibitem{DeMaio19SPL}
A.~{De Maio}, ``Invariance theory for adaptive radar detection in heterogeneous
  environment,'' \emph{IEEE Signal Processing Letters}, vol.~26, no.~7, pp.
  996--1000, 2019.

\bibitem{PascalChitour08}
F.~{Pascal}, Y.~{Chitour}, J.~{Ovarlez}, P.~{Forster}, and P.~{Larzabal},
  ``Covariance structure maximum-likelihood estimates in compound {Gaussian}
  noise: Existence and algorithm analysis,'' \emph{IEEE Transactions on Signal
  Processing}, vol.~56, no.~1, pp. 34--48, Jan 2008.

\bibitem{ConteLops95}
E.~Conte, M.~Lops, and G.~Ricci, ``Asymptotically optimum radar detection in
  compound-{Gaussian} clutter,'' \emph{IEEE Transactions on Aerospace and
  Electronic Systems}, vol.~31, no.~2, pp. 617--625, 1995.

\bibitem{Gini97a}
F.~Gini, ``Sub-optimum coherent radar detection in a mixture of {K}-distributed
  and {Gaussian} clutter,'' \emph{IEE Proceedings on Radar, Sonar and
  Navigation}, vol. 144, no.~1, pp. 39--48, 1997.

\bibitem{DeMaioConte10b}
A.~De~Maio and E.~Conte, ``Uniformly most powerful invariant detection in
  spherically invariant random vector distributed clutter,'' \emph{IET Radar,
  Sonar and Navigation}, vol.~4, no.~4, pp. 560--563, 2010.

\bibitem{ConteLops96}
E.~Conte, M.~Lops, and G.~Ricci, ``Adaptive matched filter detection in
  spherically invariant noise,'' \emph{IEEE Signal Processing Letters}, vol.~3,
  no.~8, pp. 248--250, 1996.

\bibitem{BidonBesson08b}
S.~Bidon, O.~Besson, and J.-Y. Tourneret, ``The adaptive coherence estimator is
  the generalized likelihood ratio test for a class of heterogeneous
  environments,'' \emph{IEEE Signal Processing Letters}, vol.~15, pp. 281--284,
  2008.

\bibitem{ConteLops98}
E.~Conte, M.~Lops, and G.~Ricci, ``Adaptive detection schemes in
  compound-{Gaussian} clutter,'' \emph{IEEE Transactions on Aerospace and
  Electronic Systems}, vol.~34, no.~4, pp. 1058--1069, 1998.

\bibitem{Rangaswamy05}
M.~Rangaswamy, ``Statistical analysis of the nonhomogeneity detector for
  non-{Gaussian} interference backgrounds,'' \emph{IEEE Transactions on Signal
  Processing}, vol.~53, no.~6, pp. 2101--2111, 2005.

\bibitem{DeMaioFoglia05}
A.~De~Maio, G.~Foglia, and E.~Conte, ``{CFAR} behavior of adaptive detectors:
  An experimental analysis,'' \emph{IEEE Transactions on Aerospace and
  Electronic Systems}, vol.~41, no.~1, pp. 233--251, 2005.

\bibitem{ConteDeMaio02b}
E.~Conte, A.~De~Maio, and G.~Ricci, ``Recursive estimation of the covariance
  matrix of a compound-{Gaussian} process and its application to adaptive
  {CFAR} detection,'' \emph{IEEE Transactions on Signal Processing}, vol.~8,
  no.~50, pp. 1908--1915, 2002.

\bibitem{ConteDeMaio04}
E.~Conte and A.~{De Maio}, ``Mitigation techniques for non-{Gaussian} sea
  clutter,'' \emph{IEEE Journal of Oceanic Engineering}, vol.~29, no.~2, pp.
  284--302, April 2004.

\bibitem{GaoAubry20AES}
Y.~{Gao}, A.~{Aubry}, A.~{De Maio}, and H.~{Ji}, ``Adaptive target separation
  detection,'' \emph{IEEE Transactions on Aerospace and Electronic Systems},
  pp. 1--1, 2020, DOI: 10.1109\/TAES.2020.3018898.

\bibitem{DeMaioAlfano03}
A.~De~Maio and G.~Alfano, ``Polarimetric adaptive detection in non-{Gaussian}
  noise,'' \emph{Signal Processing}, vol.~83, no.~2, pp. 297 -- 306, 2003.

\bibitem{DeMaioAlfano04}
A.~De~Maio, G.~Alfano, and E.~Conte, ``Polarization diversity detection in
  compound-{Gaussian} clutter,'' \emph{IEEE Transactions on Aerospace and
  Electronic Systems}, vol.~40, no.~1, pp. 114--131, 2004.

\bibitem{AlfanoDeMaio04}
G.~Alfano, A.~De~Maio, and E.~Conte, ``Polarization diversity detection of
  distributed targets in compound-{Gaussian} clutter,'' \emph{IEEE Transactions
  on Aerospace and Electronic Systems}, vol.~40, no.~2, pp. 755--765, 2004.

\bibitem{LiuZhang12c}
J.~Liu, Z.-J. Zhang, and Y.~Yang, ``Performance enhancement of subspace
  detection with a diversely polarized antenna,'' \emph{IEEE Signal Processing
  Letters}, vol.~19, no.~1, pp. 4--7, 2012.

\bibitem{HaoGazor16AES}
C.~Hao, S.~Gazor, X.~Ma, S.~Yan, C.~Hou, and D.~Orlando, ``Polarimetric
  detection and range estimation of a point-like target,'' \emph{IEEE
  Transactions on Aerospace and Electronic Systems}, vol.~52, no.~2, pp.
  603--616, 2016.

\bibitem{ParkLiWang95}
H.-R. Park, J.~Li, and H.~Wang, ``Polarization-space-time domain generalized
  likelihood ratio detection of radar targets,'' \emph{Signal Processing},
  vol.~41, no.~2, pp. 153--164, 1995.

\bibitem{DeMaioRicci01}
A.~De~Maio and G.~Ricci, ``A polarimetric adaptive matched filter,''
  \emph{Signal Processing}, vol.~81, no.~12, pp. 2583--2589, 2001.

\bibitem{RaghavanPulsone96}
R.~S. Raghavan, N.~Pulsone, and D.~J. McLaughlin, ``Performance of the {GLRT}
  for adaptive vector subspace detection,'' \emph{IEEE Transactions on
  Aerospace and Electronic Systems}, vol.~32, no.~4, pp. 1473--1487, 1996.

\bibitem{LombardoPastina01}
P.~Lombardo and D.~Pastina, ``Adaptive polarimetric target detection with
  coherent radar part {II}: Detection against non-{Gaussian} background,''
  \emph{IEEE Transactions on Aerospace and Electronic Systems}, vol.~37, no.~4,
  pp. 1207--1220, 2001.

\bibitem{LiuZhang12b}
J.~Liu, Z.-J. Zhang, and Y.~Yang, ``Optimal waveform design for generalized
  likelihood ratio and adaptive matched filter detectors using a diversely
  polarized antenna,'' \emph{Signal Processing}, vol.~92, no.~4, pp.
  1126--1131, 2012.

\bibitem{LiuXie14b}
W.~Liu, W.~Xie, J.~Liu, and Y.~Wang, ``Adaptive double subspace signal
  detection in {Gaussian} background--part {I}: Homogeneous environments,''
  \emph{IEEE Transactions on Signal Processing}, vol.~62, no.~9, pp.
  2345--2357, 2014.

\bibitem{KrautScharf01}
S.~Kraut and L.~L. Scharf, ``Adaptive subspace detectors,'' \emph{IEEE
  Transactions on Signal Processing}, vol.~49, no.~1, pp. 1--16, 2001.

\bibitem{PastinaLombardo01}
D.~Pastina, P.~Lombardo, and T.~Bucciarelli, ``Adaptive polarimetric target
  detection with coherent radar part {I}: Detection against {Gaussian}
  background,'' \emph{IEEE Transactions on Aerospace and Electronic Systems},
  vol.~37, no.~4, pp. 1194--1206, 2001.

\bibitem{LiuZhang11}
J.~Liu, Z.-J. Zhang, Y.~Yang, and H.~Liu, ``A {CFAR} adaptive subspace detector
  for first-order or second-order {Gaussian} signals based on a single
  observation,'' \emph{IEEE Transactions on Signal Processing}, vol.~59,
  no.~11, pp. 5126--5140, 2011.

\bibitem{LiuZhang12a}
J.~Liu, Z.-J. Zhang, P.-L. Shui, and H.~Liu, ``Exact performance analysis of an
  adaptive subspace detector,'' \emph{IEEE Transactions on Signal Processing},
  vol.~60, no.~9, pp. 4945--4950, 2012.

\bibitem{LiuWang16AES}
W.~Liu, Y.-L. Wang, J.~Liu, W.~Xie, R.~Li, and H.~Chen, ``Design and
  performance analysis of adaptive subspace detectors in orthogonal
  interference and {Gaussian} noise,'' \emph{IEEE Transactions on Aerospace and
  Electronic Systems}, vol.~52, no.~5, pp. 2068--2079, 2016.

\bibitem{Hughes83}
P.~K. {Hughes}, ``A high-resolution radar detection strategy,'' \emph{IEEE
  Transactions on Aerospace and Electronic Systems}, vol.~19, no.~5, pp.
  663--667, Sep. 1983.

\bibitem{ShuaiKong12}
X.~Shuai, L.~Kong, and J.~Yang, ``Adaptive detection for distributed targets in
  {Gaussian} noise with {Rao} and {Wald} tests,'' \emph{Science China:
  Information Sciences}, vol.~55, no.~6, pp. 1290--1300, 2012.

\bibitem{HaoMa12}
C.~Hao, X.~Ma, X.~Shang, and L.~Cai, ``Adaptive detection of distributed
  targets in partially homogeneous environment with {Rao} and {Wald} tests,''
  \emph{Signal Processing}, vol.~92, no.~4, pp. 926--930, 2012.

\bibitem{WangCai91}
H.~Wang and L.~Cai, ``On adaptive multiband signal detection with {GLR}
  algorithm,'' \emph{IEEE Transactions on Aerospace and Electronic Systems},
  vol.~27, no.~2, pp. 225--233, 1991.

\bibitem{ConteDeMaio03}
E.~Conte, A.~De~Maio, and C.~Galdi, ``{CFAR} detection of multidimensional
  signals: An invariant approach,'' \emph{IEEE Transactions on Signal
  Processing}, vol.~51, no.~1, pp. 142--151, 2003.

\bibitem{LiuXie13b}
W.~Liu, W.~Xie, and Y.~Wang, ``{Rao} and {Wald} tests for distributed targets
  detection with unknown signal steering,'' \emph{IEEE Signal Processing
  Letters}, vol.~20, no.~11, pp. 1086--1089, 2013.

\bibitem{Raghavan20TSP}
R.~S. {Raghavan}, ``A generalized version of {ACE} and performance analysis,''
  \emph{IEEE Transactions on Signal Processing}, vol.~68, pp. 2574--2585, 2020.

\bibitem{BessonScharf06b}
O.~Besson, L.~L. Scharf, and S.~Kraut, ``Adaptive detection of a signal known
  only to lie on a line in a known subspace, when primary and secondary data
  are partially homogeneous,'' \emph{IEEE Transactions on Signal Processing},
  vol.~54, no.~12, pp. 4698--4705, 2006.

\bibitem{BoseSteinhardt96a}
S.~Bose and A.~O. Steinhardt, ``Adaptive array detection of uncertain rank one
  waveforms,'' \emph{IEEE Transactions on Signal Processing}, vol.~44, no.~11,
  pp. 2801--2809, 1996.

\bibitem{LiuXie14c}
W.~Liu, W.~Xie, J.~Liu, D.~Zou, H.~Wang, and Y.~Wang, ``Detection of a
  distributed target with direction uncertainty,'' \emph{IET Radar, Sonar and
  Navigation}, vol.~8, no.~9, pp. 1177--1183, 2014.

\bibitem{LiuLiu2015b}
W.~Liu, J.~Liu, L.~Huang, K.~Yan, and Y.~Wang, ``Robust {GLRT} approaches to
  signal detection in the presence of spatial-temporal uncertainty,''
  \emph{Signal Processing}, vol. 118, pp. 272--284, 2016.

\bibitem{Liu19SCIS}
W.~Liu, F.~Gao, Y.~Luo, J.~Liu, and Y.-L. Wang, ``{GLRT}-based generalized
  direction detector in partially homogeneous environment,'' \emph{Science
  China: Information Sciences}, vol.~62, no.~11, p. 209303, 2019.

\bibitem{KellyForsythe89}
E.~J. Kelly and K.~M. Forsythe, ``Adaptive detection and parameter estimation
  for multidimensional signal models,'' Lincoln Laboratory, Lexington,
  Technical Report, 1989.

\bibitem{LiuXie14a}
W.~Liu, W.~Xie, J.~Liu, and Y.~Wang, ``Adaptive double subspace signal
  detection in {Gaussian} background--part {II}: Partially homogeneous
  environments,'' \emph{IEEE Transactions on Signal Processing}, vol.~62,
  no.~9, pp. 2358--2369, 2014.

\bibitem{Raghavan13a}
R.~S. Raghavan, ``Maximal invariants and performance of some invariant
  hypothesis tests for an adaptive detection problem,'' \emph{IEEE Transactions
  on Signal Processing}, vol.~61, no.~14, pp. 3607--3619, 2013.

\bibitem{Raghavan13b}
------, ``Analysis of steering vector mismatch on adaptive noncoherent
  integration,'' \emph{IEEE Transactions on Aerospace and Electronic Systems},
  vol.~49, no.~4, pp. 2496--2508, 2013.

\bibitem{LiuLiu15SPL}
J.~Liu, W.~Liu, B.~Chen, H.~Liu, and H.~Li, ``Detection probability of a {CFAR}
  matched filter with signal steering vector errors,'' \emph{IEEE Signal
  Processing Letters}, vol.~22, no.~12, pp. 2474--2478, 2015.

\bibitem{Kelly89}
E.~J. Kelly, ``Performance of an adaptive detection algorithm: rejection of
  unwanted signals,'' \emph{IEEE Transactions on Aerospace and Electronic
  Systems}, vol.~25, no.~2, pp. 122--133, 1989.

\bibitem{Richmond00b}
C.~D. Richmond, ``Performance of the adaptive sidelobe blanker detection
  algorithm in homogeneous environments,'' \emph{IEEE Transactions on Signal
  Processing}, vol.~48, no.~5, pp. 1235--1247, 2000.

\bibitem{LiuLiu16SP}
W.~Liu, J.~Liu, C.~Zhang, and H.~Li, ``Performance prediction of subspace-based
  adaptive detectors with signal mismatch,'' \emph{Signal Processing}, vol.
  123, pp. 122--126, 2016.

\bibitem{ZeiraFriedlander97}
A.~Zeira and B.~Friedlander, ``Robust subspace detectors,'' in \emph{the 31th
  Asilomar Conference on Signals, Systems and Computers}, vol.~1, 1997, pp.
  778--782.

\bibitem{ZeiraFriedlander98}
------, ``Robust adaptive subspace detectors for space time processing,'' in
  \emph{IEEE International Conference on Acoustics, Speech and Signal
  Processing (ICASSP)}, vol.~4, 1998, pp. 1965--1968.

\bibitem{DeMaioDeNicola10}
A.~De~Maio, S.~D. Nicola, A.~Farina, and S.~Iommelli, ``Adaptive detection of a
  signal with angle uncertainty,'' \emph{IET Radar, Sonar and Navigation},
  vol.~4, no.~4, pp. 537--547, 2010.

\bibitem{LeeNguyen19}
S.~{Lee}, M.~{Nguyen}, I.~{Song}, J.~{Bae}, and S.~{Yoon}, ``Detection schemes
  for range-spread targets based on the semidefinite problem,'' \emph{IEEE
  Transactions on Aerospace and Electronic Systems}, vol.~55, no.~1, pp.
  57--69, 2019.

\bibitem{DeMaio05}
A.~De~Maio, ``Robust adaptive radar detection in the presence of steering
  vector mismatches,'' \emph{IEEE Transactions on Aerospace and Electronic
  Systems}, vol.~41, no.~4, pp. 1322--1337, 2005.

\bibitem{Besson06}
O.~Besson, ``Detection of a signal in linear subspace with bounded mismatch,''
  \emph{IEEE Transactions on Aerospace and Electronic Systems}, vol.~42, no.~3,
  pp. 1131--1139, 2006.

\bibitem{Besson07a}
------, ``Adaptive detection with bounded steering vectors mismatch angle,''
  \emph{IEEE Transactions on Signal Processing}, vol.~55, no.~4, pp. 1560 --
  1564, 2007.

\bibitem{DeMaioHuang11}
A.~{De Maio}, Y.~{Huang}, D.~P. {Palomar}, S.~{Zhang}, and A.~{Farina},
  ``Fractional {QCQP} with applications in ml steering direction estimation for
  radar detection,'' \emph{IEEE Transactions on Signal Processing}, vol.~59,
  no.~1, pp. 172--185, Jan 2011.

\bibitem{HaoBandiera12}
C.~Hao, F.~Bandiera, J.~Yang, D.~Orlando, S.~Yan, and C.~Hou, ``Adaptive
  detection of multiple point-like targets under conic constraints,''
  \emph{Progress In Electromagnetics Research}, vol. 129, pp. 231--250, 2012.

\bibitem{ColucciaRicci19}
A.~{Coluccia}, G.~{Ricci}, and O.~{Besson}, ``Design of robust radar detectors
  through random perturbation of the target signature,'' \emph{IEEE
  Transactions on Signal Processing}, vol.~67, no.~19, pp. 5118--5129, 2019.

\bibitem{PulsoneRader01}
N.~B. Pulsone and C.~M. Rader, ``Adaptive beamformer orthogonal rejection
  test,'' \emph{IEEE Transactions on Signal Processing}, vol.~49, no.~3, pp.
  521--529, 2001.

\bibitem{BandieraBesson07WABORT}
F.~Bandiera, O.~Besson, D.~Orlando, and G.~Ricci, ``Derivation and analysis of
  an adaptive detector with enhanced mismatched signals rejection
  capabilities,'' in \emph{41st Asilomar Conference on Signals, Systems and
  Computers}, 2007, pp. 2182--2186.

\bibitem{HaoShang12}
C.~Hao, X.~Shang, F.~Bandiera, and L.~Cai, ``{Bayesian} radar detection with
  orthogonal rejection,'' \emph{IEICE Transactions on Fundamentals of
  Electronics, Communications and Computer Sciences}, vol. E95-A, no.~2, pp.
  596--599, 2012.

\bibitem{ColucciaRicci15b}
A.~Coluccia and G.~Ricci, ``A tunable {W-ABORT}-like detector with improved
  detection vs rejection capabilities trade-off,'' \emph{IEEE Signal Processing
  Letters}, vol.~22, no.~6, pp. 713--717, 2015.

\bibitem{LiuZhao17}
J.~Liu, H.-Y. Zhao, W.~Liu, H.~Li, and H.~Liu, ``Adaptive detection using both
  the test and training data for disturbance correlation estimation,''
  \emph{Signal Processing}, vol. 137, pp. 309--318, 2017.

\bibitem{LiuLiu2017AES}
W.~Liu, J.~Liu, L.~Huang, Z.~Yang, H.~Yang, and Y.-L. Wang, ``Distributed
  target detectors with capabilities of mismatched subspace signal rejection,''
  \emph{IEEE Transactions on Aerospace and Electronic Systems}, vol.~53, no.~2,
  pp. 888--900, 2017.

\bibitem{HaoYang12}
C.~Hao, J.~Yang, X.~Ma, C.~Hou, and D.~Orlando, ``Adaptive detection of
  distributed targets with orthogonal rejection,'' \emph{IET Radar, Sonar and
  Navigation}, vol.~6, no.~6, pp. 483--493, 2012.

\bibitem{LiuLiu18TSPPHE}
W.~Liu, J.~Liu, Q.~Du, and Y.~L. Wang, ``Distributed target detection in
  partially homogeneous environment when signal mismatch occurs,'' \emph{IEEE
  Transactions on Signal Processing}, vol.~66, no.~14, pp. 3918--3928, 2018.

\bibitem{OrlandoRicci10}
D.~Orlando and G.~Ricci, ``A {Rao} test with enhanced selectivity properties in
  homogeneous scenarios,'' \emph{IEEE Transactions on Signal Processing},
  vol.~58, no.~10, pp. 5385--5390, 2010.

\bibitem{Kalson92}
S.~Z. Kalson, ``An adaptive array detector with mismatched signal rejection,''
  \emph{IEEE Transactions on Aerospace and Electronic Systems}, vol.~28, no.~1,
  pp. 195--207, 1992.

\bibitem{HaoLiu10a}
C.~Hao, B.~Liu, S.~Yan, and L.~Cai, ``Parametric adaptive radar detector with
  enhanced mismatched signals rejection capabilities,'' \emph{{EURASIP} Journal
  on Advances in Signal Processing}, vol. 2010, 2010.

\bibitem{LiuXie14e}
W.~Liu, W.~Xie, and Y.~Wang, ``Parametric detector in the situation of
  mismatched signals,'' \emph{IET Radar, Sonar and Navigation}, vol.~8, no.~1,
  pp. 48--53, 2014.

\bibitem{BandieraOrlando09c}
F.~Bandiera, D.~Orlando, and G.~Ricci, ``One- and two-stage tunable
  receivers*,'' \emph{IEEE Transactions on Signal Processing}, vol.~57, no.~8,
  pp. 3264--3273, 2009.

\bibitem{RaghavanQiu95}
R.~S. Raghavan, H.~F. Qiu, and D.~J. Mclaughlin, ``{CFAR} detection in clutter
  with unknown correlation properties,'' \emph{IEEE Transactions on Aerospace
  and Electronic Systems}, vol.~31, no.~2, pp. 647--657, 1995.

\bibitem{LiuXie15b}
W.~Liu, W.~Xie, R.~Li, F.~Gao, X.~Hu, and Y.~Wang, ``Adaptive detection in the
  presence of signal mismatch,'' \emph{Journal of Systems Engineering and
  Electronics}, vol.~26, no.~1, pp. 38--43, 2015.

\bibitem{LiuXie15a}
W.~Liu, W.~Xie, Q.~Zhang, R.~Li, K.~Duan, and Y.~Wang, ``A doubly parameterized
  detector for mismatched signals,'' \emph{Chinese Journal of Electronics},
  vol.~24, no.~1, pp. 152--156, 2015.

\bibitem{LiuLiu16TSP2}
J.~Liu, W.~Liu, B.~Chen, H.~Liu, H.~Li, and C.~Hao, ``Modified {Rao} test for
  multichannel adaptive signal detection,'' \emph{IEEE Transactions on Signal
  Processing}, vol.~64, no.~3, pp. 714--725, 2016.

\bibitem{LiuZhouLiu18TSP}
J.~Liu, S.~Zhou, W.~Liu, J.~Zheng, H.~Liu, and J.~Li, ``Tunable adaptive
  detection in colocated {MIMO} radar,'' \emph{IEEE Transactions on Signal
  Processing}, vol.~66, no.~4, pp. 1080--1092, 2018.

\bibitem{PulsoneZatman00}
N.~B. Pulsone and M.~A. Zatman, ``A computationally efficient two-step
  implementation of the {GLRT},'' \emph{IEEE Transactions on Signal
  Processing}, vol.~48, no.~3, pp. 609--616, 2000.

\bibitem{BandieraBesson08a}
F.~Bandiera, O.~Besson, D.~Orlando, and G.~Ricci, ``An improved adaptive
  sidelobe blanker,'' \emph{IEEE Transactions on Signal Processing}, vol.~56,
  no.~9, pp. 4152--4161, 2008.

\bibitem{BandieraOrlando08}
F.~Bandiera, D.~Orlando, and G.~Ricci, ``A subspace-based adaptive sidelobe
  blanker,'' \emph{IEEE Transactions on Signal Processing}, vol.~56, no.~9, pp.
  4141--4151, 2008.

\bibitem{HaoLiu11}
C.~Hao, B.~Liu, and L.~Cai, ``Performance analysis of a two-stage {Rao}
  detector,'' \emph{Signal Processing}, vol.~91, no.~8, pp. 2141--2146, 2011.

\bibitem{DuanLiu17GRSL}
K.~Duan, M.~Liu, H.~Dai, F.~Xu, and W.~Liu, ``A two-stage detector for
  mismatched subspace signals,'' \emph{IEEE Geoscience and Remote Sensing
  Letters}, vol.~14, no.~12, pp. 2270--2274, 2017.

\bibitem{BandieraOrlando09a}
F.~Bandiera, D.~Orlando, and G.~Ricci, ``Advanced radar detection schemes under
  mismatched signal models,'' \emph{Synthesis Lectures on Signal Processing},
  vol.~8, pp. 1 -- 105, 2009.

\bibitem{DeMaioOrlando16AESm}
A.~De~Maio and D.~Orlando, ``A survey on two-stage decision schemes for
  point-like targets in {Gaussian} interference,'' \emph{IEEE Aerospace and
  Electronic Systems Magazine}, vol.~31, no.~4, pp. 20--29, 2016.

\bibitem{LiuLi17_SP_MisSubs}
J.~Liu, K.~Li, X.~Zhang, M.~Liu, and W.~Liu, ``A weighted detector for
  mismatched subspace signals,'' \emph{Signal Processing}, vol. 140, pp.
  110--115, 2017.

\bibitem{BandieraBesson07TSP_WABORT}
F.~Bandiera, O.~Besson, and G.~Ricci, ``An {ABORT}-like detector with improved
  mismatched signals rejection capabilities,'' \emph{IEEE Transactions on
  Signal Processing}, vol.~56, no.~1, pp. 14--25, 2007.

\bibitem{Briggs04}
J.~N. Briggs, \emph{Target Detection by Marine Radar}.\hskip 1em plus 0.5em
  minus 0.4em\relax London: The Institution of Electrical Engineers, 2004.

\bibitem{ScharfFriedlander94}
L.~L. Scharf and B.~Friedlander, ``Matched subspace detectors,'' \emph{IEEE
  Transactions on Signal Processing}, vol.~42, no.~8, pp. 2146--2156, 1994.

\bibitem{BehrensScharf94}
R.~T. Behrens and L.~L. Scharf, ``Signal processing applications of oblique
  projection operators,'' \emph{IEEE Transactions on Signal Processing},
  vol.~42, no.~6, pp. 1413--1424, 1994.

\bibitem{ScharfMcCloud02}
L.~L. Scharf and M.~L. McCloud, ``Blind adaptation of zero forcing projections
  and oblique pseudo-inverses for subspace detection and estimation when
  interference dominates noise,'' \emph{IEEE Transactions on Signal
  Processing}, vol.~50, no.~12, pp. 2938--2946, 2002.

\bibitem{BessonScharf05}
O.~Besson, L.~L. Scharf, and F.~Vincent, ``Matched direction detectors and
  estimators for array processing with subspace steering vector
  uncertainties,'' \emph{IEEE Transactions on Signal Processing}, vol.~53,
  no.~12, pp. 4453--4463, 2005.

\bibitem{BessonScharf06a}
O.~Besson and L.~L. Scharf, ``{CFAR} matched direction detector,'' \emph{IEEE
  Transactions on Signal Processing}, vol.~54, no.~7, pp. 2840--2844, 2006.

\bibitem{WangFang12}
P.~Wang, J.~Fang, H.~Li, and B.~Himed, ``Detection with target-induced subspace
  interference,'' \emph{IEEE Signal Processing Letters}, vol.~19, no.~7, pp.
  403--406, 2012.

\bibitem{LiuZhang14}
J.~Liu, Z.-J. Zhang, Y.~Cao, and M.~Wang, ``Distributed target detection in
  subspace interference plus {Gaussian} noise,'' \emph{Signal Processing},
  vol.~85, pp. 88--100, 2014.

\bibitem{BandieraDeMaio07a}
F.~Bandiera, A.~De~Maio, A.~S. Greco, and G.~Ricci, ``Adaptive radar detection
  of distributed targets in homogeneous and partially homogeneous noise plus
  subspace interference,'' \emph{IEEE Transactions on Signal Processing},
  vol.~55, no.~4, pp. 1223--1237, 2007.

\bibitem{LiuLi19}
J.~{Liu} and J.~{Li}, ``False alarm rate of the {GLRT} for subspace signals in
  subspace interference plus {Gaussian} noise,'' \emph{IEEE Transactions on
  Signal Processing}, vol.~67, no.~11, pp. 3058--3069, June 2019.

\bibitem{LiuLiu2015d}
W.~Liu, J.~Liu, L.~Huang, D.~Zou, and Y.~Wang, ``{Rao} tests for distributed
  target detection in interference and noise,'' \emph{Signal Processing}, vol.
  117, pp. 333--342, 2015.

\bibitem{LiuLiu19TAESWald}
W.~Liu, J.~Liu, H.~Li, Q.~Du, and Y.~Wang, ``Multichannel signal detection
  based on {Wald} test in subspace interference and {Gaussian} noise,''
  \emph{IEEE Transactions on Aerospace and Electronic Systems}, vol.~55, no.~3,
  pp. 1370--1381, 2019.

\bibitem{Wang20SP}
Z.~Wang, ``Modified {Rao} test for distributed target detection in interference
  and noise,'' \emph{Signal Processing}, vol. 172, p. 107578, 2020.

\bibitem{LiuLi20}
J.~{Liu} and J.~{Li}, ``Analytical performance of rank-one signal detection in
  subspace interference plus {Gaussian} noise,'' \emph{IEEE Transactions on
  Aerospace and Electronic Systems}, vol.~56, no.~2, pp. 1595--1601, 2020.

\bibitem{LiuLiu18AES}
W.~Liu, J.~Liu, L.~Huang, C.~Hao, and Y.-L. Wang, ``Performance analysis of
  adaptive detectors for point targets in subspace interference and {Gaussian}
  noise,'' \emph{IEEE Transactions on Aerospace and Electronic Systems},
  vol.~54, no.~1, pp. 429--441, 2018.

\bibitem{LiuLiu20a}
W.~Liu, J.~Liu, Y.~Gao, G.~Wang, , and Y.-L. Wang, ``Multichannel signal
  detection in interference and noise when signal mismatch happens,''
  \emph{Signal Processing}, vol. 166, 2020.

\bibitem{AubryCarotenuto16SPL}
A.~Aubry, V.~Carotenuto, A.~De~Maio, and D.~Orlando, ``Coincidence of maximal
  invariants for two adaptive radar detection problems,'' \emph{IEEE Signal
  Processing Letters}, vol.~23, no.~9, pp. 1193--1196, 2016.

\bibitem{DeMaioOrlando2016Invariance}
A.~De~Maio and D.~Orlando, ``Adaptive radar detection of a subspace signal
  embedded in subspace structured plus {Gaussian} interference via
  invariance,'' \emph{IEEE Transactions on Signal Processing}, vol.~64, no.~8,
  pp. 2156--2167, 2016.

\bibitem{CiuonzoDeMaio16a}
D.~Ciuonzo, A.~De~Maio, and D.~Orlando, ``A unifying framework for adaptive
  radar detection in homogeneous plus structured interference---part {I}: On
  the maximal invariant statistic,'' \emph{IEEE Transactions on Signal
  Processing}, vol.~64, no.~11, pp. 2894--2906, 2016.

\bibitem{CiuonzoDeMaio16b}
------, ``A unifying framework for adaptive radar detection in homogeneous plus
  structured interference---part {II}: Detectors design,'' \emph{IEEE
  Transactions on Signal Processing}, vol.~64, no.~11, pp. 2907--2919, 2016.

\bibitem{CiuonzoDeMaio17TSP}
------, ``On the statistical invariance for adaptive radar detection in
  partially homogeneous disturbance plus structured interference,'' \emph{IEEE
  Transactions on Signal Processing}, vol.~65, no.~5, pp. 1222--1234, 2017.

\bibitem{BandieraBesson07}
F.~Bandiera, O.~Besson, D.~Orlando, G.~Ricci, and L.~L. Scharf, ``{GLRT}-based
  direction detectors in homogeneous noise and subspace interference,''
  \emph{IEEE Transactions on Signal Processing}, vol.~55, no.~6, pp.
  2386--2394, 2007.

\bibitem{LiLiu18MSSP}
W.~Li, H.~Tong, K.~Li, Y.~Yang, and W.~Liu, ``{Wald} tests for direction
  detection in noise and interference,'' \emph{Multichannel Systems and Signal
  Processing}, vol.~29, no.~4, pp. 1563--1577, 2018.

\bibitem{DongLiu17SPL_DD_PHE}
Y.~Dong, M.~Liu, K.~Li, Z.~Tang, and W.~Liu, ``Adaptive direction detection in
  deterministic interference and partially homogeneous noise,'' \emph{IEEE
  Signal Processing Letters}, vol.~24, no.~5, pp. 599--603, 2017.

\bibitem{BandieraBesson13}
F.~Bandiera, O.~Besson, and G.~Ricci, ``Direction detector for distributed
  targets in unknown noise and interference,'' \emph{Electronics Letters},
  vol.~49, no.~1, pp. 68--69, 2013.

\bibitem{Richmond00c}
C.~D. Richmond, ``Statistics of adaptive nulling and use of the generalized
  eigenrelation ({GER}) for modeling inhomogeneities in adaptive processing,''
  \emph{IEEE Transactions on Signal Processing}, vol.~48, no.~5, pp.
  1263--1273, 2000.

\bibitem{Richmond00a}
------, ``Performance of a class of adaptive detection algorithms in
  nonhomogeneous environments,'' \emph{IEEE Transactions on Signal Processing},
  vol.~48, no.~5, pp. 1248--1263, 2000.

\bibitem{RabideauSteinhardt99}
D.~J. Rabideau and A.~O. Steinhardt, ``Improved adaptive clutter cancellation
  through data-adaptive training,'' \emph{IEEE Transactions on Aerospace and
  Electronic Systems}, vol.~35, no.~3, pp. 879--891, 1999.

\bibitem{BandieraDeMaio07b}
F.~Bandiera, A.~De~Maio, and G.~Ricci, ``Adaptive {CFAR} radar detection with
  conic rejection,'' \emph{IEEE Transactions on Signal Processing}, vol.~55,
  no.~6, pp. 2533--2541, 2007.

\bibitem{DeMaioDeNicola09b}
A.~De~Maio, S.~De~Nicola, Y.~Huang, S.~Zhang, and A.~Farina, ``Adaptive
  detection and estimation in the presence of useful signal and interference
  mismatches,'' \emph{IEEE Transactions on Signal Processing}, vol.~57, no.~2,
  pp. 436--450, 2009.

\bibitem{SvenssonJakobsson11}
A.~Svensson and A.~Jakobsson, ``Adaptive detection of a partly known signal
  corrupted by strong interference,'' \emph{IEEE Signal Processing Letters},
  vol.~18, no.~12, pp. 729--732, 2011.

\bibitem{LiuLiu2015c}
W.~Liu, J.~Liu, L.~Wang, K.~Duan, Z.~Chen, and Y.~Wang, ``Adaptive array
  detection in noise and completely unknown jamming,'' \emph{Digital Signal
  Processing}, vol.~46, pp. 41--48, 2015.

\bibitem{LiuLiu_16SPL}
W.~Liu, J.~Liu, X.~Hu, Z.~Tang, L.~Huang, and Y.-L. Wang, ``Statistical
  performance analysis of the adaptive orthogonal rejection detector,''
  \emph{IEEE Signal Processing Letters}, vol.~23, no.~6, pp. 873--877, 2016.

\bibitem{Besson07b}
O.~Besson, ``Detection in the presence of surprise or undernulled
  interference,'' \emph{IEEE Signal Processing Letters}, vol.~14, no.~5, pp.
  352--354, 2007.

\bibitem{LiuHan17SCIS}
W.~Liu, H.~Han, J.~Liu, H.~Li, K.~Li, and Y.-L. Wang, ``Multichannel radar
  adaptive signal detection in interference and structure nonhomogeneity,''
  \emph{Science China: Information Sciences}, vol.~60, no.~11, p.~1, Sep. 2017.

\bibitem{BessonOrlando07}
O.~Besson and D.~Orlando, ``Adaptive detection in nonhomogeneous environments
  using the generalized eigenrelation,'' \emph{IEEE Signal Processing Letters},
  vol.~14, no.~10, pp. 731--734, 2007.

\bibitem{HaoOrlando12a}
C.~Hao, D.~Orlando, and C.~Hou, ``{Rao} and {Wald} tests for nonhomogeneous
  scenarios,'' \emph{Sensors}, vol.~12, no.~4, pp. 4730--4736, 2012.

\bibitem{ShangLiu18GER}
Z.~Shang, Q.~Du, Z.~Tang, T.~Zhang, and W.~Liu, ``Multichannel adaptive signal
  detection in structural nonhomogeneous environment characterized by the
  generalized eigenrelation,'' \emph{Signal Processing}, vol. 148, pp.
  214--222, 2018.

\bibitem{TangWang20}
P.~Tang, Y.-L. Wang, W.~Liu, Q.~Du, B.~Li, and W.~Chen, ``Adaptive subspace
  signal detection in a type of structure-nonhomogeneity environment,''
  \emph{Signal Processing}, vol. 173, p. 107600, 2020.

\bibitem{Orlando17}
D.~Orlando, ``A novel noise jamming detection algorithm for radar
  applications,'' \emph{IEEE Signal Processing Letters}, vol.~24, no.~2, pp.
  206--210, 2017.

\bibitem{AddabboBesson19}
P.~Addabbo, O.~Besson, D.~Orlando, and G.~Ricci, ``Adaptive detection of
  coherent radar targets in the presence of noise jamming,'' \emph{IEEE
  Transactions on Aerospace and Electronic Systems}, vol.~67, no.~24, pp.
  6498--6510, 2019.

\bibitem{YanAddabbo20}
L.~{Yan}, P.~{Addabbo}, C.~{Hao}, D.~{Orlando}, and A.~{Farina}, ``New eccm
  techniques against noiselike and/or coherent interferers,'' \emph{IEEE
  Transactions on Aerospace and Electronic Systems}, vol.~56, no.~2, pp.
  1172--1188, 2020.

\bibitem{Raghavan19}
R.~S. Raghavan, ``A {CFAR} detector for mismatched eigenvalues of training
  sample covariance matrix,'' \emph{IEEE Transactions on Aerospace and
  Electronic Systems}, vol.~67, no.~17, pp. 4624--4635, September 2019.

\bibitem{Besson19}
O.~Besson, A.~Coluccia, E.~Chaumette, G.~Ricci, and F.~Vincent, ``Detection of
  {Gaussian} signal using adaptively whitened data,'' \emph{IEEE Signal
  Processing Letters}, vol.~26, no.~3, pp. 430--434, 2019.

\bibitem{Besson20SPL}
O.~{Besson}, ``Adaptive detection using whitened data when some of the training
  samples undergo covariance mismatch,'' \emph{IEEE Signal Processing Letters},
  vol.~27, pp. 795--799, 2020.

\bibitem{LiuLiu2017RMB}
J.~{Liu}, W.~{Liu}, and H.~{Liu}, ``A simpler proof of rapid convergence rate
  in adaptive arrays,'' \emph{IEEE Transactions on Aerospace and Electronic
  Systems}, vol.~53, no.~1, pp. 135--136, Feb 2017.

\bibitem{Gerlach02}
K.~Gerlach, ``Outlier resistant adaptive matched filtering,'' \emph{IEEE
  Transactions on Aerospace and Electronic Systems}, vol.~38, no.~3, pp. 885 --
  901, 2002.

\bibitem{RangaswamyMichels04}
J.~H.~M. M.~Rangaswamy and B.~Himed, ``Statistical analysis of the
  non-homogeneity detector for {STAP} applications,'' \emph{Digital Signal
  Processing}, vol.~14, no.~3, pp. 253--267, 2004.

\bibitem{HanDeMaio19AES}
S.~{Han}, A.~{De Maio}, V.~{Carotenuto}, L.~{Pallotta}, and X.~{Huang},
  ``Censoring outliers in radar data: An approximate {ML} approach and its
  analysis,'' \emph{IEEE Transactions on Aerospace and Electronic Systems},
  vol.~55, no.~2, pp. 534--546, 2019.

\bibitem{BessonTourneret07}
O.~Besson, J.-Y. Tourneret, and S.~Bidon, ``Knowledge-aided {Bayesian}
  detection in heterogeneous environments,'' \emph{IEEE Signal Processing
  Letters}, vol.~14, no.~5, pp. 355--358, 2007.

\bibitem{BidonBesson08a}
S.~Bidon, O.~Besson, and J.-Y. Tourneret, ``A {Bayesian} approach to adaptive
  detection in nonhomogeneous environments,'' \emph{IEEE Transactions on Signal
  Processing}, vol.~56, no.~1, pp. 205--217, 2008.

\bibitem{LiuHan18}
J.~{Liu}, J.~{Han}, Z.~{Zhang}, and J.~{Li}, ``Bayesian detection for {MIMO}
  radar in {Gaussian} clutter,'' \emph{IEEE Transactions on Signal Processing},
  vol.~66, no.~24, pp. 6549--6559, Dec 2018.

\bibitem{DeMaioFarina10}
A.~De~Maio, A.~Farina, and G.~Foglia, ``Knowledge-aided {Bayesian} radar
  detectors {\&} their application to live data,'' \emph{IEEE Transactions on
  Aerospace and Electronic Systems}, vol.~46, no.~1, pp. 170--183, 2010.

\bibitem{WangSahinoglu11}
P.~Wang, Z.~Sahinoglu, M.-O. Pun, H.~Li, and B.~Himed, ``Knowledge-aided
  adaptive coherence estimator in stochastic partially homogeneous
  environments,'' \emph{IEEE Signal Processing Letters}, vol.~18, no.~3, pp.
  193--196, 2011.

\bibitem{ZhouZhang12}
Y.~Zhou and L.-R. Zhang, ``Knowledge-aided {Bayesian} radar adaptive detection
  in heterogeneous environment: {GLRT}, {Rao} and {Wald} test,''
  \emph{International Journal of Electronics and Communications (AEU)},
  vol.~66, no.~3, pp. 239--243, 2012.

\bibitem{FrancescoBesson15}
F.~Bandiera, O.~Besson, A.~Coluccia, and G.~Ricci, ``{ABORT}-like detectors: A
  {Bayesian} approach,'' \emph{IEEE Transactions on Signal Processing},
  vol.~63, no.~19, pp. 5274--5284, 2015.

\bibitem{BandieraBesson11}
F.~Bandiera, O.~Besson, and G.~Ricci, ``Adaptive detection of distributed
  targets in compound-{Gaussian} noise without secondary data: A {Bayesian}
  approach,'' \emph{IEEE Transactions on Signal Processing}, vol.~59, no.~12,
  pp. 5698--5708, 2011.

\bibitem{RomanRangaswamy00}
J.~R. Roman, M.~Rangaswamy, D.~W. Davis, Q.~Zhang, B.~Himed, and J.~H. Michels,
  ``Parametric adaptive matched filter for airborne radar applications,''
  \emph{IEEE Transactions on Aerospace and Electronic Systems}, vol.~36, no.~2,
  pp. 677--692, 2000.

\bibitem{SohnLi07b}
K.~J. Sohn, H.~Li, and B.~Himed, ``Parametric {Rao} test for multichannel
  adaptive signal detection,'' \emph{IEEE Transactions on Aerospace and
  Electronic Systems}, vol.~43, no.~3, pp. 920--933, 2007.

\bibitem{SohnLi07a}
------, ``Parametric {GLRT} for multichannel adaptive signal detection,''
  \emph{IEEE Transactions on Signal Processing}, vol.~55, no.~11, pp.
  5351--5360, 2007.

\bibitem{LiMichels06}
H.~B. Li and J.~H. Michels, ``Parametric adaptive signal detection for
  hyperspectral imaging,'' \emph{IEEE Transactions on Signal Processing},
  vol.~54, no.~7, pp. 2704--2715, 2006.

\bibitem{MichelsHimed00}
J.~H. Michels, B.~Himed, and M.~Rangaswamy, ``Performance of {STAP} tests in
  {Gaussian} and compound-{Gaussian} clutter,'' \emph{Digital Signal
  Processing}, vol.~10, no.~4, pp. 309--324, 2000.

\bibitem{AlfanoRSN04}
G.~{Alfano}, A.~{De Maio}, and A.~{Farina}, ``Model-based adaptive detection of
  range-spread targets,'' \emph{IEE Proceedings - Radar, Sonar and Navigation},
  vol. 151, no.~1, pp. 2--10, Feb 2004.

\bibitem{SohnLi08}
K.~J. Sohn, H.~Li, and B.~Himed, ``Recursive parametric tests for multichannel
  adaptive signal detection,'' \emph{IET Radar, Sonar and Navigation}, vol.~2,
  no.~1, pp. 63--70, 2008.

\bibitem{AbramovichJohnson08a}
Y.~I. {Abramovich}, B.~A. {Johnson}, and N.~K. {Spencer}, ``Two-dimensional
  multivariate parametric models for radar applications--part {I}:
  {Maximum}-{Entropy} extensions for {Toeplitz}-block matrices,'' \emph{IEEE
  Transactions on Signal Processing}, vol.~56, no.~11, pp. 5509--5526, Nov
  2008.

\bibitem{AbramovichJohnson08b}
------, ``Two-dimensional multivariate parametric models for radar
  applications--part {II}: {Maximum}-{Entropy} extensions for
  {Hermitian}-{Block} matrices,'' \emph{IEEE Transactions on Signal
  Processing}, vol.~56, no.~11, pp. 5527--5539, Nov 2008.

\bibitem{AbramovichSpencer10}
Y.~I. Abramovich, N.~K. Spencer, and B.~A. Johnson, ``Band-inverse {TVAR}
  covariance matrix estimation for adaptive detection,'' \emph{IEEE
  Transactions on Aerospace and Electronic Systems}, vol.~46, no.~1, pp.
  375--396, 2010.

\bibitem{WangLi10a}
P.~Wang, H.~Li, and B.~Himed, ``A new parametric {GLRT} for multichannel
  adaptive signal detection,'' \emph{IEEE Transactions on Signal Processing},
  vol.~58, no.~1, pp. 317--325, 2010.

\bibitem{WangLi10b}
------, ``A {Bayesian} parametric test for multichannel adaptive signal
  detection in nonhomogeneous environments,'' \emph{IEEE Signal Processing
  Letters}, vol.~17, no.~4, pp. 351--354, 2010.

\bibitem{AbramovichRangaswamy11}
Y.~I. {Abramovich}, M.~{Rangaswamy}, B.~A. {Johnson}, P.~M. {Corbell}, and
  N.~K. {Spencer}, ``Performance analysis of two-dimensional parametric {STAP}
  for airborne radar using {KASSPER} data,'' \emph{IEEE Transactions on
  Aerospace and Electronic Systems}, vol.~47, no.~1, pp. 118--139, January
  2011.

\bibitem{JiangLi12b}
C.~Jiang, H.~Li, and M.~Rangaswamy, ``Conjugate gradient parametric detection
  of multichannel signals,'' \emph{IEEE Transactions on Aerospace and
  Electronic Systems}, vol.~48, no.~2, pp. 1521--1536, 2012.

\bibitem{JianHe13}
T.~Jian, Y.~He, F.~Su, X.~Huang, and D.~Ping, ``Adaptive detection of
  range-spread targets without secondary data in multichannel autoregressive
  process,'' \emph{Digital Signal Processing}, vol.~23, no.~5, pp. 1686--1694,
  2013.

\bibitem{WangWang14AR}
P.~Wang, Z.~Wang, H.~Li, and B.~Himed, ``Knowledge-aided parametric adaptive
  matched filter with automatic combining for covariance estimation,''
  \emph{IEEE Transactions on Signal Processing}, vol.~62, no.~18, pp.
  4713--4722, 2014.

\bibitem{ShiHao15}
B.~Shi, C.~Hao, C.~Hou, X.~Ma, and C.~Peng, ``Parametric {Rao} test for
  multichannel adaptive detection of range-spread target in partially
  homogeneous environments,'' \emph{Signal Processing}, vol. 108, pp. 421--429,
  2015.

\bibitem{Mennad2017AR}
A.~Mennad, A.~Younsi, M.~N.~E. Korso, and A.~M. Zoubir, ``Adaptive detection of
  range-spread target in compound-{Gaussian} clutter without secondary data,''
  \emph{Digital Signal Processing}, vol.~60, pp. 90--98, 2017.

\bibitem{GaoLi18TSP}
Y.~{Gao}, H.~{Li}, and B.~{Himed}, ``Adaptive subspace tests for multichannel
  signal detection in auto-regressive disturbance,'' \emph{IEEE Transactions on
  Signal Processing}, vol.~66, no.~21, pp. 5577--5587, Nov 2018.

\bibitem{YanHao20a}
L.~{Yan}, C.~{Hao}, D.~{Orlando}, A.~{Farina}, and C.~{Hou}, ``Parametric
  space-time detection and range estimation of point-like targets in partially
  homogeneous environment,'' \emph{IEEE Transactions on Aerospace and
  Electronic Systems}, vol.~56, no.~2, pp. 1228--1242, 2020.

\bibitem{Fuhrmann91}
D.~R. Fuhrmann, ``Application of {Toeplitz} covariance estimation to adaptive
  beamforming and detection,'' \emph{IEEE Transactions on Signal Processing},
  vol.~39, no.~10, pp. 2194--2198, 1991.

\bibitem{Raghavan17Kronecker}
R.~S. Raghavan, ``{CFAR} detection in clutter with a {Kronecker} covariance
  structure,'' \emph{IEEE Transactions on Signal Processing}, vol.~53, no.~2,
  pp. 619--629, 2017.

\bibitem{WangXia17Kronecker}
Y.~{Wang}, W.~{Xia}, Z.~{He}, H.~{Li}, and A.~P. {Petropulu}, ``Polarimetric
  detection in compound {Gaussian} clutter with {Kronecker} structured
  covariance matrix,'' \emph{IEEE Transactions on Signal Processing}, vol.~65,
  no.~17, pp. 4562--4576, Sep. 2017.

\bibitem{HaimovichBarNess91}
A.~M. Haimovich and Y.~Bar-Ness, ``An eigenanalysis interference canceler,''
  \emph{IEEE Transactions on Signal Processing}, vol.~39, no.~1, pp. 76--84,
  1991.

\bibitem{WangLiu14}
Y.~Wang, W.~Liu, W.~Xie, and Y.~Zhao, ``Reduced-rank space-time adaptive
  detection for airborne radar,'' \emph{Science China: Information Sciences},
  vol.~57, no.~8, pp. 1--11, 2014.

\bibitem{GoldsteinReed1997}
J.~S. Goldstein and I.~S. Reed, ``Reduced-rank adaptive filtering,'' \emph{IEEE
  Transactions on Signal Processing}, vol.~2, no.~45, pp. 492--496, 1997.

\bibitem{GoldsteinReed98}
J.~S. Goldstein, I.~S. Reed, and L.~L. Scharf, ``A multistage representation of
  the {Wiener} filter based on orthogonal projections,'' \emph{IEEE
  Transactions on Information Theory}, vol.~44, no.~7, pp. 2943--2959, 1998.

\bibitem{GoldsteinReed99}
J.~S. Goldstein, I.~S. Reed, and P.~A. Zulch, ``Multistage partially adaptive
  {STAP} {CFAR} detection algorithm,'' \emph{IEEE Transactions on Aerospace and
  Electronic Systems}, vol.~35, no.~2, pp. 645--662, 1999.

\bibitem{PadosKarystinos2001}
D.~A. Pados and G.~N. Karystinos, ``An iterative algorithm for the computation
  of the {MVDR} filter,'' \emph{IEEE Transactions on Signal Processing},
  vol.~2, no.~49, pp. 290--300, 2001.

\bibitem{FaDeLamare11}
R.~Fa and R.~C. De~Lamare, ``Reduced-rank {STAP} algorithms using joint
  iterative optimization of filters,'' \emph{IEEE Transactions on Aerospace and
  Electronic Systems}, vol.~47, no.~3, pp. 1668--1684, 2011.

\bibitem{ChenLi15}
Z.~Chen, H.~Li, and M.~Rangaswamy, ``Conjugate gradient adaptive matched
  filter,'' \emph{IEEE Transactions on Aerospace and Electronic Systems},
  vol.~51, no.~1, pp. 178--191, 2015.

\bibitem{ChenMitra02}
W.~Chen, U.~Mitra, and P.~Schniter, ``On the equivalence of three reduced rank
  linear estimators with applications to {DS-CDMA},'' \emph{IEEE Transactions
  on Information Theory}, vol.~48, no.~9, pp. 2609--2614, 2002.

\bibitem{ScharfChong08}
L.~L. Scharf, E.~K.~P. Chong, M.~D. Zoltowski, J.~S. Goldstein, and I.~S. Reed,
  ``Subspace expansion and the equivalence of conjugate direction and
  multistage {Wiener} filters,'' \emph{IEEE Transactions on Signal Processing},
  vol.~56, no.~10, pp. 5013--5019, 2008.

\bibitem{BroydenVespucci04}
C.~G. Broyden and M.~T. Vespucci, \emph{{Krylov} Solvers for Linear Algebraic
  Systems}.\hskip 1em plus 0.5em minus 0.4em\relax London: Elsevier, 2004.

\bibitem{Dietl07Krylov}
G.~K.~E. Dietl, \emph{Linear Estimation and Detection in {Krylov}
  Subspaces}.\hskip 1em plus 0.5em minus 0.4em\relax Berlin: Springer, 2007.

\bibitem{LiuXie14d}
W.~Liu, W.~Xie, R.~Li, Z.~Wang, and Y.~Wang, ``Adaptive detectors in the
  {Krylov} subspace,'' \emph{Science China: Information Sciences}, vol.~57,
  no.~10, pp. 1--11, 2014.

\bibitem{LiuXie15Krylov}
W.~Liu, W.~Xie, and Y.~Wang, ``Adaptive coherence estimator based on the
  {Krylov} subspace technique for airborne radar,'' \emph{Journal of Systems
  Engineering and Electronics}, vol.~26, no.~4, pp. 705--712, 2015.

\bibitem{GauReed98}
Y.-L. Gau and I.~S. Reed, ``An improved reduced-rank {CFAR} space-time adaptive
  radar detection algorithm,'' \emph{IEEE Transactions on Signal Processing},
  vol.~46, no.~8, pp. 2139--2146, 1998.

\bibitem{ReedGau99a}
I.~S. Reed and Y.-L. Gau, ``A fast {CFAR} detection space-time adaptive
  processing algorithm,'' \emph{IEEE Transactions on Signal Processing},
  vol.~47, no.~4, pp. 1151--1154, 1999.

\bibitem{ReedGau99b}
I.~S. Reed and Y.~L. Gau, ``Noncoherent summation of multiple reduced-rank test
  statistics for frequency-hopped {STAP},'' \emph{IEEE Transactions on Signal
  Processing}, vol.~47, no.~6, pp. 1708--1711, 1999.

\bibitem{LiuXie14f}
W.~Liu, W.~Xie, and Y.~Wang, ``Adaptive detection based on orthogonal partition
  of the primary and secondary data,'' \emph{Journal of Systems Engineering and
  Electronics}, vol.~25, no.~1, pp. 34--42, 2014.

\bibitem{LiSongLiu18RSN}
H.~{Li}, W.~{Song}, W.~{Liu}, and R.~{Wu}, ``Moving target detection with
  limited training data based on the subspace orthogonal projection,''
  \emph{IET Radar, Sonar Navigation}, vol.~12, no.~7, pp. 679--684, 2018.

\bibitem{Liu11DLSTAD}
W.~Liu, W.~Xie, and Y.-L. Wang, ``Diagonally loaded space-time adaptive
  detection,'' in \emph{Proceedings of 2011 IEEE CIE International Conference
  on Radar}, vol.~2, Oct 2011, pp. 1115--1119.

\bibitem{DeMaioOrlando16TSP}
A.~De~Maio, D.~Orlando, C.~Hao, and G.~Foglia, ``Adaptive detection of
  point-like targets in spectrally symmetric interference,'' \emph{IEEE
  Transactions on Signal Processing}, vol.~64, no.~12, pp. 3207--3220, 2016.

\bibitem{Nitzberg80}
R.~Nitzberg, ``Application of maximum likelihood estimation of persymmetric
  covariance matrices to adaptive processing,'' \emph{IEEE Transactions on
  Aerospace and Electronic Systems}, vol.~16, no.~1, pp. 124--127, 1980.

\bibitem{DeMaio03}
A.~{De Maio}, ``Maximum likelihood estimation of structured persymmetric
  covariance matrices,'' \emph{Signal Processing}, vol.~83, no.~3, pp.
  633--640, March 2003.

\bibitem{LiuLiu16TSP1}
J.~Liu, W.~Liu, H.~Liu, B.~Chen, X.-G. Xia, and F.~Dai, ``Average {SINR}
  calculation of a persymmetric sample matrix inversion beamformer,''
  \emph{IEEE Transactions on Signal Processing}, vol.~64, no.~8, pp.
  2135--2145, 2016.

\bibitem{LiuOrlando19}
J.~{Liu}, D.~{Orlando}, P.~{Addabbo}, and W.~{Liu}, ``{SINR} distribution for
  the persymmetric {SMI} beamformer with steering vector mismatches,''
  \emph{IEEE Transactions on Signal Processing}, vol.~67, no.~5, pp.
  1382--1392, March 2019.

\bibitem{PaillouxForster11}
G.~Pailloux, P.~Forster, J.-P. Ovarlez, and F.~Pascal, ``Persymmetric adaptive
  radar detectors,'' \emph{IEEE Transactions on Aerospace and Electronic
  Systems}, vol.~47, no.~4, pp. 2376--2390, 2011.

\bibitem{Hao15TAESPersymmetric}
C.~Hao, S.~Gazor, G.~Foglia, and B.~Liu, ``Persymmetric adaptive detection and
  range estimation of a small target,'' \emph{IEEE Transactions on Aerospace
  and Electronic Systems}, vol.~51, no.~4, pp. 2590--2604, 2015.

\bibitem{LiuCui15AES}
J.~Liu, G.~Cui, H.~Li, and B.~Himed, ``On the performance of a persymmetric
  adaptive matched filter,'' \emph{IEEE Transactions on Aerospace and
  Electronic Systems}, vol.~51, no.~4, pp. 2605--2614, 2015.

\bibitem{DeMaioOrlando15}
A.~{De Maio} and D.~Orlando, ``An invariant approach to adaptive radar
  detection under covariance persymmetry,'' \emph{IEEE Transactions on Signal
  Processing}, vol.~63, no.~5, pp. 1297--1309, March 2015.

\bibitem{DeMaioOrlando16b}
A.~{De Maio}, D.~Orlando, I.~Soloveychik, and A.~Wiesel, ``Invariance theory
  for adaptive detection in interference with group symmetric covariance
  matrix,'' \emph{IEEE Transaction on Signal Processing}, vol.~23, no.~64, pp.
  6299--6312, December 1 2016.

\bibitem{CaiWang91PersMSMI}
L.~Cai and H.~Wang, ``A persymmetric modified-{SMI} algorithm,'' \emph{Signal
  Processing}, vol.~23, no.~1, pp. 27--34, January 1991.

\bibitem{CaiWang92}
------, ``A persymmetric multiband {GLR} algorithm,'' \emph{IEEE Transactions
  on Aerospace and Electronic Systems}, vol.~28, no.~3, pp. 806--906, 1992.

\bibitem{LiuLi19TSP}
J.~Liu and J.~Li, ``Mismatched signal rejection performance of the persymmetric
  {GLRT} detector,'' \emph{IEEE Transactions on Signal Processing}, vol.~67,
  no.~6, pp. 1610--1619, March 15 2019.

\bibitem{LiuLiu19TSP1}
J.~Liu, W.~Liu, B.~Tang, J.~Zheng, and S.~Xu, ``Distributed target detection
  exploiting persymmetry in {G}aussian clutter,'' \emph{IEEE Transactions on
  Signal Processing}, vol.~67, no.~4, pp. 1022--1033, February 2019.

\bibitem{HaoOrlando14a}
C.~Hao, D.~Orlando, X.~Ma, S.~Yan, and C.~Hou, ``Persymmetric detectors with
  enhanced rejection capabilities,'' \emph{IET Radar, Sonar and Navigation},
  vol.~8, no.~5, pp. 557--563, 2014.

\bibitem{CasilloDeMaio07}
M.~Casillo, A.~De~Maio, and L.~Landi, ``A persymmetric {GLRT} for adaptive
  detection in partially-homogeneous environment,'' \emph{IEEE Signal
  Processing Letters}, vol.~14, no.~12, pp. 1016--1019, 2007.

\bibitem{HaoOrlando12b}
C.~Hao, D.~Orlando, X.~Ma, and C.~Hou, ``Persymmetric {Rao} and {Wald} tests
  for partially homogeneous environment,'' \emph{IEEE Signal Processing
  Letters}, vol.~19, no.~9, pp. 587--590, 2012.

\bibitem{GaoLiao14}
Y.~Gao, G.~Liao, S.~Zhu, X.~Zhang, and D.~Yang, ``Persymmetric adaptive
  detectors in homogeneous and partially homogeneous environments,'' \emph{IEEE
  Transactions on Signal Processing}, vol.~62, no.~2, pp. 331--342, 2014.

\bibitem{HaoOrlando14b}
C.~Hao, D.~Orlando, G.~Foglia, X.~Ma, S.~Yan, and C.~Hou, ``Persymmetric
  adaptive detection of distributed targets in partially-homogeneous
  environment,'' \emph{Digital Signal Processing}, vol.~24, pp. 42--51, 2014.

\bibitem{WangLi16SP}
Z.~Wang, M.~Li, H.~Chen, Y.~Lu, R.~Cao, P.~Zhang, L.~Zuo, and Y.~Wu,
  ``Persymmetric detectors of distributed targets in partially homogeneous
  disturbance,'' \emph{Signal Processing}, vol. 128, pp. 382--388, 2016.

\bibitem{CiuonzoOrlando6SPL}
D.~Ciuonzo, D.~Orlando, and L.~Pallotta, ``On the maximal invariant statistic
  for adaptive radar detection in partially homogeneous disturbance with
  persymmetric covariance,'' \emph{IEEE Signal Processing Letters}, vol.~23,
  no.~12, pp. 1830--1834, 2016.

\bibitem{LiuLiu18TSPPsmtrc}
J.~{Liu}, W.~{Liu}, Y.~{Gao}, S.~{Zhou}, and X.~{Xia}, ``Persymmetric adaptive
  detection of subspace signals: Algorithms and performance analysis,''
  \emph{IEEE Transactions on Signal Processing}, vol.~66, no.~23, pp.
  6124--6136, Dec 2018.

\bibitem{LiuSun19}
J.~Liu, S.~Sun, and W.~Liu, ``One-step persymmetric {GLRT} for subspace
  signals,'' \emph{IEEE Transaction on Signal Processing}, vol.~14, no.~67, pp.
  3639--3648, July 15 2019.

\bibitem{MaoGao19}
L.~Mao, Y.~Gao, S.~Yan, and L.~Xu, ``Persymmetric subspace detection in
  structured interference and non-homogeneous disturbance,'' \emph{IEEE Signal
  Processing Letters}, vol.~6, no.~26, pp. 928--932, June 2019.

\bibitem{LiuLiu20b}
J.~Liu, W.~Liu, B.~Tang, and D.~Orlando, ``Persymmetric adaptive detection in
  subspace interference plus {Gaussian} noise,'' \emph{Signal Processing}, vol.
  167, 2020.

\bibitem{LiuJian20}
J.~Liu, T.~Jian, W.~Liu, C.~Hao, and D.~Orlando, ``Persymmetric adaptive
  detection with improved robustness to steering vector mismatches,''
  \emph{Signal Processing}, vol. 176, p. 107669, 2020.

\bibitem{LiuLiu20AES}
J.~{Liu}, W.~{Liu}, C.~{Hao}, and D.~{Orlando}, ``Persymmetric subspace
  detectors with multiple observations in homogeneous environments,''
  \emph{IEEE Transactions on Aerospace and Electronic Systems}, vol.~56, no.~4,
  pp. 3276--3284, 2020.

\bibitem{ConteDeMaio03a}
E.~Conte and A.~De~Maio, ``Distributed target detection in compound-{Gaussian}
  noise with {Rao} and {Wald} tests,'' \emph{IEEE Transactions on Aerospace and
  Electronic Systems}, vol.~39, no.~2, pp. 568--582, 2003.

\bibitem{GaoLiao13}
Y.~Gao, G.~Liao, S.~Zhu, and D.~Yang, ``A persymmetric {GLRT} for adaptive
  detection in compound-{Gaussian} clutter with random texture,'' \emph{IEEE
  Signal Processing Letters}, vol.~20, no.~6, pp. 615--618, 2013.

\bibitem{GuoTao17}
X.~Guo, H.~Tao, H.-Y. Zhao, and J.~Liu, ``Persymmetric {Rao} and {Wald} tests
  for adaptive detection of distributed targets in compound-{G}aussian noise,''
  \emph{IET Radar, Sonar and Navigation}, vol.~11, no.~3, pp. 453--458, 2017.

\bibitem{LiuLiu18SP}
J.~Liu, S.~Liu, W.~Liu, S.~Zhou, S.~Zhua, and Z.-J. Zhang, ``Persymmetric
  adaptive detection of distributed targets in compound-{Gaussian} sea clutter
  with {Gamma} texture,'' \emph{Signal Processing}, vol. 152, pp. 340--349,
  2018.

\bibitem{LiuLi15}
J.~Liu, H.~Li, and B.~Himed, ``Persymmetric adaptive target detection with
  distributed {MIMO} radar,'' \emph{IEEE Transactions on Aerospace and
  Electronic Systems}, vol.~51, no.~1, pp. 372--382, January 2015.

\bibitem{LiuLiu18b}
J.~Liu, W.~Liu, J.~Han, B.~Tang, Y.~Zhao, and H.~Yang, ``Persymmetric {GLRT}
  detection in {MIMO} radar,'' \emph{IEEE Transactions on Vehicular
  Technology}, vol.~67, no.~12, pp. 11\,913--11\,923, December 2018.

\bibitem{LiuHan19}
J.~Liu, J.~Han, Z.-J. Zhang, and J.~Li, ``Target detection exploiting
  covariance matrix structures in {MIMO} radar,'' \emph{Signal Processing},
  vol. 154, pp. 174--181, January 2019.

\bibitem{LiuHan19a}
J.~Liu, J.~Han, W.~Liu, S.~Xu, and Z.-J. Zhang, ``Persymmetric {Rao} test for
  {MIMO} radar in {Gaussian} disturbance,'' \emph{Signal Processing}, vol. 165,
  pp. 30--36, 2019.

\bibitem{BillingsleyFarina99}
J.~B. Billingsley, A.~Farina, F.~Gini, M.~V. Greco, and L.~Verrazzani,
  ``Statistical analyses of measured radar ground clutter data,'' \emph{IEEE
  Transactions on Aerospace and Electronic Systems}, vol.~35, no.~2, pp.
  579--593, April 1999.

\bibitem{ConteDeMaio05}
E.~Conte, A.~{De Maio}, and A.~Farina, ``Statistical tests for higher order
  analysis of radar clutter: their application to {L}-band measured data,''
  \emph{IEEE Transactions on Aerospace and Electronic Systems}, vol.~41, no.~1,
  pp. 205--218, January 2005.

\bibitem{YanMassaro17}
S.~Yan, D.~Massaro, D.~Orlando, C.~Hao, and A.~Farina, ``Adaptive detection and
  range estimation of point-like targets with symmetric spectrum,'' \emph{IEEE
  Signal Processing Letters}, vol.~24, no.~11, pp. 1744--1748, November 2017.

\bibitem{FogliaHao17AES}
G.~Foglia, C.~Hao, A.~Farina, G.~Giunta, D.~Orlando, and C.~Hou, ``Adaptive
  detection of point-like targets in partially homogeneous clutter with
  symmetric spectrum,'' \emph{IEEE Transactions on Aerospace and Electronic
  Systems}, vol.~53, no.~4, pp. 2110--2119, 2017.

\bibitem{WangLi11a}
P.~Wang, H.~Li, and B.~Himed, ``Knowledge-aided parametric tests for
  multichannel adaptive signal detection,'' \emph{IEEE Transactions on Signal
  Processing}, vol.~12, no.~59, pp. 5970--5982, 2011.

\bibitem{WangSahinoglu12}
P.~Wang, Z.~Sahinoglu, M.~O. Pun, and H.~B. Li, ``Persymmetric parametric
  adaptive matched filter for multichannel adaptive signal detection,''
  \emph{IEEE Transactions on Signal Processing}, vol.~60, no.~6, pp.
  3322--3328, 2012.

\bibitem{GaoLiao15}
Y.~Gao, G.~Liao, S.~Zhou, and D.~Yang, ``Generalised persymmetric parametric
  adaptive coherence estimator for multichannel adaptive signal detection,''
  \emph{IET Radar, Sonar and Navigation}, vol.~9, no.~5, pp. 550--558, 2015.

\bibitem{GinolhacForster12}
G.~Ginolhac, P.~Forster, F.~Pascal, and J.-P. Ovarlez, ``Exploiting persymmetry
  for low-rank space time adaptive processing,'' in \emph{IEEE Proceedings of
  the 20th European Signal Processing Conference (EUSIPCO)}, 2012, pp. 81--85.

\bibitem{HaoOrlando16SPL}
C.~Hao, D.~Orlando, G.~Foglia, and G.~Giunta, ``Knowledge-based adaptive
  detection: {Joint} exploitation of clutter and system symmetry properties,''
  \emph{IEEE Signal Processing Letters}, vol.~23, no.~10, pp. 1489--1493, 2016.

\bibitem{FogliaHao17DSP_PHE}
G.~Foglia, C.~Hao, G.~Giunta, and D.~Orlando, ``Knowledge-aided adaptive
  detection in partially homogeneous clutter: {Joint} exploitation of
  persymmetry and symmetric spectrum,'' \emph{Digital Signal Processing},
  vol.~67, pp. 131 -- 138, 2017.

\bibitem{CarotenutoDeMaio19}
V.~Carotenuto, A.~{De Maio}, D.~Orlando, , and P.~Stoica, ``Radar detection
  architecture based on interference covariance structure classification,''
  \emph{IEEE Transactions on Aerospace and Electronic Systems}, vol.~55, no.~2,
  pp. 607--618, April 2019.

\bibitem{Klemm87}
R.~Klemm, ``Adaptive airborne {MTI}: An auxiliary channel approach,'' \emph{IEE
  Proceedings}, vol. 134, no.~3, pp. 269--276, 1987.

\bibitem{DiPietro92}
R.~C. DiPietro, ``Extended factored space-time processing for airborne radar
  systems,'' in \emph{the 25th Asilomar Conference on Signals, Systems and
  Computers}, 1992, pp. 425--430.

\bibitem{WangChen03}
Y.-L. Wang, J.-W. Chen, Z.~Bao, and Y.-N. Peng, ``Robust space-time adaptive
  processing for airborne radar in nonhomogeneous clutter environments,''
  \emph{IEEE Transactions on Aerospace and Electronic Systems}, vol.~39, no.~1,
  pp. 70--81, 2003.

\bibitem{BrownSchneible00}
R.~D. Brown, R.~A. Schneible, M.~C. Wicks, H.~Wang, and Y.~Zhang, ``{STAP} for
  clutter suppression with sum and difference beams,'' \emph{IEEE Transactions
  on Aerospace and Electronic Systems}, vol.~36, no.~2, pp. 634--646, 2000.

\bibitem{ZhangHe14}
W.~Zhang, Z.~He, J.~Li, H.~Liu, and Y.~Sun, ``A method for finding best
  channels in beam-space post-{Doppler} reduced-dimension {STAP},'' \emph{IEEE
  Transactions on Aerospace and Electronic Systems}, vol.~50, no.~1, pp.
  254--264, 2014.

\bibitem{CaiWu18AES}
Y.~{Cai}, X.~{Wu}, M.~{Zhao}, R.~C. {de Lamare}, and B.~{Champagne},
  ``Low-complexity reduced-dimension space-time adaptive processing for
  navigation receivers,'' \emph{IEEE Transactions on Aerospace and Electronic
  Systems}, vol.~54, no.~6, pp. 3160--3168, Dec 2018.

\bibitem{YangWang19SP}
Z.~Yang, Z.~Wang, W.~Liu, and R.~C. de~Lamare, ``Reduced-dimension space-time
  adaptive processing with sparse constraints on beam-doppler selection,''
  \emph{Signal Processing}, vol. 157, pp. 78--87, 2019.

\bibitem{WangCai94}
H.~Wang and L.~Cai, ``On adaptive spatial-temporal processing for airborne
  surveillance radar systems,'' \emph{IEEE Transactions on Aerospace and
  Electronic Systems}, vol.~30, no.~3, pp. 660--670, 1994.

\bibitem{AyoubHaimovich00}
T.~F. Ayoub and A.~M. Haimovich, ``Modified {GLRT} signal detection
  algorithm,'' \emph{IEEE Transactions on Aerospace and Electronic Systems},
  vol.~36, no.~3, pp. 810--818, 2000.

\bibitem{JinFriedlander05b}
Y.~Jin and B.~Friedlander, ``Reduced-rank adaptive detection of distributed
  sources using subarrays,'' \emph{IEEE Transactions on Signal Processing},
  vol.~53, no.~1, pp. 13--25, 2005.

\bibitem{Besson20SP}
O.~Besson, ``Adaptive detection using randomly reduced dimension generalized
  likelihood ratio test,'' \emph{Signal Processing}, vol. 166, p. 107265, 2020.

\bibitem{WangZhao19a}
Z.~Wang, Z.~Zhao, C.~Ren, Z.~Nie, and W.~Yang, ``Adaptive detection of
  point-like targets based on a reduced-dimensional data model,'' \emph{Signal
  Processing}, vol. 158, pp. 36--47, 2019.

\bibitem{Wang20b}
Z.~Wang, ``Distributed target detection using samples filtered with normalized
  conjugate signal steering vector,'' \emph{Circuits, Systems, and Signal
  Processing}, vol.~39, no.~9, pp. 4762--4774, 2020.

\bibitem{LiuLiu19JFI}
W.~Liu, J.~Liu, L.~Huang, Q.~Du, and Y.-L. Wang, ``Performance analysis of
  reduced-dimension subspace signal filtering and detection in sample-starved
  environment,'' \emph{Journal of the Franklin Institute}, vol. 356, no.~1, pp.
  629--653, 2019.

\bibitem{Wang20SPb}
Z.~Wang, ``Adaptive detection of multichannel signals without training data,''
  \emph{Signal Processing}, vol. 176, p. 107710, 2020.

\bibitem{LiStoica09book}
J.~Li and P.~Stoica, \emph{{MIMO} Radar Signal Processing}.\hskip 1em plus
  0.5em minus 0.4em\relax Hoboken: Wiley, 2009.

\bibitem{HaimovichBlum08SPM}
A.~M. Haimovich, R.~S. Blum, and L.~J. Cimini, ``{MIMO} radar with widely
  separated antennas,'' \emph{IEEE Signal Processing Magazine}, vol.~25, no.~1,
  pp. 116--129, 2008.

\bibitem{LiStoica07SPM}
J.~Li and P.~Stoica, ``{MIMO} radar with colocated antennas,'' \emph{IEEE
  Signal Processing Magazine}, vol.~24, no.~5, pp. 106--114, 2007.

\bibitem{FishlerHaimovich06TSP}
E.~Fishler, A.~Haimovich, R.~S. Blum, J.~Leonard J.~Cimini, D.~Chizhik, and
  R.~A. Valenzuela, ``Spatial diversity in radars--models and detection
  performance,'' \emph{IEEE Transactions on Signal Processing}, vol.~54, no.~3,
  pp. 823--838, 2006.

\bibitem{DuThompson08}
C.~Du, J.~S. Thompson, and Y.~Petillot, ``Predicted detection performance of
  {MIMO} radar,'' \emph{IEEE Signal Processing Letters}, vol.~15, pp. 83--86,
  2008.

\bibitem{TajerJajamovich10}
A.~Tajer, G.~H. Jajamovich, X.~Wang, and G.~V. Moustakides, ``Optimal joint
  target detection and parameter estimation by {MIMO} radar,'' \emph{IEEE
  Journal of Selected Topics in Signal Processing}, vol.~4, no.~1, pp.
  127--145, 2010.

\bibitem{AkcakayaNehorai10}
M.~Akcakaya and A.~Nehorai, ``{MIMO} radar detection and adaptive design under
  a phase synchronization mismatch,'' vol.~58, no.~10, pp. 4994--5005, October
  2010.

\bibitem{GogineniNehorai10TSP}
S.~Gogineni and A.~Nehorai, ``Polarimetric {MIMO} radar with distributed
  antennas for target detection,'' \emph{IEEE Transactions on Signal
  Processing}, vol.~58, no.~3, pp. 1689--1697, 2010.

\bibitem{SheikhiZamani08}
A.~Sheikhi and A.~Zamani, ``Temporal coherent adaptive target detection for
  multi-input multi-output radars in clutter,'' \emph{IET Radar, Sonar and
  Navigation}, vol.~2, no.~2, pp. 86--96, 2008.

\bibitem{LiuZhang13SP}
J.~Liu, Z.~J. Zhang, Y.~H. Cao, and S.~Y. Yang, ``A closed-form expression for
  false alarm rate of adaptive {MIMO-GLRT} detector with distributed {MIMO}
  radar,'' \emph{Signal Processing}, vol.~93, no.~9, pp. 2771--2776, 2013.

\bibitem{LiCui14}
N.~Li, G.~Cui, L.~Kong, and X.~Yang, ``{MIMO} radar moving target detection
  against compound-{Gaussian} clutter,'' \emph{Circuits, Systems, and Signal
  Processing}, vol.~33, no.~6, pp. 1819--1839, Jun 2014.

\bibitem{LiCui15SP}
N.~Li, G.~Cui, H.~Yang, L.~Kong, Q.~H. Liu, and S.~Iommelli, ``Adaptive
  detection of moving target with {MIMO} radar in heterogeneous environments
  based on {Rao} and {Wald} tests,'' \emph{Signal Processing}, vol. 114, pp.
  198--208, 2015.

\bibitem{LiCui15c}
N.~{Li}, G.~{Cui}, L.~{Kong}, and Q.~H. {Liu}, ``Moving target detection for
  polarimetric multiple-input multiple-output radar in {Gaussian} clutter,''
  \emph{IET Radar, Sonar Navigation}, vol.~9, no.~3, pp. 285--298, 2015.

\bibitem{HeLehmann10AES}
Q.~He, N.~H. Lehmann, R.~S. Blum, and A.~M. Haimovich, ``{MIMO} radar moving
  target detection in homogeneous clutter,'' \emph{IEEE Transactions on
  Aerospace and Electronic Systems}, vol.~46, no.~3, pp. 1290--1301, 2010.

\bibitem{WangLi11b}
P.~Wang, H.~Li, and B.~Himed, ``Moving target detection using distributed
  {MIMO} radar in clutter with nonhomogeneous power,'' \emph{IEEE Transactions
  on Signal Processing}, vol.~59, no.~10, pp. 4809--4820, October 2011.

\bibitem{AkcakayaNehorai11}
M.~{Akcakaya} and A.~{Nehorai}, ``{MIMO} radar sensitivity analysis for target
  detection,'' \emph{IEEE Transactions on Signal Processing}, vol.~59, no.~7,
  pp. 3241--3250, July 2011.

\bibitem{DeMaioLops08}
A.~De~Maio, M.~Lops, and L.~Venturino, ``Diversity-integration tradeoffs in
  {MIMO} detection,'' \emph{IEEE Transactions on Signal Processing}, vol.~56,
  no.~10, pp. 5051--5061, 2008.

\bibitem{DeMaioLops07}
A.~{De Maio} and M.~{Lops}, ``Design principles of {MIMO} radar detectors,''
  \emph{IEEE Transactions on Aerospace and Electronic Systems}, vol.~43, no.~3,
  pp. 886--898, July 2007.

\bibitem{NaghshModarresHashemi12}
M.~Naghsh and M.~Modarres-Hashemi, ``Exact theoretical performance analysis of
  optimum detector in statistical multi-input multi-output radars,'' \emph{IET
  Radar, Sonar and Navigation}, vol.~6, no.~2, pp. 99--111, 2012.

\bibitem{LiCui12}
N.~{Li}, G.~{Cui}, L.~{Kong}, and X.~{Yang}, ``Rao and {Wald} tests design of
  multiple-input multiple-output radar in compound-{Gaussian} clutter,''
  \emph{IET Radar, Sonar Navigation}, vol.~6, no.~8, pp. 729--738, October
  2012.

\bibitem{ZhangCui13SPL}
T.~Zhang, G.~Cui, L.~Kong, and X.~Yang, ``Adaptive {Bayesian} detection using
  {MIMO} radar in spatially heterogeneous clutter,'' \emph{IEEE Signal
  Processing Letters}, vol.~20, no.~6, pp. 547--550, 2013.

\bibitem{LiYang19SP}
N.~Li, H.~Yang, G.~Cui, L.~Kong, and Q.~H. Liu, ``Adaptive two-step {Bayesian}
  {MIMO} detectors in compound-{Gaussian} clutter,'' \emph{Signal Processing},
  vol. 161, pp. 1--13, 2019.

\bibitem{CuiKong12d}
G.~Cui, L.~Kong, and X.~Yang, ``{GLRT}-based detection algorithm for
  polarimetric {MIMO} radar against {SIRV} clutter,'' \emph{Circuits, Systems,
  and Signal Processing}, vol.~31, no.~3, pp. 1033--1048, 2012.

\bibitem{CuiKong12c}
G.~Cui, L.~Kong, X.~Yang, and J.~Yang, ``Distributed target detection with
  polarimetric {MIMO} radar in compound-{Gaussian} clutter,'' \emph{Digital
  Signal Processing}, vol.~22, no.~3, pp. 430--438, 2012.

\bibitem{KongCui11}
L.~{Kong}, G.~{Cui}, X.~{Yang}, and J.~{Yang}, ``Rao and {Wald} tests design of
  polarimetric multiple-input multiple-output radar in compound-{Gaussian}
  clutter,'' \emph{IET Signal Processing}, vol.~5, no.~1, pp. 85--96, Feb 2011.

\bibitem{CuiKong12}
G.~Cui, L.~Kong, X.~Yang, and J.~Yang, ``The {Rao} and {Wald} tests designed
  for distributed targets with polarization {MIMO} radar in compound-{Gaussian}
  clutter,'' \emph{Circuits, Systems, and Signal Processing}, vol.~31, no.~1,
  pp. 237--254, 2012.

\bibitem{BekkermanTabrikian06}
I.~Bekkerman and J.~Tabrikian, ``Target detection and localization using {MIMO}
  radars and sonars,'' \emph{IEEE Transactions on Signal Processing}, vol.~54,
  no.~10, pp. 3873--3883, 2006.

\bibitem{CuiKong12b}
G.~Cui, L.~Kong, and X.~Yang, ``Performance analysis of colocated {MIMO} radars
  with randomly distributed arrays in compound-{Gaussian} clutter,''
  \emph{Circuits, Systems, and Signal Processing}, vol.~31, no.~4, pp.
  1407--1422, 2012.

\bibitem{LiXu08TSP}
J.~Li, L.~Z. Xu, P.~Stoica, K.~W. Forsythe, and D.~W. Bliss, ``Range
  compression and waveform optimization for {MIMO} radar: A {Cramer}-{Rao}
  bound based study,'' \emph{IEEE Transactions on Signal Processing}, vol.~56,
  no.~1, pp. 218--232, 2008.

\bibitem{XuLi08AES}
L.~Xu, J.~Li, and P.~Stoica, ``Target detection and parameter estimation for
  {MIMO} radar systems,'' \emph{IEEE Transactions on Aerospace and Electronic
  Systems}, vol.~44, no.~3, pp. 927--939, 2008.

\bibitem{LiuWang15MIMO}
W.~Liu, Y.~Wang, J.~Liu, W.~Xie, H.~Chen, and W.~Gu, ``Adaptive detection
  without training data in colocated {MIMO} radar,'' \emph{IEEE Transactions on
  Aerospace and Electronic Systems}, vol.~51, no.~3, pp. 2469--2479, 2015.

\bibitem{LiuLi19b}
J.~{Liu} and J.~{Li}, ``Robust detection in {MIMO} radar with steering vector
  mismatches,'' \emph{IEEE Transactions on Signal Processing}, vol.~67, no.~20,
  pp. 5270--5280, Oct 2019.

\bibitem{FortunatiSanguinetti20}
S.~{Fortunati}, L.~{Sanguinetti}, F.~{Gini}, M.~S. {Greco}, and B.~{Himed},
  ``Massive {MIMO} radar for target detection,'' \emph{IEEE Transactions on
  Signal Processing}, vol.~68, pp. 859--871, 2020.

\bibitem{LanMarino20early}
L.~{Lan}, A.~{Marino}, A.~{Aubry}, A.~{De Maio}, G.~{Liao}, J.~{Xu}, and
  Y.~{Zhang}, ``{GLRT}-based adaptive target detection in {FDA-MIMO} radar,''
  \emph{IEEE Transactions on Aerospace and Electronic Systems}, 2020,
  DOI:10.1109/TAES.2020.3028485.

\bibitem{HassanienVorobyov10}
A.~{Hassanien} and S.~A. {Vorobyov}, ``Phased-{MIMO} radar: {A} tradeoff
  between phased-array and {MIMO} radars,'' \emph{IEEE Transactions on Signal
  Processing}, vol.~58, no.~6, pp. 3137--3151, June 2010.

\bibitem{FuhrmannBrowning10}
D.~R. {Fuhrmann}, J.~P. {Browning}, and M.~{Rangaswamy}, ``Signaling strategies
  for the hybrid {MIMO} phased-array radar,'' \emph{IEEE Journal of Selected
  Topics in Signal Processing}, vol.~4, no.~1, pp. 66--78, Feb 2010.

\bibitem{LiHimed10}
H.~{Li} and B.~{Himed}, ``Transmit subaperturing for {MIMO} radars with
  co-located antennas,'' \emph{IEEE Journal of Selected Topics in Signal
  Processing}, vol.~4, no.~1, pp. 55--65, Feb 2010.

\bibitem{XuDai11}
J.~{Xu}, X.~{Dai}, X.~{Xia}, L.~{Wang}, J.~{Yu}, and Y.~{Peng}, ``Optimizations
  of multisite radar system with {MIMO} radars for target detection,''
  \emph{IEEE Transactions on Aerospace and Electronic Systems}, vol.~47, no.~4,
  pp. 2329--2343, October 2011.

\bibitem{ChenZheng17}
P.~{Chen}, L.~{Zheng}, X.~{Wang}, H.~{Li}, and L.~{Wu}, ``Moving target
  detection using colocated {MIMO} radar on multiple distributed moving
  platforms,'' \emph{IEEE Transactions on Signal Processing}, vol.~65, no.~17,
  pp. 4670--4683, Sep. 2017.

\bibitem{ChaoChen12AEU}
S.~Chao, B.~Chen, and C.~Li, ``Grid cell based detection strategy for {MIMO}
  radar with widely separated subarrays,'' \emph{AEU-International Journal of
  Electronics and Communications}, vol.~66, no.~9, pp. 741--751, 2012.

\bibitem{WangLi13TSP}
P.~Wang, H.~Li, and B.~Himed, ``A parametric moving target detector for
  distributed {MIMO} radar in non-homogeneous environment,'' \emph{IEEE
  Transactions on Signal Processing}, vol.~61, no.~9, pp. 2282--2294, 2013.

\bibitem{LiWang15}
H.~{Li}, Z.~{Wang}, J.~{Liu}, and B.~{Himed}, ``Moving target detection in
  distributed {MIMO} radar on moving platforms,'' \emph{IEEE Journal of
  Selected Topics in Signal Processing}, vol.~9, no.~8, pp. 1524--1535, Dec
  2015.

\bibitem{ZhangLiu15}
Z.-J. Zhang, J.~Liu, Y.~Zhao, and Y.~Cao, ``False alarm rate of the {GLRT-LQ}
  detector in non-{Gaussian} and heterogeneous clutter,'' \emph{Aerospace
  Science and Technology}, vol.~47, pp. 191 -- 194, 2015.

\bibitem{XuLi07TSP}
L.~Xu and J.~Li, ``Iterative generalized-likelihood ratio test for {MIMO}
  radar,'' \emph{IEEE Transactions on Signal Processing}, vol.~55, no.~6, pp.
  2375--2385, 2007.

\bibitem{KongCui10}
L.~Kong, G.~Cui, X.~Yang, and J.~Yang, ``Adaptive detector design of {MIMO}
  radar with unknown covariance matrix,'' \emph{Journal of Systems Engineering
  and Electronics}, vol.~21, no.~6, pp. 954--960, 2010.

\bibitem{CuiKong10RSN}
G.~Cui, L.~Kong, and X.~Yang, ``Multiple-input multiple-output radar detectors
  design in non-{Gaussian} clutter,'' \emph{IET Radar, Sonar and Navigation},
  vol.~4, no.~5, pp. 724--732, 2010.

\bibitem{WangJiang11}
J.~{Wang}, S.~{Jiang}, J.~{He}, Z.~{Liu}, and C.~J. {Baker}, ``Adaptive
  subspace detector for multi-input multi-output radar in the presence of
  steering vector mismatch,'' \emph{IET Radar, Sonar Navigation}, vol.~5,
  no.~1, pp. 23--31, January 2011.

\bibitem{GerlachSteiner97}
K.~Gerlach, M.~Steiner, and F.~C. Lin, ``Detection of a spatially distributed
  target in white noise,'' \emph{IEEE Signal Processing Letters}, vol.~4,
  no.~7, pp. 198 -- 200, 1997.

\bibitem{GiniBordoni04}
F.~Gini, F.~Bordoni, and A.~Farina, ``Multiple radar targets detection by
  exploiting induced amplitude modulation,'' \emph{IEEE Transactions on Signal
  Processing}, vol.~52, no.~4, pp. 903--913, 2004.

\bibitem{CarotenutoDeMaio17}
V.~{Carotenuto}, A.~{De Maio}, D.~{Orlando}, and P.~{Stoica}, ``Model order
  selection rules for covariance structure classification in radar,''
  \emph{IEEE Transactions on Signal Processing}, vol.~65, no.~20, pp.
  5305--5317, Oct 2017.

\bibitem{LiuWang14b}
W.~Liu, L.~Wang, Y.~Di, T.~Jian, and D.~Xie, ``Adaptive energy detector and its
  application for mismatched signal detection,'' \emph{Journal of Radars},
  vol.~4, no.~2, pp. 149--159, 2014.

\bibitem{Muirhead05}
R.~J. Muirhead, \emph{Aspects of Multivariate Statistical Theory},
  2nd~ed.\hskip 1em plus 0.5em minus 0.4em\relax Hoboken: Wiley, 2005.

\bibitem{Anderson03}
T.~W. Anderson, \emph{An Introduction to Multivariate Statistical Analysis},
  3rd~ed.\hskip 1em plus 0.5em minus 0.4em\relax Hoboken: Wiley, 2003.

\bibitem{LongLiang19}
T.~Long, Z.~Liang, and Q.~Liu, ``Advanced technology of high-resolution radar:
  target detection, tracking, imaging, and recognition,'' \emph{Science China:
  Information Sciences}, vol.~62, no.~4, p. 40301, 2019.

\bibitem{Liu20SCIS}
W.~Liu, F.~Gao, Y.~Luo, J.~Liu, and Y.-L. Wang, ``Wald tests for signal
  detection when uncertainty exists in a target's spatial-temporal steering
  vector,'' \emph{Science China: Information Sciences}, vol.~63, no.~11, p.
  189304, 2020.

\bibitem{LiSong12}
Y.~Li, R.~Song, and W.~Wang, ``Particle swarm optimization of compression
  measurement for signal detection,'' \emph{Circuits, Systems, and Signal
  Processing}, vol.~31, no.~3, pp. 1109--1126, Jun 2012.

\bibitem{WangLiu15}
Y.-G. Wang, Z.~Liu, L.~Yang, and W.-L. Jiang, ``Generalized compressive
  detection of stochastic signals using {Neyman}--{Pearson} theorem,''
  \emph{Signal, Image and Video Processing}, vol.~9, no.~1, pp. 111--120, Dec
  2015.

\bibitem{RazaviValkama16}
A.~{Razavi}, M.~{Valkama}, and D.~{Cabric}, ``Compressive detection of random
  subspace signals,'' \emph{IEEE Transactions on Signal Processing}, vol.~64,
  no.~16, pp. 4166--4179, Aug 2016.

\bibitem{WimalajeewaVarshney17}
T.~{Wimalajeewa} and P.~K. {Varshney}, ``Sparse signal detection with
  compressive measurements via partial support set estimation,'' \emph{IEEE
  Transactions on Signal and Information Processing over Networks}, vol.~3,
  no.~1, pp. 46--60, March 2017.

\bibitem{WimalajeewaVarshney18}
------, ``Compressive sensing-based detection with multimodal dependent data,''
  \emph{IEEE Transactions on Signal Processing}, vol.~66, no.~3, pp. 627--640,
  Feb 2018.

\bibitem{MaGan19}
J.~Ma, L.~Gan, H.~Liao, and I.~Zahid, ``Sparse signal detection without
  reconstruction based on compressive sensing,'' \emph{Signal Processing}, vol.
  162, pp. 211 -- 220, 2019.

\bibitem{ZhangSward19}
X.~{Zhang}, J.~{Sw{\"a}rd}, H.~{Li}, A.~{Jakobsson}, and B.~{Himed}, ``A
  sparsity-based passive multistatic detector,'' \emph{IEEE Transactions on
  Aerospace and Electronic Systems}, vol.~55, no.~6, pp. 3658--3666, 2019.

\bibitem{CarotenutoOrlando19}
V.~{Carotenuto}, D.~{Orlando}, and A.~{Farina}, ``Interference covariance
  matrix structure classification in heterogeneous environment,'' \emph{IEEE
  Signal Processing Letters}, vol.~26, no.~10, pp. 1491--1495, Oct 2019.

\bibitem{LiuBiondi19}
J.~{Liu}, F.~{Biondi}, D.~{Orlando}, and A.~{Farina}, ``Training data
  classification algorithms for radar applications,'' \emph{IEEE Signal
  Processing Letters}, vol.~26, no.~10, pp. 1446--1450, Oct 2019.

\bibitem{ColucciaFascista20TSP}
A.~{Coluccia}, A.~{Fascista}, and G.~{Ricci}, ``{CFAR} feature plane: {A} novel
  framework for the analysis and design of radar detectors,'' \emph{IEEE
  Transactions on Signal Processing}, vol.~68, pp. 3903--3916, 2020.

\bibitem{ColucciaFascista20SP}
A.~Coluccia, A.~Fascista, and G.~Ricci, ``A {K}-nearest neighbors approach to
  the design of radar detectors,'' \emph{Signal Processing}, vol. 174, p.
  107609, 2020.

\bibitem{ZaimbashiLi20}
A.~{Zaimbashi} and J.~{Li}, ``Tunable adaptive target detection with kernels in
  colocated {MIMO} radar,'' \emph{IEEE Transactions on Signal Processing},
  vol.~68, pp. 1500--1514, 2020.

\bibitem{Gerlach99}
K.~Gerlach, ``Spatially distributed target detection in non-{Gaussian}
  clutter,'' \emph{IEEE Transactions on Aerospace and Electronic Systems},
  vol.~35, no.~3, pp. 926 -- 934, 1999.

\bibitem{GiniGreco02a}
F.~Gini and M.~Greco, ``Covariance matrix estimation for {CFAR} detection in
  correlated heavy tailed clutter,'' \emph{Signal Processing}, vol.~82, no.~12,
  pp. 1847--1859, 2002.

\bibitem{SangstonGini12}
K.~J. Sangston, F.~Gini, and M.~S. Greco, ``Coherent radar target detection in
  heavy-tailed compound-{Gaussian} clutter,'' \emph{IEEE Transactions on
  Aerospace and Electronic Systems}, vol.~64, no.~76, pp. 64--76, 2012.

\end{thebibliography}
}
\end{document}